\documentclass[acmsmall,screen,nonacm]{acmart}

\setcopyright{none}
\renewcommand\footnotetextcopyrightpermission[1]{}

\usepackage{url}
\usepackage{xspace}   %

\usepackage{thmtools}

\usepackage{stmaryrd}
\usepackage{mathtools}  %

\usepackage{listings}
\lstdefinelanguage{js}{
	keywords={const, extends, super, class, export, boolean, throw, implements, import, this, typeof, new, true, false, catch, function, return, null, catch, switch, var, if, in, while, do, else, case, break, let, type, box, module, as, sig, val, end, struct},
	identifierstyle=\color{black},
	commentstyle=\color{black}\ttfamily,
	comment=[l]{//},
	morecomment=[s]{/*}{*/},
	morestring=[b]',
	xleftmargin = 2mm,
	morestring=[b]"
}

\usepackage[euler-digits,euler-hat-accent]{eulervm}

\usepackage{mathpartir}  %

\usepackage{trimclip}

\usepackage{xcolor}

\usepackage{graphicx}
\usepackage{tikz}
\usetikzlibrary{arrows.meta}

\newcommand{\fatcomma}{\mathbin{\clipbox*{0 -1ex {\width} .66ex}{$\fatsemi$}}}

\newcommand{\eframe}{[ \; ]}
\newcommand{\size}[2]{|#1|^{#2}}
\newcommand{\mkctx}[2]{#1[#2]}

\newcommand{\bra}[1]{\llbracket #1 \rrbracket}

\newcommand{\rl}[1]{{\normalfont\scshape #1}}

\newcommand{\ranchor}[1]{\hypertarget{rule:#1}{\rl{#1}}}

\makeatletter
\newcommand{\drule}{\@ifstar\drule@plain\drule@anchored}
\newcommand{\drule@anchored}[3]{%
  \inferrule*[narrower=0.3,lab=\ranchor{#1}\space]{ #2}{ #3}}
\newcommand{\drule@plain}[3]{%
  \inferrule*[narrower=0.3,lab=\rl{#1}\space]{ #2}{ #3}}
\makeatother

\newenvironment{drulepar}[2]
  {\trivlist\item
   \fbox{$#1$}\hfill\emph{(#2)}\par
   \nopagebreak[4]%
   \noindent
   \begin{mathparpagebreakable}}
  {\end{mathparpagebreakable}%
   \endtrivlist}

\makeatletter
\newcommand{\rref}{\@ifstar{\rref@{}}{\rref@{rule~}}}
\newcommand{\Rref}{\@ifstar{\rref@{}}{\rref@{Rule~}}}
\newcommand{\rref@}[2]{#1\hyperlink{rule:#2}{\rl{#2}}}
\makeatother

\newcommand{\tint}{\mathsf{Int}}                 %
\newcommand{\tstar}{\mathsf{\star}}              %
\newcommand{\tshift}[1]{{\uparrow}#1}            %
\newcommand{\tlist}[1]{\mathsf{List}\langle#1\rangle}
\newcommand{\envplus}{\, \texttt{+} \,}          %
\newcommand{\eclos}[2]{\langle#1 \mid #2 \rangle} %

\newcommand{\numv}[1]{\lVert #1 \rVert}
\newcommand{\sig}[1]{\mathsf{sig}(#1)}
\newcommand{\uniq}[1]{\mathsf{uniq}(#1)}
\newcommand{\decl}[3]{\mathsf{decl}(#1; #2; #3)}
\newcommand{\sstruct}[1]{\mathsf{struct} \; #1}
\newcommand{\ssbox}[1]{\mathsf{box} \; #1}
\newcommand{\selab}[4]{#1 \vdash #2 \rightsquigarrow #3 : #4}
\newcommand{\sexp}[8]{#1 ;\, #2 \,{:}\, #3 \vdash #4 : #5
  \xrightarrow{\; #6 \;} #7 : #8}

\newcommand{\hforall}[2]{\forall{\color{gray}#1.\,}#2}          %
\newcommand{\hLam}[2]{\Lambda{\color{gray}#1.\,}#2}              %
\newcommand{\hlam}[2]{\lambda{\color{gray}#1.\,}#2}              %
\newcommand{\hmani}[3]{[ {\color{gray}#1 \,\texttt{=}\, } #2 ] #3} %
\newcommand{\hdef}[2]{[ {\color{gray} #1 \,\texttt{=} \,} #2 ]}   %
\newcommand{\heq}[2]{{\color{gray} #1 \,\texttt{=} \,} #2}        %
\newcommand{\hdecl}[2]{{\color{gray}#1:\,}#2}                     %

\usepackage{etoolbox}
\csundef{newlist}
\csundef{renewlist}
\usepackage{enumitem}
\hypersetup{hypertexnames=false}
\setitemize{leftmargin=*,noitemsep,topsep=0pt,parsep=0pt,partopsep=0pt}

\appto\normalsize{%
  \setlength\abovedisplayskip{3pt plus 1pt minus 2pt}%
  \setlength\belowdisplayskip{3pt plus 1pt minus 2pt}%
  \setlength\abovedisplayshortskip{1.5pt plus 1pt minus 1pt}%
  \setlength\belowdisplayshortskip{2pt plus 1pt minus 1pt}%
}
\makeatletter
\def\MathparLineskip{\lineskiplimit=0.35em\lineskip=0.35em plus 0.1em minus 0.05em}
\makeatother
\makeatletter
\newcommand\FEthmskip{%
  \thm@preskip=1.5\p@ \thm@postskip=1.5\p@}
\apptocmd{\th@acmplain}{\FEthmskip}{}{}
\apptocmd{\th@acmdefinition}{\FEthmskip}{}{}
\makeatother

\newcommand{\name}{$\mathit{F}_{\mathit{E}}$\xspace}
\newcommand{\lambdae}{$\lambda_{E}$\xspace}
\newcommand{\eical}{$\mathsf{E}_{i}$\xspace}
\newcommand{\jlambdae}{$\lambda\varepsilon$\xspace}
\newcommand{\lambdaenv}{$\lambda^{\to}_{\text{env}}$\xspace}

\newcommand{\appHints}{\ref*{sec:hints}}
\newcommand{\appAux}{\ref*{sec:aux}}
\newcommand{\appErase}{\ref*{sec:erase}}
\newcommand{\appDecide}{\ref*{sup:decide}}
\newcommand{\appSem}{\ref*{sec:sem_sound}}
\newcommand{\appSource}{\ref*{sec:source}}

\begin{document}

\title[Polymorphic First-Class Environments]{Polymorphic First-Class Environments}       %

\author{Jinhao Tan}
\email{jhtan@cs.hku.hk}
\orcid{0009-0005-5779-8409}
\affiliation{%
 \institution{The University of Hong Kong}
 \city{Hong Kong}
 \country{China}
}

\author{Jam Kabeer Ali Khan}
\email{jamkhan@connect.hku.hk}
\orcid{0009-0007-8505-9667}
\affiliation{%
  \institution{The University of Hong Kong}
  \city{Hong Kong}
  \country{China}
}

\author{Bruno C. d. S. Oliveira}
\email{bruno@cs.hku.hk}
\orcid{0000-0002-1846-7210}
\affiliation{%
 \institution{The University of Hong Kong}
 \city{Hong Kong}
 \country{China}
}

\begin{abstract}
First-class environments expose runtime environments as ordinary values and
support a substitution-free, environment-based operational semantics. They also
support applications such as first-class modules, dynamic composition of
environment fragments, and sandboxing via local environments. Existing
environment-based calculi, however, are mostly studied in a simply typed setting and
stop short of System~F and
polymorphism.

We present \name, a polymorphic calculus with first-class environments that
conservatively extends System~F with a small-step
environment semantics. \name eliminates all forms of substitution in System~F.
Term and type abstractions evaluate to closures, boxes run code under local
environments, and box types expose the local typing assumptions of both.
Manifest types, together with a two-sided type equivalence, play a dual role:
they model type-level instantiation, and enable
transparent type components in environments and modules. We prove type soundness,
normalization, decidability and standard properties of type equivalence,
and conservativity over System~F. The metatheory of \name is based on \emph{instantiation lemmas}
rather than substitution lemmas. This foundation supports polymorphic
first-class modules with dynamic composition and sandboxing. We study \name in
two forms. Our main presentation is an explicit, type-passing calculus that
follows conventional System~F. We also develop an implicitly-typed variant aligned with
the semantics of polymorphic functional languages. All
metatheory is mechanized in the Rocq theorem prover.
\end{abstract}

\maketitle

\section{Introduction}

Runtime environments are central to programming language
implementations. They record the bindings used to execute code, but
programmers usually cannot manipulate them directly in the language. \emph{First-class
environments}~\cite{GelernterJL87,MillerR91,QueinnecDeRoure1996,Shutt2010Fexprs,SatoSakuraiBurstall2001,SatoSakuraiKameyama2002,Nishizaki1994,DBLP:journals/lisp/Nishizaki00,tan2023dependent}
turn these environments into ordinary values that can be created,
passed, inspected, and composed. Once environments become first-class,
they can serve not only as a runtime device, but also as a programming
construct for controlling the context in which code executes.

Recent work shows that first-class environments already support several useful forms of
modularity. In \lambdae~\cite{DBLP:journals/pacmpl/TanO24}, environment
fragments can act as first-class modules and can be composed dynamically
through dependent merges~\cite{tan2023dependent}. Boxes of the form
$e_1 \triangleright e_2$ evaluate $e_2$ under the local environment $e_1$,
giving a direct account of sandboxing. Because the supplied environment
determines which bindings and resources are available during execution, the
same mechanism also supports \emph{capability-style modularity}, making
authority explicit~\cite{DBLP:conf/fc/MillerMF00,miller2006robust,miller2008safe,DBLP:conf/ecoop/MelicherSPA17}.

First-class environments are also attractive semantically. Standard
presentations of the lambda calculus and System~F organize reduction and
metatheory around capture-avoiding substitution, while implementations execute
programs with closures and environments. This mismatch has motivated several
lines of work that aim to eliminate or internalize meta-substitution, from
\emph{explicit substitutions}~\cite{abadiexplicit,DBLP:journals/jacm/CurienHL96} to
explicit environments and environment-based
semantics~\cite{SatoSakuraiBurstall2001,Nishizaki1994,DBLP:journals/lisp/Nishizaki00,DBLP:journals/pacmpl/TanO24}.
In an environment-based account, bindings are carried explicitly at runtime
rather than propagated by substitution. Such a semantics is closer to
implementations and can also be attractive for mechanization.

For calculi with first-class environments, the simply typed case is now
reasonably well understood, but the polymorphic case remains a key
challenge. \lambdae, for instance, is conservative over the Simply Typed
Lambda Calculus and stops short of System~F.
Extending first-class environments from the simply typed setting to System~F
requires more than adding universal quantification. In a standard presentation
of System~F, substitution appears in three places: term beta-reduction,
type beta-reduction, and type instantiation at type application. An
environment-based semantics already gives a natural account of term
substitution through closures, and the same operational idea can be used for
type abstractions. But this polymorphic setting introduces further typing
problems as well: closures and boxes must now carry enough information to track
local typing assumptions compositionally.

This paper presents \name: a calculus with polymorphism and first-class environments.
\name introduces two central devices. \emph{Box types} expose the local typing
assumptions needed to type closures and boxes in the presence of polymorphism,
so polymorphic environments can be typed compositionally. \emph{Manifest
types} have a dual role. Semantically, they model type-level
instantiation. The same mechanism is
also what supports manifest type components in modules, as found in ML-style module
systems~\cite{MacQueen1984Modules,Milner1997Definition,LeroyPOPL94Manifest,RossbergDreyerMixML,Rossberg2018OneML}
and related designs with type members~\cite{AminRompfOderskyOOPSLA14}.
Unlike ML modules, \name supports only \emph{transparent} type components,
while sealing is left to future work (Section~\ref{subsec:sealing}).
Manifest types are the \emph{sole} mechanism for type instantiation
in \name, avoiding a second, meta-level mechanism that would have to
interact with type equivalence.
\name preserves
a substitution-free, environment-based account
and remains conservative over System~F.

The resulting metatheory differs substantially from standard
developments of System~F. Traditional System~F metatheory relies
heavily on substitution lemmas. In \name these are replaced by
\emph{instantiation lemmas}, which describe how typing is preserved
when moving between universal types and their manifest
instantiations. Combined with de Bruijn indices and an
environment-based semantics, this yields a direct mechanization in
Rocq. We develop \name in two forms. The first form is an explicit,
type-passing calculus, which is our main focus in the paper. It follows the
conventional presentation of System~F, with explicit type abstractions
and applications. Although this calculus passes types at run time,
types are computationally irrelevant: we prove that a type-erasing
semantics reaches the same results. The second form is an
implicitly-typed variant that
aligns with the semantics of polymorphic functional languages.

\name supports capability-style
modules that combine dynamic module composition with polymorphism and
manifest type components. Environment fragments remain first-class
and compositional, supporting dynamic assembly, while boxes run code
under explicit local environments, supporting sandboxing. Manifest
type components make type information transparently available across
module boundaries, so this form of capability-style modularity extends
beyond the simply typed setting.
\name isolates how first-class environments scale to
polymorphism through explicit local typing assumptions, manifest type
components, and type equivalence. \name is thus a System~F-level core for
polymorphic instantiation, dynamic composition, and sandboxing through
first-class environments.

\textbf{Contributions.} This paper makes the following contributions:
\begin{itemize}
\item A polymorphic first-class-environment calculus that gives a uniform
      object-level account of polymorphic instantiation, transparent type
      components, and environment-based execution, while conservatively
      extending System~F with a substitution-free small-step semantics.
\item Box types for explicit local typing assumptions, and manifest types with
      two-sided type equivalence, which express type-level instantiation
      without substitution and support transparent type components in
      environments and modules.
\item Type soundness, normalization, and decidable type equivalence (with reflexivity,
      symmetry, and transitivity), conservativity over System~F, and a
      substitution-free metatheory based on instantiation lemmas
      rather than substitution lemmas.
\item A full Rocq mechanization of the metatheory, covering the
      explicit calculus, an implicitly-typed variant aligned with the
      semantics of polymorphic functional languages, and type-preserving
      elaboration of simple form of first-class modules
      into \name, plus a prototype implementation.
\end{itemize}

\smallskip
\noindent
The mechanization and implementation are available
online.\footnote{\url{https://github.com/tjhao/PolyFirstClassEnvs}}
The examples can be run and modified in our online
playground:
\url{https://jam-khan.github.io/EnvML/}

\section{Overview}

This section provides an overview of our work.

\subsection{Background: First-class Environments and \lambdae}
The \lambdae calculus~\cite{DBLP:journals/pacmpl/TanO24} is a call-by-value statically typed calculus that provides first-class environments for the Simply Typed Lambda Calculus (STLC). The principle of introducing first-class environments in \lambdae is to model typing contexts as types, and runtime environments as values. Therefore, the \emph{box} expression $e_{\mathrm{1}} \triangleright e_{\mathrm{2}}$, which runs an expression $e_{\mathrm{2}}$ under an environment $e_{\mathrm{1}}$, can be easily supported in \lambdae.
Moreover, unlike conventional calculi that typically employ a small-step semantics with substitution, \lambdae adopts a substitution-free environment-based semantics. As a result, de Bruijn indices~\cite{de1972lambda} for variable representation work well in \lambdae, as one does not need to define or reason about substitution. To incorporate the environment-based semantics, \lambdae also supports \emph{function closures} $\eclos{\delta}{\lambda e}$, which represents a lambda abstraction $\lambda e$ closed under the environment $\delta$. Since de Bruijn indices are used in \lambdae the binders are \emph{nameless} (i.e. without an explicit variable name). The typing rules of boxes and closures are defined as follows:
\begin{mathpar}
  \inferrule [Typ-box]
          {\Gamma \vdash e_{\mathrm{1}} : \Gamma_{\mathrm{1}} \\ \Gamma_{\mathrm{1}} \vdash e_{\mathrm{2}} : A }
          {\Gamma \vdash e_{\mathrm{1}} \triangleright e_{\mathrm{2}} : A}
  \quad \quad \quad \quad \quad
  \inferrule [Typ-clos]
          {\cdot \vdash \delta : \Gamma_{\mathrm{1}} \\ \Gamma_{\mathrm{1}}, A \vdash e : B }
          {\Gamma \vdash \eclos{\delta}{\lambda e} : A \rightarrow B}
\end{mathpar}
To type check a box expression $e_{\mathrm{1}} \triangleright e_{\mathrm{2}}$, we first type check the
environment $e_{\mathrm{1}}$ against its type $\Gamma_{\mathrm{1}}$, which represents the
typing context for $e_{\mathrm{2}}$. Then, we type check the body expression
$e_{\mathrm{2}}$ under the context $\Gamma_{\mathrm{1}}$. Similarly, the type of the value
environment $\delta$ in a closure $\eclos{\delta}{\lambda e}$ provides the typing
context for the function $\lambda e$. Note that in a box or closure, the
body expression may bind variables from the local context $\Gamma_{\mathrm{1}}$.
However, in a setting with simple types, the type of the body
expression does not depend on $\Gamma_{\mathrm{1}}$, as types do not contain type
variables in \lambdae.

\paragraph{Small-step environment semantics}
While environment-based semantics is commonly presented with a
big-step semantics~\cite{kahn1987natural}, \lambdae employs a
small-step semantics. In particular, just like term application in
STLC, box evaluation in \lambdae also follows left-to-right
reduction:
\begin{mathpar}
  \inferrule [\ranchor{Step-boxl}]
          {\delta \vdash e_{\mathrm{1}} \hookrightarrow e'_{\mathrm{1}} }
          {\delta \vdash e_{\mathrm{1}} \triangleright e_{\mathrm{2}} \hookrightarrow e'_{\mathrm{1}} \triangleright e_{\mathrm{2}}}
  \quad \quad \quad
  \inferrule [\ranchor{Step-boxr}]
          {\delta_{\mathrm{1}} \vdash e_{\mathrm{2}} \hookrightarrow e'_{\mathrm{2}} }
          {\delta \vdash \delta_{\mathrm{1}} \triangleright e_{\mathrm{2}} \hookrightarrow \delta_{\mathrm{1}} \triangleright e'_{\mathrm{2}}}
  \quad \quad \quad
  \inferrule [Step-boxv]
          { }
          {\delta \vdash \delta_{\mathrm{1}} \triangleright v \hookrightarrow v}
\end{mathpar}
The evaluation of boxes fits well with the environment-based reduction
$\delta \vdash e \hookrightarrow e'$. The local environment $e_{\mathrm{1}}$ is evaluated to a
value environment $\delta_{\mathrm{1}}$ under the global environment $\delta$ first (\rref{Step-boxl}).
As seen in \rref{Step-boxr} the global environment
$\delta$ is replaced by $\delta_{\mathrm{1}}$, when evaluating $e_{\mathrm{2}}$. Therefore $e_{\mathrm{2}}$ cannot refer to bindings
in $\delta$ and boxes provide a \emph{sandboxing} mechanism
where computations execute within an isolated and controlled
environment $\delta_{\mathrm{1}}$.
In \rref{Step-boxv}, when both the environment and body
are values, the box reduces to the body's value.

Importantly, besides enabling first-class environments, the box construct is the key device that naturally enables the environment-based semantics to be small-step in \lambdae. A standard big-step environment-based semantics specifies the
entire transition from an expression to a final value in one big step.
Instead, the beta reduction in \lambdae breaks down the evaluation into smaller steps with the help of the box construct:
\vspace{-8pt}
\begin{mathpar}
  \inferrule*[right=]{}{\delta \vdash \eclos{\delta_{\mathrm{1}}}{\lambda e} \, v \hookrightarrow (\delta_{\mathrm{1}}, v) \triangleright e}
\vspace{-8pt}
\end{mathpar}
Instead of evaluating the application directly to a value, \lambdae
stores the intermediate result as a box: the body expression $e$
is boxed by the environment that extends $\delta_{\mathrm{1}}$ with the argument
value $v$. The box will then be further reduced through the box
reduction rules. This design allows \lambdae to maintain a small-step
semantics, while avoiding substitution.

\subsection{Three Forms of Substitution in System F}\label{subsec:threeforms}
System F~\cite{girard1972interpretation,reynolds1974towards} exhibits greater expressive power than the STLC by
introducing parametric polymorphism through type abstraction and type
application. Moreover, System F is also more interesting regarding
formalization. This is primarily due to its inclusion of
type variables, which are bound within the environment. When type
checking a type abstraction, one type variable binding is added to the
context. Therefore, the typing context in System F contains both term
variable bindings and type variable bindings:
\begin{center}
(Named representation) $\Gamma \Coloneqq \cdot ~|~ \Gamma, x :A ~|~ \Gamma, X$
\quad \quad
(De Bruijn representation) $\Gamma \Coloneqq \cdot ~|~ \Gamma, A ~|~ \Gamma, \tstar$
\end{center}
Above are two representations of typing contexts in System F, where $\Gamma$ unifies term and type bindings in a single context structure. The conventional named representation contains term variable bindings $\Gamma, x:A$ and type variable bindings $\Gamma, X$. In the de Bruijn representation, variables are nameless: $\Gamma, A$ is a term variable binding with type $A$, and a type variable binding is just a tag $\tstar$. \name and its metatheory adopt the de Bruijn representation throughout.

The introduction of polymorphism gives rise to more forms of substitution.
The distinct forms of substitution required in System F are
illustrated by the following rules (written in a de Bruijn style):
\begin{mathpar}
  \inferrule [Red-beta]
          { }
          {(\lambda e) \, v \hookrightarrow e [\mathit{0} \mapsto v]}
  \quad \quad \quad
  \inferrule [Red-tapp]
          { }
          {(\Lambda e) \, A \hookrightarrow e [\mathit{0} \mapsto A]}
  \quad \quad \quad
  \inferrule [Typ-tapp]
          {\Gamma \vdash e : \forall B \\ \Gamma \vdash A }
          {\Gamma \vdash e \, A : B [\mathit{0} \mapsto A]}
\end{mathpar}
To summarize, in System F we have:
\begin{itemize}
  \item {\bf term-into-term substitution}: replacing a term variable with a term, as in $e [\mathit{0} \mapsto v]$, which occurs during beta reduction.
  \item {\bf type-into-term substitution}: replacing a type variable with a type in a term, as in $e [\mathit{0} \mapsto A]$, which occurs during type application when reducing $(\Lambda e) \, A$.
  \item {\bf type-into-type substitution}: replacing a type variable with a type within a type, as in $B [\mathit{0} \mapsto A]$, which is needed to compute the result type of a type application.
\end{itemize}
The first form of substitution arises from term-level beta reduction, and
the \lambdae calculus already provides a solution to eliminate this form of substitution.
Nevertheless, the second and third forms of substitution come from type application and they are unique to polymorphic calculi.
This proliferation of substitution forms significantly complicates the metatheory of polymorphic calculi.
Our first goal is to eliminate all three forms of substitution from the calculus, replacing them with an environment-based approach that enables first-class environments.

\subsection{Bringing First-class Environments to System F}\label{subsec:bringing}

\paragraph{Contexts as types}
Following \lambdae, which views typing contexts as types,
to extend System F with first-class environments, we first extend the syntax of types to include
typing contexts:
\vspace{-4pt}
\begin{align*}
& \text{Types} &A, B &\Coloneqq \tint ~|~ \mathit{N} ~|~ A \rightarrow B ~|~ \forall A ~|~ \cdot ~|~ \Gamma,A ~|~ \Gamma,\tstar \\
& \text{Contexts} &\Gamma  &\Coloneqq \cdot ~|~ \Gamma,
  A ~|~ \Gamma, \tstar
\end{align*}
In the above syntax, a type variable is represented by the de Bruijn
index $\mathit{N}$. The type syntax now also includes contexts: the empty context $\cdot$, a term binding
$\Gamma,A$, and a type-variable binding $\Gamma,\tstar$. The
typing contexts of System F then form a subset of the types.
Including contexts into types is a first-step to incorporate
first-class environments into System F. However, more changes
are necessary to eliminate all forms of substitution in System F.

\paragraph{A hint notation for de Bruijn indices}
The formal calculus uses de Bruijn indices for both term and type variables.
Raw de Bruijn examples are hard to read, so we use a lightweight hint notation
in examples. Greek letters hint type indices: $\alpha$ denotes index $0$,
$\beta$ denotes index $1$, and so on; roman letters play the analogous role
for term indices. These hints are not names and are not part of the formal
syntax. Binder annotations are shown in gray only to help the reader see which
binder an occurrence is intended to refer to. Deleting the gray annotations and
replacing each hinted variable by its corresponding index recovers the formal
de Bruijn term.

Because the hints follow de Bruijn indices rather than binders, the notation is
partial. A single binder may be represented by different indices in different
scopes, in which case no single hint can name it consistently. Conversely, the
same index may refer to different binders in different scopes. In examples we
may use subscripts such as ${\alpha}_{1}$ and ${\alpha}_{2}$ only to
distinguish these occurrences for the reader, although both are still index
$0$ formally. The table below shows the cases that matter for the examples in
the paper.
\begin{center}
\begin{tabular}{lcc}
\hline
 & pure de Bruijn notation & hint notation \\
\hline
unambiguous     & $\forall \forall (0 \rightarrow 1)$ & $\hforall{\beta}{\hforall{\alpha}{(\alpha \rightarrow \beta)}}$ \\
irrelevant      & $\forall \forall (0 \rightarrow 0)$ & $\hforall{\beta}{\hforall{\alpha}{(\alpha \rightarrow \alpha)}}$ \\
disambiguated   & $\forall (0 \rightarrow \forall 0)$ & $\hforall{{\alpha}_{2}}{({\alpha}_{2} \rightarrow \hforall{{\alpha}_{1}}{{\alpha}_{1}})}$ \\
irrepresentable & $\forall (0 \rightarrow \forall 1)$ & --- \\
\hline
\end{tabular}
\end{center}
The unambiguous case is the usual one: each hinted occurrence refers to a
binder at a consistent index. In the irrelevant case, the outer binder is
unused, so any gray hint for it is harmless. The disambiguated case shows that
one index can refer to different binders in different scopes. The subscripts
separate these reader hints, not formal indices. The irrepresentable case shows
the limitation of the notation: the outer binder is seen as index $0$ before the
inner binder and as index $1$ under it, so it would need to be displayed both as
$\alpha$ and as $\beta$. Every hinted example used in the paper is representable.
A table in Appendix~\appHints{} gives the corresponding raw
de Bruijn forms.

\paragraph{From term-level substitutions to environments}
The elimination of term-into-term substitution is already achieved by
\lambdae.
Now we consider type-into-term substitution. It is natural to apply the
same recipe that \lambdae uses for term abstractions/applications
also for type abstractions and type applications. Specifically,
for type abstractions $\Lambda e$, we introduce \emph{type abstraction
  closures} $\eclos{\delta}{\Lambda e}$ that capture the current runtime
environment $\delta$.
Furthermore, type application reduces to a box whose local
environment combines the closure's environment $\delta$ with the
\emph{type} argument. For instance, suppose that
$\hLam{\alpha}{\mathsf{id}}$ is a polymorphic identity
function and $\mathsf{id}$ is a lambda expression of type $\alpha \rightarrow \alpha$.
The evaluation
of type application $(\hLam{\alpha}{\mathsf{id}}) \, \tint$
proceeds:
\[
  \cdot \vdash
  (\hLam{\alpha}{\mathsf{id}}) \, \tint
  \hookrightarrow \eclos{\cdot}{\hLam{\alpha}{\mathsf{id}}} \, \tint
  \hookrightarrow (\cdot, [\heq{\alpha}{\tint}]) \triangleright \mathsf{id}
  \hookrightarrow^* \eclos{\cdot, [\heq{\alpha}{\tint}]}{\mathsf{id}}
\]
Here type argument $\tint$ is pushed into the box as part of the local environment. At this point, the local environment contains all the concrete type information that the body relies on. The box will then be further evaluated to a closure, where $\mathsf{id}$ is closed with the environment containing the type argument $[\heq{\alpha}{\tint}]$. Since type application introduces concrete type arguments into environments, the syntax of types and contexts needs to be extended for concrete type information:
\vspace{-4pt}
\[
\text{Types} \quad A, B \Coloneqq \dots ~|~ \Gamma, [A] \qquad \qquad
\text{Contexts} \quad \Gamma \Coloneqq \cdot ~|~ \Gamma,
  A ~|~ \Gamma, \tstar ~|~ \Gamma, [A]
\]
Besides term bindings $(\Gamma, A)$, now we have two kinds of type variable bindings: $\Gamma, \tstar$ represents an \emph{abstract type binding} (like a universally quantified type parameter with no definition), and $\Gamma, [A]$ represents a \emph{manifest type binding} (a type variable bound to a concrete definition $A$). Manifest type bindings serve as explicit type equalities, enabling visible type definitions in a typing context.

\paragraph{Typing boxes and closures}
While the dynamic semantics for applications can be adapted from
\lambdae, typing for closures and boxes requires some care. Now that
runtime environments contain type entries, the types of closures and
boxes need to be closed under their corresponding typing contexts,
which may contain type variable bindings. Take $\eclos{\cdot, [\heq{\alpha}{\tint}]}{\mathsf{id}}$ as an example.  Expression $\mathsf{id}$ has type $\alpha \rightarrow \alpha$, and type variable $\alpha$ should refer to the entry $[\heq{\alpha}{\tint}]$ stored in the environment. Therefore, the type of
the closure $\eclos{\cdot, [\heq{\alpha}{\tint}]}{\mathsf{id}}$ is supposed to be closed with
respect to this type information. To achieve this, we extend the
syntax of types by introducing \emph{box types}:
\[
A, B \Coloneqq \dots ~|~ \Gamma \triangleright A
\]
Intuitively, box types act analogously to function closures, in the
sense that they are enclosed over their defining context $\Gamma$.
For example, closure
$\eclos{\cdot, [\heq{\alpha}{\tint}]}{\mathsf{id}}$ has the box type
$(\cdot, \hdef{\alpha}{\tint}) \triangleright (\alpha \rightarrow \alpha)$, indicating that within the box,
type variable $\alpha$ is defined to be $\tint$.

\paragraph{Manifest types}
Since we have manifest type binding $[A]$ entries in the environment, it is
natural to have a type that corresponds to such entries. Such a type is
a \emph{manifest type} $[A]B$, extending the syntax of types as follows:
\[
A, B \Coloneqq \dots \mid \Gamma \triangleright A \mid [A]B
\]
The manifest type $[\heq{\alpha}{A}] B$ represents ``$B$ where type variable
$\alpha\triangleq0$ is instantiated to $A$'' and is the type-level
analogue of extending a context with the manifest binding $[A]$ and then
reading $B$ under that extended context.  We will see in the next section how
this construct eliminates type-level substitution from type application.

\subsection{Eliminating Type-level Substitutions}\label{subsec:elim}

This subsection explains how manifest types record type instantiations at the object level,
while type equivalence recovers the usual instantiated types.

\paragraph{Typing type application with manifest types}
With manifest types, the typing rule for type application uses no
meta-level substitution:
\[
\inferrule* [right=Typ-tapp]
  {\Gamma \vdash e : \forall B \\ \Gamma \vdash A}
  {\Gamma \vdash e \, A : [A] B}
\]
Similar to universal types, manifest types reserve a slot for the
type argument. However, while universal types bind type variables
abstractly (with $\tstar$), manifest types bind them concretely
(with $[A]$). Thus type application can instantiate type variable
$\alpha$ directly within the body type. We examine the above typing rule
with type application $(\hLam{{\alpha}_{2}}{(\hLam{{\alpha}_{1}}{\mathsf{id}})} \, {\alpha}_{2}) \, \tint$, where the type of $\mathsf{id}$ is ${\alpha}_{1} \rightarrow {\alpha}_{1}$. The outer type application instantiates
${\alpha}_{2}$ to $\tint$, and the inner type application
$(\hLam{{\alpha}_{1}}{\mathsf{id}}) \, {\alpha}_{2}$ instantiates ${\alpha}_{1}$ to ${\alpha}_{2}$.
The typing derivation of this example is:
\[
\inferrule*[left=Typ-tapp]
  {
    \inferrule*[left=Typ-tabs]
    {
      \inferrule*[left=Typ-tapp]
        {\cdot, \hdecl{{\alpha}_{2}}{\tstar} \vdash \hLam{{\alpha}_{1}}{\mathsf{id}} : \hforall{{\alpha}_{1}}{{\alpha}_{1}} \rightarrow {\alpha}_{1}}
        {\cdot, \hdecl{{\alpha}_{2}}{\tstar} \vdash (\hLam{{\alpha}_{1}}{\mathsf{id}}) \, {\alpha}_{2} : \hmani{{\alpha}_{1}}{{\alpha}_{2}}{({\alpha}_{1} \rightarrow {\alpha}_{1})}}
    }
     {\cdot \vdash \hLam{{\alpha}_{2}}{(\hLam{{\alpha}_{1}}{\mathsf{id}})} \, {\alpha}_{2} : \hforall{{\alpha}_{2}}{\hmani{{\alpha}_{1}}{{\alpha}_{2}}{({\alpha}_{1} \rightarrow {\alpha}_{1})}}}
  }
  {\cdot \vdash (\hLam{{\alpha}_{2}}{(\hLam{{\alpha}_{1}}{\mathsf{id}})} \, {\alpha}_{2}) \, \tint : \hmani{{\alpha}_{2}}{\tint}{\hmani{{\alpha}_{1}}{{\alpha}_{2}}{({\alpha}_{1} \rightarrow {\alpha}_{1})}}}
  \]
From the outer type application, we know that ${\alpha}_{2}$ will
be instantiated to $\tint$. However, when typing the inner type
application $(\hLam{{\alpha}_{1}}{\mathsf{id}}) \, {\alpha}_{2}$, type variable
${\alpha}_{2}$ is \emph{open}, in that it is abstractly bound by
$\hdecl{{\alpha}_{2}}{\tstar}$ instead of being associated with any concrete
type instantiation. In the second line of the derivation, manifest
type $\hmani{{\alpha}_{1}}{{\alpha}_{2}}{({\alpha}_{1} \rightarrow {\alpha}_{1})}$
keeps ${\alpha}_{2}$ open and assigns ${\alpha}_{2}$ to
${\alpha}_{1}$. Then, after the outer type application, manifest
type $\hmani{{\alpha}_{2}}{\tint}{\hmani{{\alpha}_{1}}{{\alpha}_{2}}{({\alpha}_{1} \rightarrow {\alpha}_{1})}}$ further instantiates ${\alpha}_{2}$ with
$\tint$. Hence, concrete type information propagates naturally,
without requiring meta-level substitution.

\paragraph{Type equivalence}
Semantically, $\hmani{{\alpha}_{2}}{\tint}{\hmani{{\alpha}_{1}}{{\alpha}_{2}}{({\alpha}_{1} \rightarrow {\alpha}_{1})}}$ contains sufficient manifest bindings
to ensure that it is essentially equivalent to $\tint \rightarrow \tint$. Nevertheless, the two types are syntactically distinct,
so recovering the instantiated type requires a notion of \emph{type
equivalence}. Importantly, this need is not created by our treatment
of type application: unlike System~F, \name needs type equivalence in
any case, since manifest type bindings and box types already relate
syntactically different types. For instance, the type variable
$\alpha$ is equivalent to $\tint$ under the context
$\cdot, \hdef{\alpha}{\tint}$, and the box type
$(\cdot, \hdef{\alpha}{\tint}) \triangleright (\alpha \rightarrow \alpha)$ of the closure in
Section~\ref{subsec:bringing} is just $\tint \rightarrow \tint$. Type
equivalence is also key for modeling modules with transparent type
components (Section~\ref{subsec:modules}).
The type equivalence judgment $\Gamma_{\mathrm{1}} \vdash A \sim B \dashv \Gamma_{\mathrm{2}}$ relates a type
$A$ under a context $\Gamma_{\mathrm{1}}$ to a type $B$ under a context
$\Gamma_{\mathrm{2}}$. The judgment is \emph{two-sided} because the
two types may be read under different contexts: discharging a manifest
type extends the context on one side with a manifest binding, and
comparing a box type switches to its local context on one side.
Concrete type variables are resolved by fetching their manifest
bindings. These rules, together with standard structural rules, form
the full relation (Section~\ref{subsec:teq}). Typing employs
equivalence through \rref{Typ-teq}, which replaces the type of an
expression with an equivalent one:
\[
\inferrule*[right=Typ-teq]
  {\Gamma \vdash e : A \\ \Gamma \vdash A \sim B \dashv \Gamma}
  {\Gamma \vdash e : B}
\]

\paragraph{Manifest types as the sole mechanism for type substitution}
Keeping System~F's meta-level substitution for type application
(Section~\ref{subsec:threeforms}) alongside manifest types would
yield two mechanisms for type instantiation in the same calculus:
manifest types $[A]B$
express ``$B$ with index~0 instantiated to $A$'' at the object level,
while meta-level substitution $B[0\mapsto A]$ expresses the same at
the meta-level. Maintaining both would require proving
that the two mechanisms interact correctly---in particular, that
substituting into equivalent types preserves equivalence:
\[
\frac{\Gamma_1,\star \vdash A \sim B \dashv \Gamma_2,\star
\qquad
\Gamma_1 \vdash C \sim D \dashv \Gamma_2}
{\Gamma_1 \vdash A [\mathit{0} \mapsto C] \sim B [\mathit{0} \mapsto D] \dashv \Gamma_2}
\]
Proving such a lemma would require the full machinery of de~Bruijn
substitution---generalized statements over decomposed contexts, weakening,
renaming, and shifting---on top of the type equivalence development.

We avoid this redundancy by using manifest types as the \emph{sole
mechanism} for type substitution. Because manifest types are an
object-level construct, they participate directly in the type
equivalence relation that \name requires anyway. Discharging a manifest type is just a rule of the
equivalence judgment, not a lemma about the interaction of two
distinct mechanisms. The result is a simpler calculus and a lighter
metatheory.

\paragraph{Substitution-free metatheory}

In System F, substitution lemmas are central to proving type
preservation and other metatheoretical properties. In particular, for
type application, we need the following substitution lemma to deal
with type substitution:
\begin{lemma}[Substitution Lemma]
  If $\Gamma, \tstar \vdash e : B$ and $\Gamma \vdash A$, then $\Gamma \vdash e [\mathit{0} \mapsto A] : B [\mathit{0} \mapsto A]$.
\end{lemma}

The first premise arises from the typing rule for type abstractions,
which extends the context with an abstract type binding when checking
a type abstraction $\Lambda e$ against a universal type $\forall B$. The
conclusion states that type application preserves typing by type
substitution in both the body expression $e$ and body type
$B$. Essentially, the substitution lemma guarantees that the
\emph{instantiation} of type variables with concrete types does not
violate type safety.

In our calculus, the substitution lemma is replaced by an \emph{instantiation lemma} that directly captures the essence of type variable instantiation without involving substitution:
\begin{lemma}[Instantiation Lemma]
  If $\Gamma, \tstar \vdash e : B$ and $\Gamma \vdash A$, then $\Gamma, [A] \vdash e : B$.
\end{lemma}
Rather than performing meta-level substitution, instantiation is
achieved directly through manifest type bindings: an expression
well-typed under an abstract type binding remains well-typed when that
binding is replaced by a manifest binding.
Due to the presence of type equivalence, we also need an instantiation lemma for type equivalence. Two types that are equivalent under an abstract binding remain equivalent when the binding is replaced by equivalent manifest bindings.
\begin{lemma}[Instantiation]
  If $\Gamma_{\mathrm{1}}, \tstar \vdash A \sim B \dashv \Gamma_{\mathrm{2}}, \tstar$ and $\Gamma_{\mathrm{1}} \vdash C \sim D \dashv \Gamma_{\mathrm{2}}$, then $\Gamma_{\mathrm{1}}, [C] \vdash A \sim B \dashv \Gamma_{\mathrm{2}}, [D]$.
\end{lemma}

Traditional approaches reason about syntactic equality between meta-level substitutions~\cite{vouillon2012solution}. \name reasons about definitional equivalence between object-level types.

\subsection{Applications: Sandboxed Modules}\label{subsec:modules}
\enlargethispage{\baselineskip}
This subsection illustrates that \name enables a form of
\emph{sandboxed modules}. Three key features are illustrated: 1)
first-class modules can be passed to and returned from functions; 2) module fragments can be composed
\emph{dynamically}, at run time, and this composition still preserves the
dependencies of later entries on earlier ones; and 3) the modules are sandboxed
and cannot access definitions in the ambient environment. The last point is what
enables a capability-style programming
discipline~\cite{DBLP:conf/fc/MillerMF00,miller2006robust,miller2008safe}: any
capability a module needs must be passed to it
explicitly~\cite{DBLP:conf/ecoop/MelicherSPA17}.
Features 2) and 3) are what set \name apart from classical module systems such
as ML modules~\cite{MacQueen1984Modules,Milner1997Definition,LeroyPOPL94Manifest,RossbergDreyerMixML,Rossberg2018OneML}.
There, composition is \emph{static}: functors are applied to arguments that
match fixed signatures. In \name, modules are first-class values composed
\emph{dynamically}. Independently written fragments are merged at run time
while preserving the dependencies of later components on earlier ones---something that
a (dependent) record, whose fields are fixed at definition time, cannot express.
Sandboxing, likewise, is not part of the classical module picture.

The \lambdae calculus already provides simple sandboxed
modules~\cite{DBLP:journals/pacmpl/TanO24}: a module that takes other modules as
arguments is \emph{boxed} under the \emph{empty} environment, so it acts only
through the arguments passed to it. \name inherits boxes---and thus
sandboxing---from \lambdae, and adds polymorphism and manifest type components.

Manifest type bindings enable modules with transparent type
components. \name's environments are $\Delta \Coloneqq \cdot ~|~ \Delta, e ~|~ \Delta, [A]$, where $e$ is a term component and $[A]$ a type component. Environments are \emph{ordered}: a later entry may depend on earlier ones, and the typing rules capture this directly:
\begin{mathpar}
  \inferrule [Typ-edef]
    {\Gamma \vdash \Delta : \Gamma_{\mathrm{1}} \\ \Gamma \envplus \Gamma_{\mathrm{1}} \vdash e : A }
    {\Gamma \vdash \Delta, e : \Gamma_{\mathrm{1}}, A}
  \quad \quad \quad \quad \quad
  \inferrule [Typ-tdef]
    {\Gamma \vdash \Delta : \Gamma_{\mathrm{1}} \\ \Gamma \envplus \Gamma_{\mathrm{1}} \vdash A }
    {\Gamma \vdash \Delta, [A] : \Gamma_{\mathrm{1}}, [A]}
\end{mathpar}
Each entry is typechecked under the combined context
$\Gamma \envplus \Gamma_{\mathrm{1}}$, where $\Gamma_{\mathrm{1}}$ collects the types of preceding entries. A type definition introduces a manifest type binding.
Later entries can refer to the defined type at the type level. Our concrete
examples use an ML-like surface syntax
(\lstinline|module type|, \lstinline|sig|, \lstinline|struct|) only
because it is familiar.  Our focus is not on ML features such as
opaque type components or sealing (which \name does not support), but on \name's
runtime-composable environments with type entries that this syntax elaborates into.

\paragraph{Modules as first-class environments}
Consider a data-access system whose \emph{access policies} control how query results are returned---which elements, in what order---without inspecting the data. A policy is polymorphic in the element type, so its module type
\begin{lstlisting}[language=js,mathescape=true]
module type ACCESS_POLICY = forall t.
  sig val reshape : List<t> -> List<t>; end;
\end{lstlisting}
translates in \name to a universal type over a one-entry context, $\hforall{\alpha}{(\cdot, \{ \mathsf{reshape} : \tlist{\alpha} \rightarrow \tlist{\alpha} \})}$.
More generally, a signature translates to a context and a module implementation to an environment. Modules are first-class: \lstinline|makeQuery| turns a policy into an integer-specialized query module,
\begin{lstlisting}[language=js,mathescape=true]
module makeQuery (policy : ACCESS_POLICY) : (sig type a = Int; val query : List<a> -> List<a>; end) = struct
  type a = Int;
  let query : List<a> -> List<a> = (policy @ a).reshape;
end;
\end{lstlisting}
whose output type is the context $\cdot, \hdef{\alpha}{\tint}, \{ \mathsf{query} : \tlist{\alpha} \rightarrow \tlist{\alpha} \}$.
Here the type component \lstinline|a| is transparently equal to \lstinline|Int|, and \lstinline|query| instantiates the policy at \lstinline|a| and takes its \lstinline|reshape|.

\paragraph{Dynamic composition}
Beyond building a single module, \name supports \emph{dynamically composing}
modules.
The module \lstinline|assemble| takes a query module \lstinline|m1| and an audit module \lstinline|m2|, concatenates them, and appends a \lstinline|report| function that feeds \lstinline|m1|'s query result to \lstinline|m2|'s \lstinline|decide|:
\begin{lstlisting}[language=js,mathescape=true]
module assemble (m1 : sig type a = Int; val query : List<a> -> List<a>; end)
                 (m2 : sig type b = Bool; val decide : List<Int> -> b; end)
= m1 + m2 + struct
    let report : List<Int> -> Bool = fun (xs) -> m2.decide(m1.query(xs));
  end;
\end{lstlisting}
Concretely, the operator \lstinline|+|
concatenates the two fragments' environments.
It appends the entries of the second environment after those of the first, forming a single environment.
The composition is dynamic: \lstinline|m1| and \lstinline|m2| are unknown when \lstinline|assemble| is defined and are supplied only at application time.
We use \lstinline|assemble| to compose two concrete modules. First, \lstinline|previewPolicy| keeps only the first two elements as a preview window, and \lstinline|makeQuery| turns it into the query module \lstinline|previewQuery|:
\begin{lstlisting}[language=js,mathescape=true]
module previewPolicy : ACCESS_POLICY = (type t) -> struct
  let reshape : List<t> -> List<t> = take(2);
end;
module previewQuery = makeQuery(previewPolicy);
\end{lstlisting}
We then define an audit module \lstinline|audit| that decides whether a result list is non-empty, and compose \lstinline|previewQuery| and \lstinline|audit| into \lstinline|previewProcessor| using \lstinline|assemble|:
\begin{lstlisting}[language=js,mathescape=true]
module audit = struct
  type b = Bool;
  let decide : List<Int> -> b = fun (xs) -> length(xs) > 0;
end;
module previewProcessor = assemble(previewQuery)(audit);
\end{lstlisting}
In \name, \lstinline|previewProcessor| corresponds to the dependent environment
\[
\cdot, [\heq{\alpha}{\tint}], \{ \mathsf{query} = \ldots \}, [\heq{\beta}{\mathsf{Bool}}], \{ \mathsf{decide} = \ldots \}, \{ \mathsf{report} = \ldots \}
\]
Dependencies are preserved after composition: \lstinline|query| still refers to the type component \lstinline|a| and \lstinline|decide| to \lstinline|b|, and the appended \lstinline|report| builds on both.

\paragraph{Sandboxing}
Composition and sandboxing work together: a sandboxed module runs with no ambient bindings, so its authority is exactly the capabilities passed as arguments. Handing the assembled \lstinline|previewProcessor| to a sandboxed module thus grants it precisely the authority it was assembled from, and nothing else.
Suppose queries run for external clients and must not reach sensitive global resources such as database credentials, audit logs, or administrative APIs.
The module \lstinline|executeQuery| below takes a query processor \lstinline|proc| as its only argument and runs its \lstinline|report|:
\begin{lstlisting}[language=js,mathescape=true]
module executeQuery (proc : sig type a = Int; val query : List<a> -> List<a>; type b = Bool; val decide : List<Int> -> b; val report : List<Int> -> Bool; end) = struct
  let input  : List<Int> = [3, 1, 2];
  let result : Bool      = proc.report(input);
end;
\end{lstlisting}
Since \lstinline|executeQuery| is sandboxed, the only binding it can reach is \lstinline|proc|; ambient resources are invisible. An ambient \lstinline|secret : Int|, say, would be out of scope, so \lstinline|proc.report([secret, 1, 2])| is statically rejected. We can now apply \lstinline|executeQuery| to the previous composed module \lstinline|previewProcessor| and obtain the result of its \lstinline|report| via \lstinline|executeQuery(previewProcessor).result|.

The authority of \lstinline|executeQuery| is fixed entirely by the processor it is handed, so swapping the processor swaps the authority.
Testing with mocks then follows immediately: we compose a \lstinline|mockProcessor| from a \lstinline|mockPolicy| that returns its list unchanged and a \lstinline|mockAudit| whose \lstinline|decide| returns \lstinline|false|, and pass it to the same \lstinline|executeQuery|.
\begin{lstlisting}[language=js,mathescape=true]
module mockProcessor = assemble(makeQuery(mockPolicy))(mockAudit);
let mockResult : Bool = executeQuery(mockProcessor).result;
\end{lstlisting}
The same sandboxed module yields a different result purely by the capabilities it is granted: composition assembles the capabilities, and the module is confined to them.

We have a prototype implementation of \name with a surface language in conventional \emph{named} syntax, inspired by OCaml's modules, that runs and type-checks all the examples above and elaborates into our de Bruijn core. The examples also use records with named fields, a lightweight extension of \name that we formalize in Section~\ref{sec:records}.
Section~\ref{subsec:module-elaboration} presents a small source language
for these module constructs and proves that its elaboration preserves typing.
Beyond the core calculus, the implementation also supports practical conveniences such as built-in lists (used in the examples). These are orthogonal to first-class environments and are not part of the formal development.

\section{The \name Calculus: Syntax, Typing and Semantics}
\label{sec:calculus}

This section introduces the \name calculus: a
polymorphic calculus with first-class environments.

\begin{figure}[t]
\setlength{\jot}{0pt}
\begin{align*}
  &\text{De Bruijn indices} & \mathit{N} &\Coloneqq \alpha, \beta, \dots \quad \quad \quad \mathit{n} \Coloneqq a, b, c, \dots \\
  &\text{Types} &A, B, C &\Coloneqq \tint ~|~ \mathit{N} ~|~ A \rightarrow B ~|~ \forall A ~|~ \Gamma \triangleright A ~|~ [A]B ~|~
  \cdot ~|~ \Gamma,
  A ~|~ \Gamma, \tstar ~|~ \Gamma, [A] \\ &\text{Contexts} &\Gamma & \Coloneqq \cdot ~|~ \Gamma,
  A ~|~ \Gamma, \tstar ~|~ \Gamma, [A] \\ &\text{Expressions}
  &e &\Coloneqq \mathtt{i} ~|~ \mathit{n} ~|~ \lambda e ~|~ e_{\mathrm{1}} \, e_{\mathrm{2}} ~|~ \Lambda e
  ~|~e \, A ~|~ \eclos{\delta}{\lambda e} ~|~ \eclos{\delta}{\Lambda e} ~|~ e_{\mathrm{1}} \triangleright e_{\mathrm{2}} ~|~ \cdot ~|~ e_{\mathrm{1}}, e_{\mathrm{2}} ~|~ e, [A] \\ &\text{Environments} &\Delta
  &\Coloneqq \cdot ~|~ \Delta, e ~|~ \Delta, [A] \\ &\text{Values} &v
  &\Coloneqq \mathtt{i} ~|~ \eclos{\delta}{\lambda e} ~|~ \eclos{\delta}{\Lambda e} ~|~ \cdot ~|~ v_{\mathrm{1}},v_{\mathrm{2}} ~|~ v, [\Gamma \triangleright A] \\ &\text{Value environments} &\delta &\Coloneqq \cdot
  ~|~ \delta, v ~|~ \delta, [\Gamma \triangleright A] \\ &\text{Frames} &F &\Coloneqq
  \eframe \, e ~|~ v \, \eframe ~|~ \eframe \, A ~|~ \eframe \triangleright e
  ~|~ \eframe , e ~|~ \eframe , [A]
\end{align*}
\caption{Syntax of the \name calculus.}
  \label{fig:syntax}
\end{figure}

\subsection{Syntax and Well-formedness}
\begin{figure}[t]
  \centering
  \begin{tikzpicture}[
    enventry/.style={inner sep=1pt},
    envarrow/.style={-{Stealth[length=2mm]}, thick, gray}
  ]
    \fill[gray!40, rounded corners=3mm] (-0.45,0.28) rectangle (0.45,-1.18);
    \fill[gray!40, rounded corners=3mm] (2.55,0.28) rectangle (3.45,-1.18);
    \fill[gray!20, rounded corners=3mm] (1.05,0.28) rectangle (1.95,-1.63);
    \fill[gray!20, rounded corners=3mm] (4.05,0.28) rectangle (5.85,-1.63);

    \node[enventry] at (-1.5,0) {$\dots$};
    \node[enventry] at (-0.8,0) {,};
    \node[enventry] (three) at (0,0) {$[\tint]$};
    \node[enventry] (two)   at (1.5,0) {$\alpha$};
    \node[enventry] (one)   at (3.0,0) {$[\alpha]$};
    \node[enventry] (zeroL) at (4.45,0) {$(\alpha$};
    \node[enventry] (zeroA) at (4.95,0) {$\rightarrow$};
    \node[enventry] (zeroR) at (5.45,0) {$\beta)$};
    \node[enventry] at (0.75,0) {,};
    \node[enventry] at (2.25,0) {,};
    \node[enventry] at (3.75,0) {,};

    \node[enventry, anchor=west] at (-6,-1) {Positions of type entries};
    \node[enventry] at (0,-1) {$\beta$};
    \node[enventry] at (3.0,-1) {$\alpha$};

    \node[enventry, anchor=west] at (-6,-1.45) {Positions of term entries};
    \node[enventry] at (1.5,-1.45) {$b$};
    \node[enventry] at (4.95,-1.45) {$a$};

    \coordinate (twoTop) at (1.5,0.45);
    \coordinate (threeTopA) at (0,0.45);
    \coordinate (oneTopA) at (3.0,0.6);
    \coordinate (threeTopB) at (0,0.6);
    \coordinate (zeroLBot) at (4.45,-0.45);
    \coordinate (oneBot) at (3.0,-0.45);
    \coordinate (zeroRBot) at (5.45,-0.6);
    \coordinate (threeBot) at (0,-0.6);

    \draw[envarrow] (two.north) -- (twoTop) -- (threeTopA) -- (three.north);
    \draw[envarrow] (one.north) -- (oneTopA) -- (threeTopB) -- (three.north);
    \draw[envarrow] (zeroL.south) -- (zeroLBot) -- (oneBot) -- (one.south);
    \draw[envarrow] (zeroR.south) -- (zeroRBot) -- (threeBot) -- (three.south);
  \end{tikzpicture}
  \caption{Visualization of an environment with de Bruijn indices.}
  \label{fig:visual}
\end{figure}
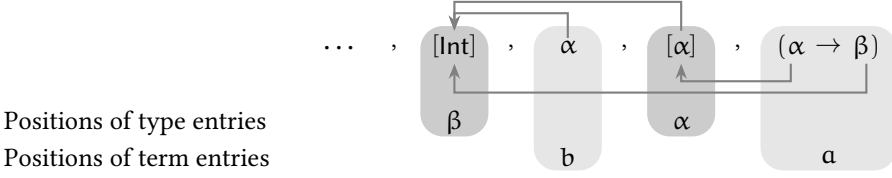

The syntax of the \name calculus is shown in Figure~\ref{fig:syntax}.
The calculus uses a de Bruijn~\cite{de1972lambda} encoding for both
term and type variables, thus all binders are nameless. To keep the de
Bruijn notation more manageable, we separate indices for type
variables (greek letters $\alpha, \beta, \dots$) and term
variables (roman letters $a, b, c, \dots$).  Note that these letters are not names but de Bruijn indices $0, 1, 2, \dots$ in
alphabetical order ($\alpha \triangleq 0$, $\beta \triangleq 1$, and so on).
Figure~\ref{fig:visual} shows how such mixed environments are read.
Term entries (lighter gray) and type entries (darker gray)
form one ordered sequence, but each variable resolves within its own
namespace, counting only entries of that kind: for example $\alpha$ (index 0)
is the nearest preceding type binding and $\beta$ (index 1) the second-nearest.
The arrows and the two rows beneath the sequence make these positions explicit.

\paragraph{Types and contexts}
Meta-variables $A,B$ and $C$ range over types. Types include base
types ($\tint$), type variables (de Bruĳn indices $\mathit{N}$),
arrow types ($A \rightarrow B$), universal types ($\forall A$), box types
($\Gamma \triangleright A$), manifest types ($[A] B$), and contexts. A context ($\Gamma$) is either empty ($\cdot$) or extends one with a term binding
($\Gamma, A$), an abstract type binding ($\Gamma, \tstar$), or a
manifest type binding ($\Gamma, [A]$).

\paragraph{Expressions}
Meta-variables $e$ range over expressions, which include: integer
literals ($\mathtt{i}$), variables (de Bruĳn indices $\mathit{n}$), lambda
abstractions ($\lambda e$), applications ($e_{\mathrm{1}} \, e_{\mathrm{2}}$), type
abstractions ($\Lambda e$), type applications ($e \, A$), closures for
abstractions $\eclos{\delta}{\lambda e}$, closures for type abstractions $\eclos{\delta}{\Lambda e}$, boxes $e_{\mathrm{1}} \triangleright e_{\mathrm{2}}$, and sequence expressions
formed by comma-separated entries. Environment expressions $\Delta$
are the subset of expressions with this sequence shape, starting
with $\cdot$ and extending with either an expression $e$ or a
manifest type definition $[A]$. Note that the environment
entries may refer to term or type variables bound in $\Delta$. Thus
environments are \emph{dependent}~\cite{tan2023dependent}.
In contrast to \lambdae where there are only term dependencies, in \name there can
be type dependencies as well.
For instance, in
$\cdot, [\heq{{\alpha}_{2}}{\tint}], (\hLam{{\alpha}_{1}}{\hlam{a}{a}}) \, {\alpha}_{2}$ the term
component instantiates the polymorphic identity at the type variable
${\alpha}_{2}$ bound by the earlier type entry $[\heq{{\alpha}_{2}}{\tint}]$,
so its type $\hmani{{\alpha}_{1}}{{\alpha}_{2}}{({\alpha}_{1} \rightarrow {\alpha}_{1})}$
depends on that entry. Composition is the concatenation of entries, and it
preserves these dependencies of later entries on earlier ones.

\paragraph{Values and frames}
Meta-variables $v$ range over values. Values include integer literals
($\mathtt{i}$), closures for abstractions $\eclos{\delta}{\lambda e}$, closures for
type abstractions $\eclos{\delta}{\Lambda e}$, and sequence values formed
by comma-separated entries. Value environments $\delta$ are the subset
of values that have this sequence shape, starting with $\cdot$
and extending with a value or a box type. A value environment can
contain both values and manifest box-type definitions.
In contrast to environment expressions, value environments $\delta$ are not
ordered: an entry in $\delta$ does not have dependencies on previous
entries. In other words, all the entries in $\delta$ are
\emph{closed} and merges in value environments are \emph{not dependent}.
Frames ($F$) specify evaluation contexts for
applications, type applications, boxes, environments.

\begin{figure}[t]
  \begin{drulepar}{\Gamma \vdash A}{Well-formed Types}
    \drule{Wft-int}{\vdash \Gamma}{\Gamma \vdash \tint}

    \drule{Wft-var}
      {\vdash \Gamma \and \mathit{N} \, \in \, \Gamma}
      {\Gamma \vdash \mathit{N}}

    \drule{Wft-vareq}
      {\vdash \Gamma \and \Gamma [\mathit{N}] = A}
      {\Gamma \vdash \mathit{N}}

    \drule{Wft-arrow}{\Gamma \vdash A \and \Gamma \vdash B}{\Gamma \vdash A \rightarrow B}

    \drule{Wft-forall}
      {\vdash \Gamma \and \Gamma, \tstar \vdash A}
      {\Gamma \vdash \forall A}

    \drule{Wft-mani}{\Gamma \vdash A \and \Gamma, [A] \vdash B}{\Gamma \vdash [A] B}

    \drule{Wft-box}
      {\vdash \Gamma \and \Gamma_{\mathrm{1}} \vdash A \and \Gamma_{\mathrm{1}} \vdash_{r} A}
      {\Gamma \vdash \Gamma_{\mathrm{1}} \triangleright A}

    \drule{Wft-nil}{\vdash \Gamma}{\Gamma \vdash \cdot}

    \drule{Wft-edef}
      {\Gamma \vdash \Gamma_{\mathrm{1}} \and \Gamma \envplus \Gamma_{\mathrm{1}} \vdash A}
      {\Gamma \vdash \Gamma_{\mathrm{1}}, A}

    \drule{Wft-star}
      {\Gamma \vdash \Gamma_{\mathrm{1}}}
      {\Gamma \vdash \Gamma_{\mathrm{1}}, \tstar}

    \drule{Wft-tdef}
      {\Gamma \vdash \Gamma_{\mathrm{1}} \and \Gamma \envplus \Gamma_{\mathrm{1}} \vdash A}
      {\Gamma \vdash \Gamma_{\mathrm{1}}, [A]}
  \end{drulepar}
  \vspace{-1em}
  \begin{drulepar}{\Gamma \vdash e : A}{Typing}
    \drule{Typ-lit}{\vdash \Gamma}{\Gamma \vdash \mathtt{i} : \tint}

    \drule{Typ-var}
      {\vdash \Gamma \and \Gamma [\mathit{n}] = A}
      {\Gamma \vdash \mathit{n} : A}

    \drule{Typ-abs}{\Gamma, A \vdash e : B}{\Gamma \vdash \lambda e : A \rightarrow B}

    \drule{Typ-app}
      {\Gamma \vdash e_{\mathrm{1}} : A \rightarrow B \and \Gamma \vdash e_{\mathrm{2}} : A}
      {\Gamma \vdash e_{\mathrm{1}} \, e_{\mathrm{2}} : B}

    \drule{Typ-tabs}{\Gamma, \tstar \vdash e : A}{\Gamma \vdash \Lambda e : \forall A}

    \drule{Typ-tapp}
      {\Gamma \vdash e : \forall B \and \Gamma \vdash A}
      {\Gamma \vdash e \, A : [A] B}

    \drule{Typ-box}
      {\Gamma \vdash e_{\mathrm{1}} : \Gamma_{\mathrm{1}} \and \Gamma_{\mathrm{1}} \vdash e_{\mathrm{2}} : A}
      {\Gamma \vdash e_{\mathrm{1}} \triangleright e_{\mathrm{2}} : \Gamma_{\mathrm{1}} \triangleright A}

    \drule{Typ-nil}{\vdash \Gamma}{\Gamma \vdash \cdot : \cdot}

    \drule{Typ-edef}
      {\Gamma \vdash \Delta : \Gamma_{\mathrm{1}} \and \Gamma \envplus \Gamma_{\mathrm{1}} \vdash e : A}
      {\Gamma \vdash \Delta, e : \Gamma_{\mathrm{1}}, A}

    \drule{Typ-tdef}
      {\Gamma \vdash \Delta : \Gamma_{\mathrm{1}} \and \Gamma \envplus \Gamma_{\mathrm{1}} \vdash A}
      {\Gamma \vdash \Delta, [A] : \Gamma_{\mathrm{1}}, [A]}

    \drule{Typ-teq}
      {\Gamma \vdash e : A \and \Gamma \vdash A \sim B \dashv \Gamma}
      {\Gamma \vdash e : B}

    \drule{Typ-clos}
      {\vdash \Gamma \and \cdot \vdash \delta : \Gamma_{\mathrm{1}} \and \Gamma_{\mathrm{1}}, A \vdash e : B}
      {\Gamma \vdash \eclos{\delta}{\lambda e} : \Gamma_{\mathrm{1}} \triangleright (A \rightarrow B)}

    \drule{Typ-bclos}
      {\vdash \Gamma \and \cdot \vdash \delta : \Gamma_{\mathrm{1}} \and \Gamma_{\mathrm{1}}, \tstar \vdash e : A}
      {\Gamma \vdash \eclos{\delta}{\Lambda e} : \Gamma_{\mathrm{1}} \triangleright \forall A}
  \end{drulepar}
  \caption{Well-formedness of types (top) and \name's type system (bottom).}\label{fig:typ}
\end{figure}
\paragraph{Well-formedness} Most well-formedness rules (Figure~\ref{fig:typ}, top) are standard, and the auxiliary judgments they use---context membership $\mathit{N} \, \in \, \Gamma$, type variable lookup $\Gamma [\mathit{N}] = A$, well-formed contexts $\vdash \Gamma$, context concatenation $\Gamma \envplus \Gamma_{\mathrm{1}}$, and rigidity $\Gamma_{\mathrm{1}} \vdash_{r} A$---are given in Appendix~\appAux. A type variable is well-formed when it points to an abstract binding $\tstar$ (\rref{Wft-var}) or a concrete one (\rref{Wft-vareq}), and environment types are checked entrywise under the concatenation $\Gamma \envplus \Gamma_{\mathrm{1}}$ of the global and local contexts (\rref{Wft-edef}, \rref{Wft-tdef}).
One interesting rule is \rref{Wft-box}. For a box type $\Gamma_{\mathrm{1}} \triangleright A$,
the body $A$
must be well-formed \emph{and} \emph{rigid} in $\Gamma_{\mathrm{1}}$ (written $\Gamma_{\mathrm{1}} \vdash_{r} A$), meaning that $A$ does not refer to abstract type variables in
$\Gamma_{\mathrm{1}}$. Rigidity mirrors environments and values, which carry only
manifest (concrete) type bindings, never abstract ones. When we
build a box $\Delta \triangleright e$, the type of $\Delta$ is a concrete
context without abstract variables, and any type under such
a context is rigid.

\subsection{Typing and Type Equivalence}\label{subsec:typing}

\paragraph{Typing}
Figure~\ref{fig:typ} (bottom) shows the type system. The typing rules for term
variables (\rref{Typ-var}, in which term variable lookup $\Gamma [\mathit{n}] = A$ is defined in Appendix~\appAux), term abstractions (\rref{Typ-abs}),
applications (\rref{Typ-app}) and type abstractions (\rref{Typ-tabs})
are standard.
For type applications, instead of using a meta-level
substitution, we utilize a manifest type $[A] B$ to represent
the instantiation of type variable $\alpha$ ($\triangleq 0$) in $B$ with type
$A$ (\rref{Typ-tapp}). Together with \rref{Typ-teq}, type
application can check against some expected type using the type
equivalence relation.  For example, by \rref{Typ-tapp} we know that
type application $(\hLam{\alpha}{\hlam{a}{a}}) \, \tint$ has type $\hmani{\alpha}{\tint}{(\alpha \rightarrow \alpha)}$.
Then, by \rref{Typ-teq}, we can conclude that it also has an equivalent type
$\tint \rightarrow \tint$. We will explain the type equivalence relation
$\Gamma_{\mathrm{1}} \vdash A \sim B \dashv \Gamma_{\mathrm{2}}$ in detail later in this subsection.

\Rref{Typ-box} is the typing rule for term-level boxes. A box
expression has a box type. To typecheck a box $e_{\mathrm{1}} \triangleright e_{\mathrm{2}}$, the local environment $e_{\mathrm{1}}$ is first typed
under the global context $\Gamma$, giving an environment type $\Gamma_{\mathrm{1}}$. The body
$e_{\mathrm{2}}$ is then typed under $\Gamma_{\mathrm{1}}$, so that if $e_{\mathrm{2}}$ has type $A$ then the box
has type $\Gamma_{\mathrm{1}} \triangleright A$.  \Rref{Typ-clos}
and \rref{Typ-bclos} are rules for closures of term abstractions and
type abstractions. These two rules are similar to \rref{Typ-abs} and
\rref{Typ-tabs} respectively, except that the term/type abstractions
are checked under the type of the local value environment $\delta$, and the
types of closures are also box types. Moreover, since $\delta$ is a sequence of
values and it is closed, the typing context for it can be empty in
these two rules.

For typing an environment, the older environment is first type-checked
under the global context $\Gamma$, and then the last element is checked
under the concatenation of $\Gamma$ and the older local context
$\Gamma_{\mathrm{1}}$. If the last element is an expression $e$, it is
type-checked under the extended context (\rref{Typ-edef}). Otherwise it is a
type $A$, which must be well-formed under the extended context, and the
environment type is extended with a concrete binding $[A]$ (\rref{Typ-tdef}).

\paragraph{Type equivalence}\label{subsec:teq}
\begin{small}
\begin{figure}[t]
  \begin{drulepar}{\Gamma_{\mathrm{1}} \vdash A \sim B \dashv \Gamma_{\mathrm{2}}}{Type Equivalence}
    \drule{Teq-int}
      {\vdash \Gamma_{\mathrm{1}} \and \vdash \Gamma_{\mathrm{2}}}
      {\Gamma_{\mathrm{1}} \vdash \tint \sim \tint \dashv \Gamma_{\mathrm{2}}}

    \drule{Teq-tvar}
      {\vdash \Gamma_{\mathrm{1}} \and \vdash \Gamma_{\mathrm{2}} \and \mathit{N} \, \in \, \Gamma_{\mathrm{1}} \and \mathit{N} \, \in \, \Gamma_{\mathrm{2}}}
      {\Gamma_{\mathrm{1}} \vdash \mathit{N} \sim \mathit{N} \dashv \Gamma_{\mathrm{2}}}

    \drule{Teq-eql}
      {\Gamma_{\mathrm{1}} \vdash A \sim B \dashv \Gamma_{\mathrm{2}} \and \Gamma_{\mathrm{1}} [\mathit{N}] = A}
      {\Gamma_{\mathrm{1}} \vdash \mathit{N} \sim B \dashv \Gamma_{\mathrm{2}}}

    \drule{Teq-eqr}
      {\Gamma_{\mathrm{1}} \vdash A \sim B \dashv \Gamma_{\mathrm{2}} \and \Gamma_{\mathrm{2}} [\mathit{N}] = B}
      {\Gamma_{\mathrm{1}} \vdash A \sim \mathit{N} \dashv \Gamma_{\mathrm{2}}}

    \drule{Teq-manil}
      {\Gamma_{\mathrm{1}}, [A] \vdash B \sim \tshift{C} \dashv \Gamma_{\mathrm{2}}, \tstar}
      {\Gamma_{\mathrm{1}} \vdash [A] B \sim C \dashv \Gamma_{\mathrm{2}}}

    \drule{Teq-manir}
      {\Gamma_{\mathrm{1}}, \tstar \vdash \tshift{B} \sim C \dashv \Gamma_{\mathrm{2}}, [A]}
      {\Gamma_{\mathrm{1}} \vdash B \sim [A] C \dashv \Gamma_{\mathrm{2}}}

    \drule{Teq-boxl}
      {\Gamma_{\mathrm{1}} \vdash \Gamma_{\mathrm{3}} \triangleright A \and \Gamma_{\mathrm{3}} \vdash A \sim B \dashv \Gamma_{\mathrm{2}}}
      {\Gamma_{\mathrm{1}} \vdash \Gamma_{\mathrm{3}} \triangleright A \sim B \dashv \Gamma_{\mathrm{2}}}

    \drule{Teq-boxr}
      {\Gamma_{\mathrm{2}} \vdash \Gamma_{\mathrm{3}} \triangleright B \and \Gamma_{\mathrm{1}} \vdash A \sim B \dashv \Gamma_{\mathrm{3}}}
      {\Gamma_{\mathrm{1}} \vdash A \sim \Gamma_{\mathrm{3}} \triangleright B \dashv \Gamma_{\mathrm{2}}}

    \drule{Teq-forall}
      {\Gamma_{\mathrm{1}}, \tstar \vdash A \sim B \dashv \Gamma_{\mathrm{2}}, \tstar}
      {\Gamma_{\mathrm{1}} \vdash \forall A \sim \forall B \dashv \Gamma_{\mathrm{2}}}

    \drule{Teq-arr}
      {\Gamma_{\mathrm{1}} \vdash A \sim C \dashv \Gamma_{\mathrm{2}} \and \Gamma_{\mathrm{1}} \vdash B \sim D \dashv \Gamma_{\mathrm{2}}}
      {\Gamma_{\mathrm{1}} \vdash A \rightarrow B \sim C \rightarrow D \dashv \Gamma_{\mathrm{2}}}

    \drule{Teq-nil}
      {\vdash \Gamma_{\mathrm{1}} \and \vdash \Gamma_{\mathrm{2}}}
      {\Gamma_{\mathrm{1}} \vdash \cdot \sim \cdot \dashv \Gamma_{\mathrm{2}}}

    \drule{Teq-star}
      {\Gamma_{\mathrm{1}} \vdash \Gamma_{\mathrm{3}} \sim \Gamma_{\mathrm{4}} \dashv \Gamma_{\mathrm{2}}}
      {\Gamma_{\mathrm{1}} \vdash \Gamma_{\mathrm{3}}, \tstar \sim \Gamma_{\mathrm{4}}, \tstar \dashv \Gamma_{\mathrm{2}}}

    \drule{Teq-edef}
      {\Gamma_{\mathrm{1}} \vdash \Gamma_{\mathrm{3}} \sim \Gamma_{\mathrm{4}} \dashv \Gamma_{\mathrm{2}} \and \Gamma_{\mathrm{1}} \envplus \Gamma_{\mathrm{3}} \vdash A \sim B \dashv \Gamma_{\mathrm{2}} \envplus \Gamma_{\mathrm{4}}}
      {\Gamma_{\mathrm{1}} \vdash \Gamma_{\mathrm{3}}, A \sim \Gamma_{\mathrm{4}}, B \dashv \Gamma_{\mathrm{2}}}

    \drule{Teq-tdef}
      {\Gamma_{\mathrm{1}} \vdash \Gamma_{\mathrm{3}} \sim \Gamma_{\mathrm{4}} \dashv \Gamma_{\mathrm{2}} \and \Gamma_{\mathrm{1}} \envplus \Gamma_{\mathrm{3}} \vdash A \sim B \dashv \Gamma_{\mathrm{2}} \envplus \Gamma_{\mathrm{4}}}
      {\Gamma_{\mathrm{1}} \vdash \Gamma_{\mathrm{3}}, [A] \sim \Gamma_{\mathrm{4}}, [B] \dashv \Gamma_{\mathrm{2}}}
  \end{drulepar}
  \caption{Type equivalence.}\label{fig:teq}
\end{figure}
\end{small}
Figure~\ref{fig:teq} shows the definition of type equivalence. The judgment is
written $\Gamma_{\mathrm{1}} \vdash A \sim B \dashv \Gamma_{\mathrm{2}}$. It compares a type $A$ under a context
$\Gamma_{\mathrm{1}}$ with a type $B$ under a context $\Gamma_{\mathrm{2}}$. The two contexts may
differ, and are needed for box types. A box type is self-contained
under its own local context. Comparing one therefore switches to that local
context on a single side (\rref{Teq-boxl} and \rref{Teq-boxr}). The two types are
then read under different contexts. A single shared context could not express
this. Everywhere else the contexts stay \emph{aligned}: a type variable
of a given de Bruijn index refers to corresponding binders on both sides. This
alignment is an important invariant for the metatheory of type equivalence
(Section~\ref{sec:meta}).
An \emph{abstract} type variable $\mathit{N}$ is equivalent to itself when it is
well-scoped in both $\Gamma_{\mathrm{1}}$ and $\Gamma_{\mathrm{2}}$ (\rref{Teq-tvar}). For \emph{concrete}
type variables, the corresponding type is fetched from the context and equivalence
is checked against that type (\rref{Teq-eql} and \rref{Teq-eqr}). \Rref{Teq-forall}
adds an abstract type binding to both sides, preserving the alignment. A manifest
type $[A] B$ creates a concrete type binding $[A]$ in the context for
$B$, so that type variable $\alpha$ in $B$ refers to this newly created
binding. Since this binder is added on only one side, \rref{Teq-manil} and
\rref{Teq-manir} restore the alignment by weakening the other side with an extra
abstract binding $\tstar$ and shifting its type, so that its de Bruijn indices
step over the added binding.
For environment types,
elements in an environment can depend on not only the global context,
but also the older local context. Thus, when checking for equivalence
of two environment types, we need to check the equivalence of each
corresponding element under the extended contexts.
Our relation for type equivalence has the following properties.
\begin{theorem}[Reflexivity]
  If $\Gamma \vdash A$ then $\Gamma \vdash A \sim A \dashv \Gamma$.
\end{theorem}
\begin{theorem}[Symmetry]
  If $\Gamma_{\mathrm{1}} \vdash A \sim B \dashv \Gamma_{\mathrm{2}}$ then $\Gamma_{\mathrm{2}} \vdash B \sim A \dashv \Gamma_{\mathrm{1}}$.
\end{theorem}
\begin{theorem}[Transitivity]
  If $\Gamma_{\mathrm{1}} \vdash A \sim B \dashv \Gamma_{\mathrm{2}}$ and $\Gamma_{\mathrm{2}} \vdash B \sim C \dashv \Gamma_{\mathrm{3}}$ then $\Gamma_{\mathrm{1}} \vdash A \sim C \dashv \Gamma_{\mathrm{3}}$.
\end{theorem}

\subsection{Environment-based Semantics}\label{subsec:sem}

Figure~\ref{fig:small} shows \name's call-by-value small-step
environment-based semantics, with the
form of $\delta \vdash e \hookrightarrow e'$ where $\delta$ denotes the runtime environment.
The semantics passes types explicitly at run time, keeping type
information in both terms and environments.
\Rref{Step-var} looks up the value of a variable from the runtime environment via the lookup function $\delta [\mathit{n}]$ defined as:
\begin{center}
  $(\delta, v) [a] = v$; \quad \quad $(\delta, v) [\mathit{S} \, \mathit{n}] = \delta [\mathit{n}]$; \quad \quad $(\delta, [A]) [\mathit{n}] = \delta [\mathit{n}]$
\end{center}
Similarly to the semantics of \lambdae~\cite{DBLP:journals/pacmpl/TanO24}, an application $e_{\mathrm{1}} \, e_{\mathrm{2}}$ evaluates $e_{\mathrm{1}}$ to a closure $\eclos{\delta_{\mathrm{1}}}{\lambda e}$, then $e_{\mathrm{2}}$ to a value $v$, and finally applies the closure to the argument by adding $v$ into the lexical environment $\delta_{\mathrm{1}}$ for the closure, using the updated environment to evaluate the function body $e$ (\rref{Step-beta}).

For type abstractions, \name uses weak reduction, following
treatments of System F in standard
textbooks~\cite{DBLP:books/daglib/0005958,Harper_2016}.  Reduction
never proceeds under the $\Lambda$ binder for $\Lambda e$, and the body
$e$ is evaluated when a type argument is supplied.
An alternative is to reduce under the $\Lambda$ binder, extending the runtime
environment with an abstract binding for the bound variable, reducing the body, and
reading the result back into a term, as in the Cr{\'e}gut
machine~\cite{cregut2007strongly} (a full-reduction variant of the Krivine
machine~\cite{krivine2007call}). This needs extra machinery such as de Bruijn
levels, so we adopt weak reduction for simplicity.

\begin{figure}[t]
  \begin{drulepar}{\delta \vdash e \hookrightarrow e'}{Small-step Operational Semantics}
    \drule{Step-eval}
      {\delta \vdash e \hookrightarrow e'}
      {\delta \vdash F [e] \hookrightarrow F [e']}

    \drule{Step-var}{\delta [\mathit{n}] = v}{\delta \vdash \mathit{n} \hookrightarrow v}

    \drule{Step-box}
      {\delta_{\mathrm{1}} \vdash e_{\mathrm{2}} \hookrightarrow e'_{\mathrm{2}}}
      {\delta \vdash \delta_{\mathrm{1}} \triangleright e_{\mathrm{2}} \hookrightarrow \delta_{\mathrm{1}} \triangleright e'_{\mathrm{2}}}

    \drule{Step-boxv}
      {}
      {\delta \vdash \delta_{\mathrm{1}} \triangleright v_{\mathrm{2}} \hookrightarrow v_{\mathrm{2}}}

    \drule{Step-tapp}
      {}
      {\delta \vdash \eclos{\delta_{\mathrm{1}}}{\Lambda e} \, A \hookrightarrow (\delta_{\mathrm{1}}, [| \delta | \triangleright A]) \triangleright e}

    \drule{Step-beta}
      {}
      {\delta \vdash \eclos{\delta_{\mathrm{1}}}{\lambda e} \, v \hookrightarrow (\delta_{\mathrm{1}}, v) \triangleright e}

    \drule{Step-clos}{}{\delta \vdash \lambda e \hookrightarrow \eclos{\delta}{\lambda e}}

    \drule{Step-bclos}
      {}
      {\delta \vdash \Lambda e \hookrightarrow \eclos{\delta}{\Lambda e}}

    \drule{Step-edef}
      {\delta \envplus \delta_{\mathrm{1}} \vdash e \hookrightarrow e'}
      {\delta \vdash \delta_{\mathrm{1}}, e \hookrightarrow \delta_{\mathrm{1}}, e'}

    \drule{Step-tdef}
      {\neg \mathsf{box}(A)}
      {\delta \vdash \delta_{\mathrm{1}}, [A] \hookrightarrow \delta_{\mathrm{1}}, [| \delta \envplus \delta_{\mathrm{1}} | \triangleright A]}
  \end{drulepar}
  \caption{Dynamic semantics.}\label{fig:small}
\end{figure}

For type applications, we use a similar strategy as in term applications, to avoid using meta-level substitution.
A type application $e \, A$ first evaluates $e$ to a
closure $\eclos{\delta_{\mathrm{1}}}{\Lambda e'}$, and then the type argument $A$ is pushed into $\delta_{\mathrm{1}}$ for evaluating the body $e'$.
However, since type $A$ can depend on the
global environment, we need to make the type argument closed before
pushing it into the lexical environment: we turn $A$
into a box type $| \delta | \triangleright A$ which is always closed, and then push
the box type into the lexical environment $\delta_{\mathrm{1}}$
(\rref{Step-tapp}). In this rule, the operation $\mid \cdot \mid$
extracts the type information from the global environment $\delta$ for
$A$. Then it constructs a concrete local context to form the box type. The definition of $| \delta |$ is given in Appendix~\appAux.

\Rref{Step-box} evaluates a box $e_{\mathrm{1}} \triangleright e_{\mathrm{2}}$: $e_{\mathrm{1}}$
is first evaluated to a value environment $\delta_{\mathrm{1}}$, and then $e_{\mathrm{2}}$ is
reduced under environment $\delta_{\mathrm{1}}$. \Rref{Step-boxv} extracts the value from the box
after evaluation reaches a value.  Evaluation of an
environment is also from left to right. The older environment
is first evaluated to a value environment $\delta_{\mathrm{1}}$. Then, if the last
element is an expression $e$, it is evaluated under the concatenation $\delta \envplus \delta_{\mathrm{1}}$ (\rref{Step-edef}).
The operation $\texttt{+}$ (defined in Appendix~\appAux) appends the entries of the second environment to the first, so the bindings of both are in scope.
When the
last element is a type $A$ then, if $A$ is a box type, the environment
is already a value. Otherwise, type $A$ may depend on the global environment or the previous elements, and thus we make the type
definition closed by turning it into a box type $| \delta \envplus \delta_{\mathrm{1}} | \triangleright A$
(\rref{Step-tdef}).

\paragraph{Computational irrelevance of types}
The semantics of Figure~\ref{fig:small} is \emph{type-passing}: type arguments
are carried at runtime and appear as entries in runtime environments
(\rref{Step-tapp}), so reduction does some computation on types. Types, however,
are only needed for static checking and do not affect the result. We make this
precise with a \emph{type-erasing} semantics that drops all types from terms and
environments, along with the associated runtime computation (Appendix~\appErase).
An erase function $\bra{e}$ removes types from expressions, following
TAPL~\citep{DBLP:books/daglib/0005958}, and we mark the type-erasing semantics with the
subscript $|\rm{i}|$. We prove that it reaches the same (erased) values as the
type-passing semantics (Theorem~\ref{thm:irrelevance}).

\begin{theorem}[Computational Irrelevance of Types]\label{thm:irrelevance}
  If $\delta \vdash e \hookrightarrow^* v'$, then $\bra{\delta} \vdash \bra{e} \hookrightarrow_{|\rm{i}|}^{*} \bra{v'}$.
\end{theorem}

The type-passing and type-erasing semantics are two presentations of
the same calculus, differing only in whether types are kept at run
time. We adopt the type-passing semantics as primary: it is the proof
vehicle for conservativity of the dynamics, which must match
System~F's runtime type substitution (Section~\ref{sec:meta}).
Section~\ref{sec:implicit-poly} further discusses an implicitly-typed
variant, where a polymorphic value is used directly without type
abstraction or application, as in Haskell and OCaml.

\section{Metatheory of \name}\label{sec:meta}

This section develops the metatheory of \name. The environment semantics yields
instantiation lemmas that replace the conventional substitution lemmas, and we
use them to prove type soundness, conservativity with respect to System F,
normalization, and decidability of type equivalence.

\subsection{Instantiation Lemmas and Decidability of Type Equivalence}
The instantiation lemma for type equivalence states that, if types $A$
and $B$ are equivalent, and if we instantiate abstract type
bindings with two equivalent types in the contexts, then type
equivalence between $A$ and $B$ still holds under the concretized
contexts.

\begin{lemma}[Instantiation]\label{lem:inst}
  If $\Gamma_{\mathrm{1}}, \tstar \vdash A \sim B \dashv \Gamma_{\mathrm{2}}, \tstar$ and $\Gamma_{\mathrm{1}} \vdash C \sim D \dashv \Gamma_{\mathrm{2}}$, then $\Gamma_{\mathrm{1}}, [C] \vdash A \sim B \dashv \Gamma_{\mathrm{2}}, [D]$.
\end{lemma}

To prove this lemma, a naive induction on $\Gamma_{\mathrm{1}}, \tstar \vdash A \sim B \dashv \Gamma_{\mathrm{2}}, \tstar$
does not work, since during the process of checking type equivalence,
it cannot be guaranteed that the last elements on both sides of the
contexts are always abstract type bindings. For example, in
\rref{Teq-manil} and \rref{Teq-manir}, a concrete type binding
is added to the context. Hence, we need to generalize the lemma to
allow additional context elements after the abstract type
bindings.

\begin{lemma}[Generalized Instantiation]\label{lem:gen-inst}
  If $(\Gamma_{\mathrm{1}}, \tstar) \envplus \Gamma_{\mathrm{3}} \vdash A \sim B \dashv (\Gamma_{\mathrm{2}}, \tstar) \envplus \Gamma_{\mathrm{4}}$ and $\lvert \Gamma_{\mathrm{3}} \rvert = \lvert \Gamma_{\mathrm{4}} \rvert$ and $\Gamma_{\mathrm{1}} \vdash C \sim D \dashv \Gamma_{\mathrm{2}}$, then $(\Gamma_{\mathrm{1}}, [C]) \envplus \Gamma_{\mathrm{3}} \vdash A \sim B \dashv (\Gamma_{\mathrm{2}}, [D]) \envplus \Gamma_{\mathrm{4}}$.
\end{lemma}

In this generalized form, $\lvert \Gamma \rvert$ denotes the number of type bindings in context $\Gamma$. The lemma
requires that both sides of the contexts have the same number of
additional type bindings. Importantly, this invariant is
maintained for most cases of the type equivalence checking, and thus
we can prove the lemma by induction on the generalized judgment.
For example, when checking the equivalence for universal types (\rref{Teq-forall}) and
manifest types (\rref{Teq-manil} and \rref{Teq-manir}), each of these three
rules extends both contexts by exactly one type binding, so the numbers of
additional type bindings on both sides stay equal. Furthermore,
when checking the equivalence of two non-empty environment types $\Gamma_{\mathrm{3}},\_$ and $\Gamma_{\mathrm{4}},\_$ (\rref{Teq-edef}, \rref{Teq-star}, and \rref{Teq-tdef}),
the equivalence of older environments $\Gamma_{\mathrm{3}}$ and $\Gamma_{\mathrm{4}}$ is checked. The following lemma ensures that the numbers of type bindings in the older
environments are equal. Hence, after the concatenation of two
environments, the invariant still holds.

\begin{lemma}
  If $\Gamma_{\mathrm{1}} \vdash \Gamma_{\mathrm{3}} \sim \Gamma_{\mathrm{4}} \dashv \Gamma_{\mathrm{2}}$, then $\lvert \Gamma_{\mathrm{3}} \rvert = \lvert \Gamma_{\mathrm{4}} \rvert$.
\end{lemma}

The box cases require special consideration. In \rref{Teq-boxl} and
\rref{Teq-boxr}, the local context replaces the global context for
checking the equivalence for the body type. This way, the invariant
requiring that the numbers of additional type bindings on
both sides are equal can fail. For example, with \rref{Teq-boxl}, to check $\cdot, \hdecl{\gamma}{\tstar}, \hdecl{\beta}{\tstar}, \hdef{\alpha}{\tint} \vdash (\cdot, \hdef{\beta}{\tint}, \hdecl{\alpha}{\tstar}) \triangleright \beta \sim \alpha \dashv \cdot, \hdecl{\gamma}{\tstar}, \hdecl{\beta}{\tstar}, \hdef{\alpha}{\tint}$, we need to check
$\cdot, \hdef{\beta}{\tint}, \hdecl{\alpha}{\tstar} \vdash \beta \sim \alpha \dashv \cdot, \hdecl{\gamma}{\tstar}, \hdecl{\beta}{\tstar}, \hdef{\alpha}{\tint}$. Now the left side has two type bindings while the right side has
three. However, this case is not problematic. The body type of a box does
not refer to any abstract type binding of its local context, as it is
required to be rigid by the well-formedness condition for box types. Moreover,
type equivalence preserves rigidity: whenever one side is rigid, so is the other. Thus it
is safe to instantiate the abstract binding on the right-hand side to a
concrete one without affecting the equivalence, as formalized in the following lemma.

\begin{lemma}
  If $\Gamma_{\mathrm{1}} \vdash A \sim B \dashv (\Gamma_{\mathrm{2}}, \tstar) \envplus \Gamma_{\mathrm{3}}$ and $\Gamma_{\mathrm{1}} \vdash_{r} A$ and $\vdash (\Gamma_{\mathrm{2}}, [C]) \envplus \Gamma_{\mathrm{3}}$, then $\Gamma_{\mathrm{1}} \vdash A \sim B \dashv (\Gamma_{\mathrm{2}}, [C]) \envplus \Gamma_{\mathrm{3}}$.
\end{lemma}

We also prove a generalized instantiation lemma for typing, by induction on the typing derivation and using the type-equivalence instantiation lemma in the \rref*{Typ-teq} case.
\begin{lemma}[Generalized Instantiation for Typing]
  If $(\Gamma, \tstar) \envplus \Gamma_{\mathrm{1}} \vdash e : B$ and $\vdash (\Gamma, [A]) \envplus \Gamma_{\mathrm{1}}$, then $(\Gamma, [A]) \envplus \Gamma_{\mathrm{1}} \vdash e : B$.
\end{lemma}

By picking $\Gamma_{\mathrm{1}}$ to be empty, we have the following form of instantiation lemma for typing.
\begin{lemma}[Instantiation for Typing]
  If $\Gamma, \tstar \vdash e : B$ and $\Gamma \vdash A$, then $\Gamma, [A] \vdash e : B$.
\end{lemma}

\paragraph{Decidability of type equivalence}
The type equivalence relation $\Gamma_{\mathrm{1}} \vdash A \sim B \dashv \Gamma_{\mathrm{2}}$ is decidable. To establish this, we define size measures $\size{A}{\Psi}$ on types and $\mkctx{\Gamma}{\Psi}$ on
contexts, both parameterized by a
measure context defined as $\Psi \Coloneqq \cdot ~|~ \Psi, k ~|~
\Psi, [k]$.
Intuitively, the measure context $\Psi$ keeps track of
the sizes of bindings in the typing context, where $k$ is for
term variable bindings and $[k]$ is for type variable
bindings.
Here $\size{A}{\Psi}$ gives the size of a type $A$, using $\Psi$ to supply the sizes of $A$'s free type variables. The context measure $\mkctx{\Gamma}{\Psi}$ computes the size of each binding of $\Gamma$, returning a list of sizes which is itself a measure context. The context measure takes $\Psi$ as a parameter because an environment type $\Gamma$ can depend on the surrounding context: a binding of $\Gamma$ may mention type variables bound outside $\Gamma$, whose sizes are supplied by $\Psi$.
The definitions of $\size{A}{\Psi}$ and $\mkctx{\Gamma}{\Psi}$ are given in Appendix~\appDecide.
These two measures yield the lemma below; decidability is then a corollary.
\begin{lemma}
  If $\size{A}{\mkctx{\Gamma_{\mathrm{1}}}{\cdot}} + \size{B}{\mkctx{\Gamma_{\mathrm{2}}}{\cdot}}\le k$, then it is decidable whether $\Gamma_{\mathrm{1}} \vdash A \sim B \dashv \Gamma_{\mathrm{2}}$ holds.
\end{lemma}

\begin{corollary}[Decidability]
  $\Gamma_{\mathrm{1}} \vdash A \sim B \dashv \Gamma_{\mathrm{2}}$ is decidable.
\end{corollary}

\subsection{Type Soundness and Normalization}
\paragraph{Syntactic type soundness} We first prove type soundness via progress and preservation~\cite{WRIGHT199438}. Since the type equivalence rule is included, the proofs of
progress and preservation cannot be based solely on case analysis of
the typing derivation. Instead, they require appropriate inversion
lemmas for values. We have the following inversion lemmas for term and type abstractions; besides
the canonical forms, they give typing derivations and type-equivalence
information.

\begin{lemma}[Inversion Lemma for Term Abstractions]
  If $\Gamma \vdash v : A \rightarrow B$, then there exist $\delta$, $e$, $\Gamma_{\mathrm{1}}$, $C$, $D$, s.t. $v = \eclos{\delta}{\lambda e}$ and $\cdot \vdash \delta : \Gamma_{\mathrm{1}}$ and $\Gamma_{\mathrm{1}}, C \vdash e : D$ and $\Gamma \vdash A \sim C \dashv \Gamma_{\mathrm{1}}$ and $\Gamma \vdash B \sim D \dashv \Gamma_{\mathrm{1}}$.
\end{lemma}

\begin{lemma}[Inversion Lemma for Type Abstractions]\label{lem:inv-tapp}
  If $\Gamma \vdash v : \forall A$, then there exist $\delta$, $e$, $\Gamma_{\mathrm{1}}$, $B$, s.t. $v = \eclos{\delta}{\Lambda e}$ and $\cdot \vdash \delta : \Gamma_{\mathrm{1}}$ and $\Gamma_{\mathrm{1}}, \tstar \vdash e : B$ and $\Gamma, \tstar \vdash A \sim B \dashv \Gamma_{\mathrm{1}}, \tstar$.
\end{lemma}

\paragraph{Lookup lemmas}

Since the semantics is environment-based, we also prove progress and
preservation for the lookup function: if the runtime environment $\delta$ is
well-typed, any variable lookup in $\delta$ succeeds and preserves the type.

\begin{lemma}[Progress of Lookup]
  If $\Gamma [\mathit{n}] = A$ and $\cdot \vdash \delta : \Gamma$, then there exists $v$ s.t. $\delta [\mathit{n}] = v$.
\end{lemma}

\begin{lemma}[Preservation of Lookup]
  If $\Gamma [\mathit{n}] = A$, $\cdot \vdash \delta : \Gamma$, and $\delta [\mathit{n}] = v$, then $\Gamma \vdash v : A$.
\end{lemma}

\paragraph{Progress and preservation}
We have the following generalized progress and preservation theorems,
which have the form similar to the ones in previous
work~\cite{DBLP:journals/pacmpl/TanO24} on small-step
environment-based semantics. The generalized form of theorems
considers an arbitrary well-typed runtime environment $\delta$ instead
of the empty environment.

\begin{theorem}[Generalized Progress]\label{lem:gprog}
  If $\Gamma \vdash e : A$ and $\cdot \vdash \delta : \Gamma$, then either $e$ is a value or $\exists e', \delta \vdash e \hookrightarrow e'$.
\end{theorem}

\begin{theorem}[Generalized Preservation]\label{lem:gpres}
  If $\Gamma \vdash e : A$ and $\cdot \vdash \delta : \Gamma$ and $\delta \vdash e \hookrightarrow e'$, then $\Gamma \vdash e' : A$.
\end{theorem}

The generalized progress theorem follows by canonical forms of
values, which handle the application and type application
cases. For other cases, the proof follows a standard structure. The
generalized preservation theorem is proved by first doing induction on
the typing derivation $\Gamma \vdash e : A$ and then inversion on the
reduction relation $\delta \vdash e \hookrightarrow e'$. The proof of
preservation follows similar reasoning to that
of~\citet{DBLP:journals/pacmpl/TanO24}. In particular, the reduction
rules for closures and boxes in \name are quite similar.
However, reasoning is more involved due to
the more powerful type system of \name. Closures and boxes now
have box types in \name, thus the preservation proof for them
requires type equivalence reasoning. In \rref{Step-clos}, a lambda
abstraction of type $A \rightarrow B$ is converted to a closure of type
$\Gamma \triangleright (A \rightarrow B)$, where $\Gamma$ is the type of the runtime
environment. Since $A \rightarrow B$ and $\Gamma \triangleright (A \rightarrow B)$ are equivalent
under $\Gamma$, we can use the typing rule for type equivalence
(\rref{Typ-teq}) to conclude that the closure preserves the type.
Similar reasoning applies to \rref{Step-bclos} for closures of type
abstraction. Moreover, \rref{Step-boxv} extracts the body value $v$
from a box. The box has type $\Gamma \triangleright A$, while $v$ has type
$A$. The following lemma ensures that the type of a value can be
boxed with the type of its runtime environment $\delta$. Thus \rref{Step-boxv} also preserves the type.

\begin{lemma}
  If $\cdot \vdash \delta : \Gamma$ and $\Gamma \vdash v : A$ and $\vdash \Gamma_{\mathrm{1}}$, then $\Gamma_{\mathrm{1}} \vdash v : \Gamma \triangleright A$.
\end{lemma}

Next we focus on the most interesting case, which is the one for type
applications (\rref{Step-tapp}). Suppose that $\cdot \vdash \delta : \Gamma$ and $\Gamma \vdash \eclos{\delta_{\mathrm{1}}}{\Lambda e} \, A : [A] B$, we want to show that $\Gamma \vdash (\delta_{\mathrm{1}}, [| \delta | \triangleright A]) \triangleright e : [A] B$.  From the typing judgment for
the type application, we have the typing judgment for the closure $\Gamma \vdash \eclos{\delta_{\mathrm{1}}}{\Lambda e} : \forall B$. Moreover, by the inversion lemma (Lemma
\ref{lem:inv-tapp}), there exist $\Gamma_{\mathrm{1}}$ and $C$ such that
\begin{center}
  (1) $\cdot \vdash \delta_{\mathrm{1}} : \Gamma_{\mathrm{1}}$; \quad (2) $\Gamma_{\mathrm{1}}, \tstar \vdash e : C$; \quad (3) $\Gamma, \tstar \vdash B \sim C \dashv \Gamma_{\mathrm{1}}, \tstar$
\end{center}
According to the typing rule for boxes (\rref{Typ-box}), we need to
make sure that $(\delta_{\mathrm{1}}, [| \delta | \triangleright A]) \triangleright e$ is well-typed, and
construct a box type that is equivalent to $[A] B$ so that the
box can typecheck via \rref{Typ-teq}. To be concrete, we need:

\begin{itemize}
  \item {\bf instantiation lemma for typing}, to fill the gap between
    the inversion lemma and the typing rule for boxes. From (1), we
    know that the local environment $\delta_{\mathrm{1}}, [| \delta | \triangleright A]$ in the
    box has type $\Gamma_{\mathrm{1}}, [| \delta | \triangleright A]$ by \rref{Typ-tdef}. Then we
    need to show that the body $e$ in the box is well-typed under the
    local context $\Gamma_{\mathrm{1}}, [| \delta | \triangleright A]$. From (2), $e$ is well-typed
    under an uninstantiated context $\Gamma_{\mathrm{1}}, \tstar$. By the
    instantiation lemma for typing, we have $\Gamma_{\mathrm{1}}, [| \delta | \triangleright A] \vdash e : C$.
  \item {\bf instantiation lemma for type equivalence}, to handle the
    equivalence between the manifest type and the box type. From
    above we know that the box has type $(\Gamma_{\mathrm{1}}, [| \delta | \triangleright A]) \triangleright C$. It remains to show that this type is equivalent to $[A] B$ under $\Gamma$. With the type equivalence rules for
    manifest types and box types, it suffices to show that $\Gamma, [A] \vdash B \sim C \dashv \Gamma_{\mathrm{1}}, [| \delta | \triangleright A]$. This comes from (3)
    and the instantiation lemma for type equivalence. Note that the
    types used for instantiation must be equivalent. Here we need to
    prove $\Gamma \vdash A \sim | \delta | \triangleright A \dashv \Gamma_{\mathrm{1}}$, which reduces to $\Gamma \vdash A \sim A \dashv | \delta |$. The context $\Gamma$ may differ from
    $|\delta|$, but the two are equivalent: they contain equivalent
    types at corresponding positions. Reflexivity therefore holds
    across them, as formalized below.
\end{itemize}

\begin{lemma}
  If $\cdot \vdash \delta : \Gamma$ and $\Gamma \vdash A$, then $\Gamma \vdash A \sim A \dashv | \delta |$.
\end{lemma}

Standard progress and
preservation theorems now follow as corollaries:

\begin{theorem}[Progress]
  If $\cdot \vdash e : A$, then either $e$ is a value or $\exists e', \cdot \vdash e \hookrightarrow e'$.
\end{theorem}

\begin{theorem}[Preservation]
  If $\cdot \vdash e : A$ and $\cdot \vdash e \hookrightarrow e'$, then $\cdot \vdash e' : A$.
\end{theorem}

\paragraph{Semantic typing and normalization}
We also provide semantic typing, through a unary logical relation
in the style of reducibility~\citep{Tait67,girard1972interpretation}. The
relation interprets every type $A$ as a set of values and every typing
context $\Gamma$ as a set of well-typed runtime environments.
A term is semantically well-typed,
written $\Gamma \models e : A$, if under every runtime environment that realizes
$\Gamma$ it reduces to a value in the interpretation of $A$. The fundamental
theorem of the relation shows that syntactic typing implies semantic typing, and
is proved by induction on the typing derivation with one compatibility lemma per
typing rule.

\begin{theorem}[Fundamental Theorem]
  If $\Gamma \vdash e : A$, then $\Gamma \models e : A$.
\end{theorem}

\noindent Our definition of semantic typing entails that
evaluation reaches a value, so the fundamental theorem yields normalization.

\begin{corollary}[Normalization]
  If $\cdot \vdash e : A$, then $\cdot \vdash e \hookrightarrow^* v$ for some value $v$.
\end{corollary}

\noindent Here reduction is call-by-value and never goes under $\lambda$ or
$\Lambda$ binders. The corollary states that closed, well-typed terms evaluate to
a value, which is weaker than strong normalization of full reduction.
The logical relation and semantic typing ($\Gamma \models e : A$)
are presented in detail in Appendix~\appSem.

\subsection{Conservativity}
Our calculus \name is capable of simulating and expressing different
aspects of the semantics of System F. In what follows, we relate \name
to System F.  For avoiding ambiguity, we use $\vdash_{F} \Gamma$, $\vdash_{F} e$,
$\vdash_{F} A$ and $\Gamma \vdash_{F} e : A$ to denote typing contexts,
expressions, types, and the typing judgment respectively in System
F. We first show that the type system of \name is complete with
respect to System F typing. That is, any typing judgment derivable in
System F is also derivable in \name.

\begin{theorem}[Completeness w.r.t. Typing]
 If $\Gamma \vdash_{F} e : A$, then $\Gamma \vdash e : A$.
\end{theorem}
The interesting case in proving this theorem is for type
application. One key difference between \name and System F is that,
for type application, System F uses meta-level type substitution,
while \name uses manifest types to delay the substitution. Hence,
we need to relate meta-level substitution $B [\mathit{0} \mapsto A]$ and
manifest type $[A] B$ via type equivalence, as formalized in the following lemma.

\begin{lemma}[Equivalence for Type Substitution]\label{lem:open-subst}
  If $\vdash_{F} \Gamma$, $\vdash_{F} B$, $\vdash_{F} C$, $\Gamma \vdash C$, and $\Gamma, \tstar \vdash A \sim B \dashv \Gamma, \tstar$, then $\Gamma \vdash [C] A \sim B [\mathit{0} \mapsto C] \dashv \Gamma$.
\end{lemma}

\paragraph{Conservativity of typing}
Now we turn to the reverse direction. For typing contexts,
expressions, and types in System F, if the typing relation is valid in
\name, then it should also be valid in System F.

\begin{theorem}[Conservativity w.r.t. Typing]\label{thm:cons}
  If $\vdash_{F} \Gamma$, $\vdash_{F} e$, $\vdash_{F} A$, and $\Gamma \vdash e : A$, then $\Gamma \vdash_{F} e : A$.
\end{theorem}
The reasoning is trickier in this direction, since there is type
equivalence in the type system of \name. When deriving $\Gamma \vdash_{F} e : A$ from $\Gamma \vdash e : A$, if we do induction on the typing
relation, then for the type equivalence case (\rref{Typ-teq}), we will
reach some type that is not in System F. Consider the following simple
example:
\begin{mathpar}
  \inferrule* [Right=Typ-teq]
          {\cdot \vdash \text{\texttt{42} } : \hmani{\alpha}{\tint}{\alpha} \\ \cdot \vdash \hmani{\alpha}{\tint}{\alpha} \sim \tint \dashv \cdot }
          {\cdot \vdash \text{\texttt{42} } : \tint}
\end{mathpar}
In this example, $\cdot \vdash \text{\texttt{42} } : \tint$ does not mention types in
\name, but the typing subderivation $\cdot \vdash \text{\texttt{42} } : \hmani{\alpha}{\tint}{\alpha}$
contains an \name type $\hmani{\alpha}{\tint}{\alpha}$. Therefore, the induction
hypothesis cannot be applied, as condition $\vdash_{F} A$ cannot be
satisfied. This problem is addressed by proving the theorem via two
steps: first proving a relaxed form of lemma that does not require
types to be in System F, and then using this lemma to derive the
original theorem. The lemma is stated as following:
\begin{lemma}[Relaxed Conservativity]\label{lem:relaxed}
  If $\vdash_{F} \Gamma$, $\vdash_{F} e$ and $\Gamma \vdash e : A$, then there exists $B$, s.t.,  $\Gamma \vdash A \sim B \dashv \Gamma$ and $\Gamma \vdash_{F} e : B$ and $\vdash_{F} B$.
\end{lemma}
Compared with Theorem \ref{thm:cons}, this lemma allows type $A$ not to be
 in System F. Instead, it finds a System F type $B$ that is
equivalent to $A$ under $\Gamma$. This way, we can do induction on $\Gamma \vdash e : A$ without worrying about encountering types that are not
in System F.  Moreover, if $A$ is already in System F, then by the
following uniqueness lemma (Lemma~\ref{lem:unique}), we know that any equivalent type $B$ in System F must
be syntactically the same as $A$. Hence, conservativity theorem
follows as a corollary of Lemma \ref{lem:relaxed}.
\begin{lemma}[Uniqueness]\label{lem:unique}
  If $\vdash_{F} \Gamma$, $\vdash_{F} A$, $\vdash_{F} B$, and $\Gamma \vdash A \sim B \dashv \Gamma$, then $A = B$.
\end{lemma}

In proving Lemma~\ref{lem:relaxed}, the interesting case is again type
application, where we rely on Lemma~\ref{lem:open-subst} to relate manifest types
and meta-level substitutions under type equivalence.

\begin{figure}[t]
\begin{drulepar}{v \sim v_{F}}{Value Correspondence}
  \drule{Vrel-lit}{}{\mathtt{i} \sim \mathtt{i}}

  \drule{Vrel-abs}
    {\mathsf{T}(\delta, e', \mathit{0}) = e_{\mathrm{1}} \and \mathsf{S}(\delta, e_{\mathrm{1}}, \mathit{1}) = e}
    {\eclos{\delta}{\lambda e'} \sim \lambda e}

  \drule{Vrel-tabs}
    {\mathsf{T}(\delta, e', \mathit{1}) = e_{\mathrm{1}} \and \mathsf{S}(\delta, e_{\mathrm{1}}, \mathit{0}) = e}
    {\eclos{\delta}{\Lambda e'} \sim \Lambda e}
\end{drulepar}
\caption{Value relation between \name and System F.}\label{fig:Vrel}
\end{figure}

\paragraph{Conservativity of the dynamic semantics}
We connect our semantics (Section~\ref{subsec:sem}) with System~F's
call-by-value, substitution-based semantics. The type-passing semantics is essential here: System F evaluates a type application such as
$(\hLam{\alpha}{e} \, \alpha) \, \tint$ by substituting $\tint$ into the body,
which contains the free variable $\alpha$. To model this without meta-level
substitution, we must keep type information in the runtime environment. A
type-erasing semantics discards it and cannot capture System F's substitution
behavior.

\citet{DBLP:journals/pacmpl/TanO24} relate \lambdae to the STLC via an operation
$S(\delta,e,k)$ that \emph{closes} a \lambdae expression $e$ under $\delta$:
it replaces each free term variable of $e$ by its value in $\delta$, treating the
innermost $k$ variables as bound. A closure $\eclos{\delta}{\lambda e'}$ then corresponds to
the STLC abstraction $\lambda e$ with $S(\delta,e',1) = e$. We extend this to type
variables and type-abstraction closures in \name with a companion operation
$T(\delta,e,k)$ that closes the \emph{types} occurring in $e$: it resolves each
free type variable against the type entries of $\delta$, and when such an entry is
a box type $\Gamma \triangleright A$ it recursively closes $A$ under $\Gamma$ to yield a
plain System~F type. With $S$ and $T$ we define the value relation $v \sim v_{F}$
between an \name value $v$ and a System~F value $v_{F}$ (Figure~\ref{fig:Vrel}).
For a closure body $e'$ we first close its types with $T$ and then its term
variables with $S$. Since type and term variables are indexed separately, for
$\eclos{\delta}{\lambda e'}$ the depth is $0$ for $T$ and $1$ for $S$ (the body sits under a
$\lambda$ binder), and for $\eclos{\delta}{\Lambda e'}$ it is $1$ for $T$ and $0$ for $S$.
We prove that \name's dynamic semantics is complete and conservative with
respect to System~F.

\begin{theorem}[Completeness w.r.t. Dynamics]
  If $\cdot \vdash_{F} e : A$ and $e \hookrightarrow^* v_{F}$, then there exists $v$ s.t. $\cdot \vdash e \hookrightarrow^* v$ and $v \sim v_{F}$.
\end{theorem}

\begin{theorem}[Conservativity w.r.t. Dynamics]
  If $\vdash_{F} e$, $\cdot \vdash e : A$ and $\cdot \vdash e \hookrightarrow^* v$, then there exists $v_{F}$ s.t. $e \hookrightarrow^* v_{F}$ and $v \sim v_{F}$.
\end{theorem}

\section{Implicit Polymorphism}\label{sec:implicit-poly}

The calculus in Section~\ref{sec:calculus} has explicit terms for type abstraction and type application.
This choice keeps the substitution-free machinery working for explicit polymorphism where there are three forms of substitution, but our environment-based approach is not tied to it.
We also develop an implicitly-typed variant of \name, whose terms carry no type information: type
abstraction and type application are no longer term constructs, but typing coercions that leave the term
unchanged. In the explicitly-typed calculus a type abstraction $\Lambda e$ blocks until a type argument
is supplied. In contrast, the implicitly-typed variant has no such term, so polymorphic expressions reduce immediately,
as in languages such as Haskell and OCaml.
Types, well-formedness, and type equivalence are as in the explicitly-typed calculus; only the term syntax, the typing judgment, and the reduction relation
change.

The implicitly-typed variant relates to \name as Curry-style System~F relates to System~F. Curry-style
System~F has no type-into-term substitution (Section~\ref{subsec:threeforms}), but it still substitutes terms
into terms in $\beta$-reduction and types into types in its instantiation rule. As in the explicit calculus,
the variant eliminates both, through its environment-based semantics and manifest types.

\subsection{Syntax}\label{subsec:implicit-syntax}

\begin{figure}[t]
\setlength{\jot}{0pt}
\begin{align*}
  &\text{Expressions} &e &\Coloneqq \mathtt{i} ~|~ \mathit{n} ~|~ \lambda e ~|~ e_{\mathrm{1}} \, e_{\mathrm{2}} ~|~ e_{\mathrm{1}} \triangleright e_{\mathrm{2}} ~|~ \eclos{\delta}{\lambda e} ~|~ \cdot ~|~ e_{\mathrm{1}}, e_{\mathrm{2}} \\
  &\text{Environments} &\Delta &\Coloneqq \cdot ~|~ \Delta, e \\
  &\text{Values} &v &\Coloneqq \mathtt{i} ~|~ \eclos{\delta}{\lambda e} ~|~ \cdot ~|~ v_{\mathrm{1}}, v_{\mathrm{2}} \\
  &\text{Value environments} &\delta &\Coloneqq \cdot ~|~ \delta, v \\
  &\text{Frames} &F &\Coloneqq \eframe \, e ~|~ v \, \eframe ~|~ \eframe \triangleright e ~|~ \eframe , e
\end{align*}
\caption{Syntax of the implicitly-typed variant. Types and contexts are as in Figure~\ref{fig:syntax}.}
\label{fig:implicit-syntax}
\end{figure}

Figure~\ref{fig:implicit-syntax} shows the syntax of the implicitly-typed variant. It is the syntax of
\name (Figure~\ref{fig:syntax}) with all type-level constructs removed from terms: there is no type
abstraction, no type application, and no type definitions in environments.

\paragraph{Types and contexts}
Types and contexts are those of \name. Universal types $\forall A$, manifest types $[A] B$, and manifest
bindings $\Gamma, [A]$ remain, but are now introduced by typing coercions (Section~\ref{subsec:implicit-typing})
rather than by terms. Well-formedness (Figure~\ref{fig:typ}, top) and type equivalence (Figure~\ref{fig:teq})
are unchanged, as are their properties.

\paragraph{Expressions}
Expressions are those of \name without type abstractions $\Lambda e$, type applications $e \, A$, closures
for type abstractions $\eclos{\delta}{\Lambda e}$, and type definitions $e, [A]$. Environments $\Delta$ thus
contain only expressions, but they remain dependent: the type of an entry may refer to type bindings that
occur only in the environment's type (Section~\ref{subsec:implicit-typing}).

\paragraph{Values and frames}
Values no longer include closures for type abstractions, and value environments $\delta$ are plain sequences of values,
since no type ever enters a runtime environment. Nor do frames include $\eframe \, A$ or $\eframe , [A]$. Unlike the
type-erasing semantics of \name (Appendix~\appErase), which removes types but keeps type abstractions and a
dummy application $e \, {\_}$, the variant has no type abstractions to block evaluation.

\subsection{Typing}\label{subsec:implicit-typing}

\begin{figure}[t]
\begin{drulepar}{\Gamma \vdash e : A}{Typing}
  \drule{ITyp-int}{\vdash \Gamma}{\Gamma \vdash \mathtt{i} : \tint}

  \drule{ITyp-var}
    {\vdash \Gamma \and \Gamma [\mathit{n}] = A}
    {\Gamma \vdash \mathit{n} : A}

  \drule{ITyp-lam}{\Gamma, A \vdash e : B}{\Gamma \vdash \lambda e : A \rightarrow B}

  \drule{ITyp-app}
    {\Gamma \vdash e_{\mathrm{1}} : A \rightarrow B \and \Gamma \vdash e_{\mathrm{2}} : A}
    {\Gamma \vdash e_{\mathrm{1}} \, e_{\mathrm{2}} : B}

  \drule*{ITyp-gen}{\Gamma, \tstar \vdash e : A}{\Gamma \vdash e : \forall A}

  \drule*{ITyp-inst}
    {\Gamma \vdash e : \forall B \and \Gamma \vdash A}
    {\Gamma \vdash e : [A] B}

  \drule{ITyp-box}
    {\Gamma \vdash e_{\mathrm{1}} : \Gamma_{\mathrm{1}} \and \Gamma_{\mathrm{1}} \vdash e_{\mathrm{2}} : A}
    {\Gamma \vdash e_{\mathrm{1}} \triangleright e_{\mathrm{2}} : \Gamma_{\mathrm{1}} \triangleright A}

  \drule{ITyp-clos}
    {\vdash \Gamma \and \cdot \vdash \delta : \Gamma_{\mathrm{1}} \and \Gamma_{\mathrm{1}}, A \vdash e : B}
    {\Gamma \vdash \eclos{\delta}{\lambda e} : \Gamma_{\mathrm{1}} \triangleright (A \rightarrow B)}

  \drule{ITyp-teq}
    {\Gamma \vdash e : A \and \Gamma \vdash A \sim B \dashv \Gamma}
    {\Gamma \vdash e : B}

  \drule{ITyp-nil}{\vdash \Gamma}{\Gamma \vdash \cdot : \cdot}

  \drule{ITyp-cons}
    {\Gamma \vdash e_{\mathrm{1}} : \Gamma_{\mathrm{1}} \and \Gamma \envplus \Gamma_{\mathrm{1}} \vdash e_{\mathrm{2}} : A}
    {\Gamma \vdash e_{\mathrm{1}}, e_{\mathrm{2}} : \Gamma_{\mathrm{1}}, A}

  \drule{ITyp-mani}
    {\Gamma \vdash e : \Gamma_{\mathrm{1}} \and \Gamma \envplus \Gamma_{\mathrm{1}} \vdash A}
    {\Gamma \vdash e : \Gamma_{\mathrm{1}}, [A]}
\end{drulepar}
\caption{Type system of the implicitly-typed variant.}
\label{fig:implicit-typing}
\end{figure}

Figure~\ref{fig:implicit-typing} shows the type system of the implicitly-typed variant.
Every rule coincides with its counterpart in Figure~\ref{fig:typ}, except for three \emph{coercions},
\rref*{ITyp-gen}, \rref*{ITyp-inst}, and \rref*{ITyp-mani}, which leave the term untouched and take the place
of the rules for the term forms $\Lambda e$, $e \, A$, and $\Delta, [A]$.

\paragraph{Generalization and instantiation}
Polymorphism is handled by the first two coercions:
\[ \drule{ITyp-gen}{\Gamma, \tstar \vdash e : A}{\Gamma \vdash e : \forall A} \qquad \qquad \qquad \drule{ITyp-inst}{\Gamma \vdash e : \forall B \and \Gamma \vdash A}{\Gamma \vdash e : [A] B} \]
\Rref{ITyp-gen} introduces a polymorphic type $\forall A$ by checking the term under one additional abstract
type binding $\tstar$, just as the explicitly-typed calculus does when typing a type abstraction.
The term needs no adjustment under the new binding: it contains no types, and since term and type variables
are indexed separately (Section~\ref{sec:calculus}), the binding $\tstar$ does not shift the indices of its
term variables.
\Rref{ITyp-inst} instantiates a polymorphic type $\forall B$ at a well-formed type $A$ while leaving the
term unchanged, since the variant has no term-level type application. Crucially, rather than substituting $A$ into $B$ at the
meta level, \rref{ITyp-inst} records the instantiation as the manifest type $[A] B$ and leaves the actual unfolding to
the type-equivalence relation. The implicitly-typed variant is
therefore substitution-free, just like the explicitly-typed calculus.

To see the two coercions at work, recall the type application
$(\hLam{{\alpha}_{2}}{(\hLam{{\alpha}_{1}}{\mathsf{id}})} \, {\alpha}_{2}) \, \tint$ from
Section~\ref{subsec:elim}, where $\mathsf{id}$ is the identity function $\hlam{a}{a}$. Its implicitly-typed
counterpart is just $\mathsf{id}$, which can be given the same type. Omitting well-formedness premises, the
derivation is:
\[
\inferrule*[left=ITyp-inst]
  {\inferrule*[left=ITyp-gen]
    {\inferrule*[left=ITyp-inst]
      {\cdot, \hdecl{{\alpha}_{2}}{\tstar} \vdash \mathsf{id} : \hforall{{\alpha}_{1}}{{\alpha}_{1} \rightarrow {\alpha}_{1}}}
      {\cdot, \hdecl{{\alpha}_{2}}{\tstar} \vdash \mathsf{id} : \hmani{{\alpha}_{1}}{{\alpha}_{2}}{({\alpha}_{1} \rightarrow {\alpha}_{1})}}}
    {\cdot \vdash \mathsf{id} : \hforall{{\alpha}_{2}}{\hmani{{\alpha}_{1}}{{\alpha}_{2}}{({\alpha}_{1} \rightarrow {\alpha}_{1})}}}}
  {\cdot \vdash \mathsf{id} : \hmani{{\alpha}_{2}}{\tint}{\hmani{{\alpha}_{1}}{{\alpha}_{2}}{({\alpha}_{1} \rightarrow {\alpha}_{1})}}}
\]
The topmost judgment follows by \rref{ITyp-gen} from typing $\mathsf{id}$ at ${\alpha}_{1} \rightarrow {\alpha}_{1}$
under one more abstract binding. The derivation has the same shape as the explicit one in
Section~\ref{subsec:elim}, with rules \rref*{ITyp-gen} and \rref*{ITyp-inst} in place of \rref*{Typ-tabs}
and \rref*{Typ-tapp}, and it yields the same manifest type, but the term stays $\mathsf{id}$ throughout. As in
the explicit calculus, \rref{ITyp-teq} then turns this type into the equivalent type $\tint \rightarrow \tint$,
so that, for instance, $\mathsf{id} \, \mathtt{1}$ has type $\tint$ by \rref{ITyp-app}. Type equivalence
thus does the work that type substitution does in the instantiation rule of Curry-style System~F.

\paragraph{Manifest bindings in environments}
The third coercion, \rref{ITyp-mani}, plays the role of \rref{Typ-tdef}: it extends the type of an
environment with a manifest binding $[A]$, which exists only in the type, as the environment has no type
entries. Later entries may refer to this binding (\rref{ITyp-cons}), so environments can still have types with
transparent type components (Section~\ref{subsec:modules}). For instance, $\cdot, \hlam{a}{a}$ has the type of
the dependent environment
$\cdot, [\heq{{\alpha}_{2}}{\tint}], (\hLam{{\alpha}_{1}}{\hlam{a}{a}}) \, {\alpha}_{2}$ of Section~\ref{sec:calculus}:
\[
\inferrule*[left=ITyp-cons]
  {\inferrule*[left=ITyp-mani]
     {\inferrule*[left=ITyp-nil]{ }{\cdot \vdash \cdot : \cdot} \\ \cdot \vdash \tint}
     {\cdot \vdash \cdot : \cdot, \hdef{{\alpha}_{2}}{\tint}}
   \\
   \cdot, \hdef{{\alpha}_{2}}{\tint} \vdash \hlam{a}{a} : \hmani{{\alpha}_{1}}{{\alpha}_{2}}{({\alpha}_{1} \rightarrow {\alpha}_{1})}}
  {\cdot \vdash \cdot, \hlam{a}{a} : \cdot, \hdef{{\alpha}_{2}}{\tint}, \hmani{{\alpha}_{1}}{{\alpha}_{2}}{({\alpha}_{1} \rightarrow {\alpha}_{1})}}
\]
Here the entry $\hlam{a}{a}$ is typed under the binding $\hdef{{\alpha}_{2}}{\tint}$ by rules
\rref*{ITyp-gen} and \rref*{ITyp-inst}, as in the previous derivation.

\paragraph{Coercions and type equivalence}
Unlike \rref{ITyp-teq}, which replaces a type with an equivalent one, the coercions change the type itself,
and they apply to terms of any shape. A term may therefore have several types that are not equivalent: the
identity $\hlam{a}{a}$, for instance, has both $\tint \rightarrow \tint$ and
$\hforall{\alpha}{\alpha \rightarrow \alpha}$. Generalization is also not limited to abstractions:
applying the identity to another identity, $(\hlam{a}{a})\,(\hlam{a}{a})$, has type
$\hforall{\alpha}{\alpha} \rightarrow \alpha$. Here the applied identity is typed at
$(\alpha \to \alpha) \to (\alpha \to \alpha)$ and its argument at $\alpha \to \alpha$, and \rref{ITyp-gen}
abstracts the free variable $\alpha$.
As with Curry-style System~F, typability in this variant is
undecidable~\cite{Wells1999Typability}, and we make no type-inference claims.

\subsection{Environment-based Semantics}\label{subsec:implicit-sem}

\begin{figure}[t]
\begin{drulepar}{\delta \vdash e \hookrightarrow e'}{Reduction}
  \drule{IStep-eval}
    {\delta \vdash e \hookrightarrow e'}
    {\delta \vdash F [e] \hookrightarrow F [e']}

  \drule{IStep-var}{\delta [\mathit{n}] = v}{\delta \vdash \mathit{n} \hookrightarrow v}

  \drule{IStep-lam}{}{\delta \vdash \lambda e \hookrightarrow \eclos{\delta}{\lambda e}}

  \drule{IStep-beta}
    {}
    {\delta \vdash \eclos{\delta_{\mathrm{1}}}{\lambda e} \, v_{\mathrm{2}} \hookrightarrow (\delta_{\mathrm{1}}, v_{\mathrm{2}}) \triangleright e}

  \drule{IStep-box}
    {\delta_{\mathrm{1}} \vdash e_{\mathrm{2}} \hookrightarrow e'_{\mathrm{2}}}
    {\delta \vdash \delta_{\mathrm{1}} \triangleright e_{\mathrm{2}} \hookrightarrow \delta_{\mathrm{1}} \triangleright e'_{\mathrm{2}}}

  \drule{IStep-boxv}
    {}
    {\delta \vdash \delta_{\mathrm{1}} \triangleright v \hookrightarrow v}

  \drule{IStep-mrgr}
    {\delta \envplus \delta_{\mathrm{1}} \vdash e_{\mathrm{1}} \hookrightarrow e_{\mathrm{2}}}
    {\delta \vdash \delta_{\mathrm{1}}, e_{\mathrm{1}} \hookrightarrow \delta_{\mathrm{1}}, e_{\mathrm{2}}}
\end{drulepar}
\caption{Dynamic semantics of the implicitly-typed variant.}
\label{fig:implicit-step}
\end{figure}

Figure~\ref{fig:implicit-step} shows the call-by-value, environment-based reduction relation
$\delta \vdash e \hookrightarrow e'$ of the implicitly-typed variant. It is the semantics of
Figure~\ref{fig:small} without the rules that compute on types (\rref*{Step-tapp}, \rref*{Step-bclos}, and
\rref*{Step-tdef}), so evaluation never blocks on a type abstraction; the remaining rules are unchanged up to
their names. Since runtime environments contain only values, the extraction $| \delta |$ is not needed, and
types play no role at run time by construction, whereas the explicitly-typed \name passes types at run time, and
Theorem~\ref{thm:irrelevance} shows that they do not affect the result.

\paragraph{Evaluation under type abstraction}
The coercions are what let evaluation proceed under type abstraction. Consider again
$(\hlam{a}{a})\,(\hlam{a}{a})$, which has type $\hforall{\alpha}{\alpha} \rightarrow \alpha$
(Section~\ref{subsec:implicit-typing}). It evaluates as follows:
\begin{align*}
  \cdot \vdash (\hlam{a}{a}) \, (\hlam{a}{a})
  &\hookrightarrow \eclos{\cdot}{\hlam{a}{a}} \, (\hlam{a}{a}) && \text{(\rref*{IStep-eval}, \rref*{IStep-lam})}\\
  &\hookrightarrow \eclos{\cdot}{\hlam{a}{a}} \, \eclos{\cdot}{\hlam{a}{a}} && \text{(\rref*{IStep-eval}, \rref*{IStep-lam})}\\
  &\hookrightarrow (\cdot, \eclos{\cdot}{\hlam{a}{a}}) \triangleright a && \text{(\rref*{IStep-beta})}\\
  &\hookrightarrow (\cdot, \eclos{\cdot}{\hlam{a}{a}}) \triangleright \eclos{\cdot}{\hlam{a}{a}} && \text{(\rref*{IStep-box}, \rref*{IStep-var})}\\
  &\hookrightarrow \eclos{\cdot}{\hlam{a}{a}} && \text{(\rref*{IStep-boxv})}
\end{align*}
To have this type in the explicitly-typed calculus, the same program must be written under a type abstraction,
$\Lambda{\color{gray}\alpha.\,}((\hlam{a}{a})\,(\hlam{a}{a}))$. This term evaluates to a closure by
\rref{Step-bclos} and stops there: its body is not evaluated until a type is supplied, and it is evaluated
again for every type argument supplied (\rref{Step-tapp}), so the inner redex stays frozen under $\Lambda$.
The implicitly-typed variant has no type abstraction in terms, so the redex is free to reduce and evaluation
is not held up by polymorphism.

\paragraph{Polymorphic entries in environments}
Polymorphic values can also be stored in first-class environments and used at different types, without any
type information at run time. In the box $(\cdot, \hlam{a}{a}) \triangleright (a \, a) \, \mathtt{1}$, the
local environment holds the identity, typed at $\hforall{\alpha}{\alpha \rightarrow \alpha}$ by
\rref{ITyp-gen}. In the body, rules \rref*{ITyp-inst} and \rref*{ITyp-teq} instantiate the first occurrence
of $a$ at $(\tint \rightarrow \tint) \rightarrow (\tint \rightarrow \tint)$ and the second at
$\tint \rightarrow \tint$. Hence the body has type $\tint$, and so does the box, by rules \rref*{ITyp-box}
and \rref*{ITyp-teq}. The box evaluates as follows:
\begin{align*}
  \cdot \vdash (\cdot, \hlam{a}{a}) \triangleright (a \, a) \, \mathtt{1}
  &\hookrightarrow (\cdot, \eclos{\cdot}{\hlam{a}{a}}) \triangleright (a \, a) \, \mathtt{1} && \text{(\rref*{IStep-eval}, \rref*{IStep-mrgr}, \rref*{IStep-lam})}\\
  &\hookrightarrow^* (\cdot, \eclos{\cdot}{\hlam{a}{a}}) \triangleright \eclos{\cdot}{\hlam{a}{a}} \, \mathtt{1} && \text{(\rref*{IStep-box})}\\
  &\hookrightarrow^* (\cdot, \eclos{\cdot}{\hlam{a}{a}}) \triangleright \mathtt{1} && \text{(\rref*{IStep-box})}\\
  &\hookrightarrow \mathtt{1} && \text{(\rref*{IStep-boxv})}
\end{align*}
Inside the box, both occurrences of $a$ are looked up in the local environment (\rref{IStep-var}), and the
same closure $\eclos{\cdot}{\hlam{a}{a}}$ serves at both types. In the explicitly-typed \name, the program would be
$(\cdot, \hLam{\alpha}{\hlam{a}{a}}) \triangleright (a \, (\tint \rightarrow \tint)) \, (a \, \tint) \, \mathtt{1}$,
and each of its two type applications would first reduce, by \rref{Step-tapp}, to a box whose local
environment stores the type argument, before the identity could be applied.

\subsection{Metatheory}\label{subsec:implicit-meta}

\paragraph{Type soundness and normalization}
Type soundness is established semantically, through a unary logical relation analogous to the one for the
explicit calculus (Appendix~\appSem). The fundamental theorem shows that syntactic typing implies semantic typing
$\Gamma \models e : A$, which yields normalization, and hence type soundness: a closed well-typed term
never gets stuck.

\begin{theorem}[Fundamental Theorem]
If $\Gamma \vdash e : A$, then $\Gamma \models e : A$.
\end{theorem}

\begin{corollary}[Normalization]
If $\cdot \vdash e : A$, then $\cdot \vdash e \hookrightarrow^* v$ for some value $v$.
\end{corollary}

The interesting case of the proof is \rref{ITyp-gen}, which leaves the term unchanged; it relies on
reduction being deterministic, so that $e$ evaluates to the same value under every interpretation of the new
abstract binding.

\paragraph{Conservativity over implicitly-typed System F}
The variant is also a conservative extension of implicitly-typed System~F, whose terms are those of the
$\lambda$-calculus with integer literals and whose reduction is the standard call-by-value reduction. With
the notation of Section~\ref{sec:meta} ($\Gamma \vdash_{F} e : A$, $e \hookrightarrow^* v_{F}$,
$\vdash_{F} \Gamma$, $\vdash_{F} e$, $\vdash_{F} A$, and $v \sim v_{F}$), we prove completeness and
conservativity with respect to implicitly-typed System~F, for both typing and the dynamic semantics, in the
same form as for the explicit calculus.

\begin{theorem}[Completeness w.r.t.\ Typing]
If $\Gamma \vdash_{F} e : A$, then $\Gamma \vdash e : A$.
\end{theorem}

\begin{theorem}[Conservativity w.r.t.\ Typing]
If $\vdash_{F} \Gamma$, $\vdash_{F} e$, $\vdash_{F} A$, and $\Gamma \vdash e : A$, then $\Gamma \vdash_{F} e : A$.
\end{theorem}

\begin{theorem}[Completeness w.r.t.\ Dynamics]
If $\cdot \vdash_{F} e : A$ and $e \hookrightarrow^* v_{F}$, then there exists $v$ s.t. $\cdot \vdash e \hookrightarrow^* v$ and $v \sim v_{F}$.
\end{theorem}

\begin{theorem}[Conservativity w.r.t.\ Dynamics]
If $\vdash_{F} e$, $\cdot \vdash e : A$ and $\cdot \vdash e \hookrightarrow^* v$, then there exists $v_{F}$ s.t. $e \hookrightarrow^* v_{F}$ and $v \sim v_{F}$.
\end{theorem}

As in Section~\ref{sec:meta}, in the proof of completeness, instantiation in System~F, which substitutes
into the type, is matched by \rref{ITyp-inst} followed by \rref{ITyp-teq}. Since values carry no
types, the value relation $v \sim v_{F}$ needs only \rref{Vrel-lit} and \rref{Vrel-abs} without the
type-closing operation $\mathsf{T}$.

This variant and all of the results above are fully formalized in Rocq.

\section{Discussion and Extensions}\label{sec:discussion}

We extend \name with records and use them to elaborate a source language of
first-class modules. We then discuss subtyping and sealing as directions for
future work.

\subsection{Extension to Records}\label{sec:records}
Similarly to \lambdae~\cite{DBLP:journals/pacmpl/TanO24},
we add \emph{extensible} (and
dependent) records as a lightweight extension, following Reynolds's
Forsythe encoding~\citep{reynolds1997design}: a multi-field record is
a \emph{concatenation} of single-field records.
Thus, records add only
labels and field selection, which are reached by name, through a
separate lookup, and on top of environments. The added rules appear in
Figure~\ref{fig:ext_rules}.
Records are useful to model programs that access components by a
nominal label, such as the modules of Section~\ref{subsec:modules}.

\paragraph{Syntax}
The syntax for records in the extended calculus is as follows:
\begingroup\setlength{\jot}{0pt}%
\begin{align*}
  &\text{Types} &A, B, C &\Coloneqq \dots ~|~ \{ \mathit{l} : A \} \\
  &\text{Expressions}
  &e &\Coloneqq \dots ~|~ \{ \mathit{l} = e \} ~|~ e.\mathit{l} \\
  &\text{Values} &v &\Coloneqq \dots ~|~ \{ \mathit{l} = v \} \\
  &\text{Frames} &F &\Coloneqq \dots ~|~ \{ \mathit{l} = \eframe \} ~|~ \eframe . \mathit{l} \\
  &\text{Syntactic sugar} & & \{l_1: A_1, \, \dots, \, l_n: A_n\} \triangleq \{l_1:A_1\}, \, \dots, \, \{l_n:A_n\} \\
  & & & \{l_1 = e_1, \, \dots, \, l_n = e_n\} \triangleq \{l_1 = e_1\}, \, \dots, \, \{l_n = e_n\}
\end{align*}\endgroup

The syntax adds single-field record types $\{ \mathit{l} : A \}$, single-field
record terms $\{ \mathit{l} = e \}$, and field selection $e.\mathit{l}$. A multi-field
record is their concatenation, at both the type and term levels.
Records leverage \name's existing concatenation mechanism. The encoding fits
almost for free, since \name's environments are \emph{already} ordered,
concatenable sequences. A multi-field record is simply such a sequence,
with both the dependencies among its fields and the means to concatenate
them coming from the existing environment rules. What is new is the \emph{kind} of dependency. \lambdae's records depend
on earlier term entries, and \name inherits this. \name's environments,
however, also carry manifest \emph{type} components. A field's type can
therefore depend on a type entry declared earlier in the same record.
Reynolds's records are not dependent at all, and \lambdae's not on type
entries. These transparent, dependent type components are thus new to \name.

A consequence of this encoding---again inherited from Reynolds---is that
the syntax does not, by itself, distinguish record fields from other
entries. A sequence such as $\{ \mathit{l} : \tint \}, \tint$, mixing a labelled
field with an ordinary type, is well-formed. We regard this as a
property of the general mechanism, not a defect: records are one
disciplined fragment of it, sitting alongside ordinary bindings and
manifest type components. The source language in
Section~\ref{subsec:module-elaboration} restricts the signatures of merge
operands to labelled fields and manifest type entries, allowing both in the
same signature while excluding unlabelled term entries.

\paragraph{Typing}
\Rref{Typ-rec} types a single-field record, while
\rref{Typ-sel} types field projection using label lookup $\mathit{l} : A \in \Gamma$. The type of
$e.\mathit{l}$ is $A$ when looking up label $\mathit{l}$ in the type of $e$ yields
$A$. Since \rref{Typ-teq} allows typing to use type equivalence, the type equivalence relation is also extended with \rref{Teq-rcd} which compares record types by comparing their fields.

\begin{figure}[t]

  \begin{drulepar}{\mathit{l} : A \in \Gamma}{Label Lookup on Environment Types}
    \drule{Lookup-rcd}
      {\mathit{l} \notin \Gamma_{\mathrm{1}}}
      {\mathit{l} : A^{| \Gamma_{\mathrm{1}} |} \in \Gamma_{\mathrm{1}}, \{ \mathit{l} : A \}}

    \drule{Lookup-andl}
      {\mathit{l} : A \in \Gamma_{\mathrm{1}} \and \mathit{l} \neq \mathit{l_{\mathrm{1}}}}
      {\mathit{l} : A \in \Gamma_{\mathrm{1}}, \{ \mathit{l_{\mathrm{1}}} : B \}}

    \drule{Lookup-andr}
      {\mathit{l} : A \in \Gamma_{\mathrm{2}} \and \mathit{l} \notin \Gamma_{\mathrm{1}}}
      {\mathit{l} : A^{| \Gamma_{\mathrm{1}} |} \in \Gamma_{\mathrm{1}}, \Gamma_{\mathrm{2}}}

    \drule{Lookup-tdef}
      {\mathit{l} : A \in \Gamma_{\mathrm{1}}}
      {\mathit{l} : A \in \Gamma_{\mathrm{1}}, [B]}
  \end{drulepar}

  \begin{drulepar}{\Gamma \vdash e : A}{Typing}
    \drule{Typ-rec}
      {\Gamma \vdash e : A}
      {\Gamma \vdash \{ \mathit{l} = e \} : \{ \mathit{l} : A \}}

    \drule{Typ-sel}
      {\Gamma \vdash e : \Gamma_{\mathrm{1}} \and \mathit{l} : A \in \Gamma_{\mathrm{1}}}
      {\Gamma \vdash e.\mathit{l} : A}
  \end{drulepar}

  \begin{drulepar}{\Gamma_{\mathrm{1}} \vdash A \sim B \dashv \Gamma_{\mathrm{2}}}{Type Equivalence}
    \drule{Teq-rcd}
      {\Gamma_{\mathrm{1}} \vdash A \sim B \dashv \Gamma_{\mathrm{2}}}
      {\Gamma_{\mathrm{1}} \vdash \{ \mathit{l} : A \} \sim \{ \mathit{l} : B \} \dashv \Gamma_{\mathrm{2}}}
  \end{drulepar}

  \begin{drulepar}{\delta.\mathit{l} \rightsquigarrow v}{Selection}
    \drule{Sel-rec}{}{(\delta, \{ \mathit{l} = v \}).\mathit{l} \rightsquigarrow v}

    \drule{Sel-defl}
      {\delta.\mathit{l} \rightsquigarrow v \and \mathit{l} \neq \mathit{l_{\mathrm{1}}}}
      {(\delta, \{ \mathit{l_{\mathrm{1}}} = v_{\mathrm{1}} \}).\mathit{l} \rightsquigarrow v}

    \drule{Sel-defr}
      {\delta_{\mathrm{1}}.\mathit{l} \rightsquigarrow v}
      {(\delta, \delta_{\mathrm{1}}).\mathit{l} \rightsquigarrow v}

    \drule{Sel-tdefl}
      {\delta.\mathit{l} \rightsquigarrow v}
      {(\delta, [A]).\mathit{l} \rightsquigarrow v}
  \end{drulepar}

  \begin{drulepar}{\delta \vdash e \hookrightarrow e'}{Operational Semantics}
    \drule{Step-sel}
      {\delta_{\mathrm{1}}.\mathit{l} \rightsquigarrow v}
      {\delta \vdash \delta_{\mathrm{1}}.\mathit{l} \hookrightarrow v}
  \end{drulepar}
  \caption{Extension to records.}\label{fig:ext_rules}
\end{figure}

Note that label lookup, used in \rref{Typ-sel} raises an interesting problem since
$\Gamma_1$ may have manifest type bindings that are \emph{not in scope}
in $\Gamma$: a naive lookup could result in an ill-scoped type.
Therefore the label lookup relation $\mathit{l} : A \in \Gamma$ does not simply yield the type
associated with label $\mathit{l}$ in $\Gamma$. Instead, since environment types
are dependent, it also collects the type information required by the
associated type. For example, for $\Gamma_{\mathrm{1}}, \{ \mathit{l} : A \}$ in \rref{Lookup-rcd}, $A$ may depend on the previous environment $\Gamma_{\mathrm{1}}$ and the label lookup needs to
account for the dependency to ensure that the result is
well-scoped. Instead of returning $A$ directly, a wrapped type $A^{| \Gamma_{\mathrm{1}} |}$ is returned. The definition of the wrapping operation $A^{| \Gamma |}$
is given as follows:
\begin{center}
  $A^{| \cdot |} = A$; \quad \quad $A^{| \Gamma, B |} = A^{| \Gamma |}$; \quad \quad $A^{| \Gamma, [B] |} = ([B] A)^{| \Gamma |}$
\end{center}
Intuitively, $A^{| \Gamma |}$ constructs a manifest type along the way of
traversing the environment type, so that all the type dependencies on
the previous environment are properly captured. For instance,
environment $\cdot, [\heq{\beta}{\mathsf{Bool}}], [\heq{\alpha}{\tint}], \{ \mathit{l} = \mathsf{toBool} \}$ can have
environment type $\cdot, \hdef{\beta}{\mathsf{Bool}}, \hdef{\alpha}{\tint}, \{ \mathit{l} : \alpha \rightarrow \beta \}$,
where function $\mathsf{toBool}$ converts an integer to a boolean and its
type $\alpha \rightarrow \beta$ depends on the previous two type
definitions. When looking up label $\mathit{l}$, the result is
$\hmani{\beta}{\mathsf{Bool}}{\hmani{\alpha}{\tint}{(\alpha \rightarrow \beta)}}$. This way, the type of
$\mathsf{toBool}$ correctly reflects its dependencies on the previous type
definitions.  Furthermore, with type equivalence, $\mathsf{toBool}$ also
has type $\tint \rightarrow \mathsf{Bool}$. Note that labels can also be accessed from the global context. Rather than introducing a separate lookup mechanism for this purpose, we use the de Bruijn index $\mathit{n}$ to retrieve the corresponding labelled entry from the context, and then apply the projection $\mathit{n}.\mathit{l}$ to the variable. This approach keeps the calculus minimal while still supporting flexible access patterns.

\paragraph{Reduction}
Dynamically, a projection $e.\mathit{l}$ first
evaluates $e$ to a value environment $\delta_{\mathrm{1}}$. The selection relation
$\delta_{\mathrm{1}}.\mathit{l} \rightsquigarrow v$ then searches this value environment for the field
labelled $\mathit{l}$ and returns its associated value. It skips unrelated
record fields and type definitions, and it continues through
value environments until the labelled value is found. Thus
\rref{Step-sel} reduces $\delta_{\mathrm{1}}.\mathit{l}$ to the value $v$ selected by this relation.

\paragraph{Metatheory}
The metatheory scales to this extension. The new
rules add record-specific cases to theorems, but they do not change the central structure of
the proofs. Thus the same
environment-based reasoning, manifest bindings, and type equivalence
infrastructure developed earlier continue to apply in
the presence of records.

\subsection{Dynamic Module Composition and Elaboration}\label{subsec:module-elaboration}

We formalize a small source language for the modules of
Section~\ref{subsec:modules}, with signatures given by core types.
The judgment $\selab{\Gamma}{M}{e}{A}$ elaborates a module $M$ to an expression
$e$ of type $A$ in \name with records; declaration blocks elaborate similarly.
Figure~\ref{fig:module-elaboration} gives the rules for adding a field and
composing two modules. In \rref{Elabd-val}, the context
$\Gamma \envplus \Gamma_{\mathrm{1}}$ includes the preceding declarations,
so the new field may refer to earlier type definitions.

\begin{figure}[ht]
\begin{mathpar}
\inferrule[\ranchor{Elabd-val}]{
  \selab{\Gamma}{D}{\Delta}{\Gamma_{\mathrm{1}}} \\
  \Gamma \envplus \Gamma_{\mathrm{1}} \vdash e : A }
  {\selab{\Gamma}{D, \mathit{l} = e}{\Delta, \{ \mathit{l} = e \}}{\Gamma_{\mathrm{1}}, \{ \mathit{l} : A \}}}
\end{mathpar}
\begin{mathpar}
\inferrule[\ranchor{Elab-merge}]{
  \selab{\Gamma}{M_1}{e_{\mathrm{1}}}{\Gamma_{\mathrm{1}}} \\
  \selab{\Gamma}{M_2}{e_{\mathrm{2}}}{\Gamma_{\mathrm{2}}} \\
  \cdot \vdash \Gamma_{\mathrm{1}} \\ \cdot \vdash \Gamma_{\mathrm{2}} \\\\
  \sig{\Gamma_{\mathrm{1}}} \\ \sig{\Gamma_{\mathrm{2}}} \\
  \uniq{\Gamma_{\mathrm{1}}} \\ \uniq{\Gamma_{\mathrm{2}}} \\
  \Gamma_{\mathrm{3}} = \Gamma, \Gamma_{\mathrm{1}}, \Gamma_{\mathrm{2}} \\\\
  \sexp{\Gamma_{\mathrm{3}}}{1}{\Gamma_{\mathrm{1}}}{\cdot}{\cdot}{\Gamma_{\mathrm{1}}}{\Delta_{\mathrm{1}}}{\Gamma_{\mathrm{4}}} \\
  \sexp{\Gamma_{\mathrm{3}}}{0}{\Gamma_{\mathrm{2}}}{\Delta_{\mathrm{1}}}{\Gamma_{\mathrm{4}}}{\Gamma_{\mathrm{2}}}{\Delta}{\Gamma_{\mathrm{5}}} }
  {\selab{\Gamma}{M_1 + M_2}{(\lambda\, \lambda\, \Delta)\; e_{\mathrm{1}}\; e_{\mathrm{2}}}{\Gamma_{\mathrm{5}}}}
\end{mathpar}
\caption{Elaboration of value declarations and module composition. }
\label{fig:module-elaboration}
\end{figure}

\Rref{Elab-merge} binds the two operands before rebuilding their components.
Each operand must have a closed signature consisting of labelled fields and
manifest type entries ($\sig{\Gamma}$), with no repeated labels
($\uniq{\Gamma}$). Inside the two abstractions, the operands occupy indices
$1$ and $0$. The expansion premises build $\Delta$ from left to right,
projecting fields and copying manifest type bindings. Label lookup wraps the
projected types with their manifest definitions, preserving their dependencies
(Section~\ref{sec:records}). The signatures determine which entries to rebuild;
the operands may be computed at run time and are each evaluated once.
We prove in Rocq that elaboration preserves typing for both modules and
declarations. The complete rules and preservation theorem appear in Appendix~\appSource{}.

\subsection{Future Work: Subtyping and Sealing}\label{subsec:sealing}

\name provides a substitution-free foundation for first-class modules, with
dynamic composition and sandboxing. ML module systems are a mature, well-developed
area~\cite{MacQueen1984Modules,Milner1997Definition,LeroyPOPL94Manifest,RossbergDreyerMixML,Rossberg2018OneML}
with a different focus. Nevertheless, their features are a rich source of ideas for extending \name. The interesting
question is how to bring these ideas into a substitution-free setting. Below we sketch
two directions, partly inspired by features of ML modules: subtyping and sealing.

\paragraph{Subtyping}
One way to incorporate subtyping into \name is to replace type
equivalence by a subtyping relation $\Gamma_{\mathrm{1}} \vdash A <: B \dashv \Gamma_{\mathrm{2}}$, and
recover type equivalence as a special case where the two types are
mutual subtypes. Accordingly, the subsumption rule in the type system
would have the form:
\[
\inferrule* [right=Typ-sub]
  {\Gamma \vdash e : A \\ \Gamma \vdash A <: B \dashv \Gamma }
  {\Gamma \vdash e : B}
\]
Since our environment types are dependent---an entry can depend on
preceding entries---traditional notions of width and depth subtyping
must be generalized to account for dependencies between type and term
components. This leads to more sophisticated subtyping rules for
environment types. We expect that the work
by~\citet{DBLP:conf/lics/Kopylov03} will help formulate
subtyping for dependent environment types, since Kopylov presents a
standard subtyping judgment of the form $\Gamma \vdash A <: B$ for
subtyping between dependent intersections. However, Kopylov's rules
rely on meta-level substitution and are not directly applicable to our
substitution-free setting. Thus, adapting subtyping to this setting
requires further study.

\paragraph{Sealing}
Moreover, \emph{sealing}~\cite{harper1994type,dreyer2005thesis,RossbergDreyerMixML} is a key abstraction mechanism in ML
modules, allowing internal type information to be hidden behind an
abstract interface. To illustrate, consider the following ML-style
module:
\begin{lstlisting}[language=ML,mathescape=true]
  signature API = sig type a; val id : a -> a end
  structure M_int :> API = struct type a = Int; fun id x = x end
\end{lstlisting}
Here, the module \lstinline|M_int| internally knows that \lstinline|a = Int|, but externally exposes only the abstract type
\lstinline|a|. This is achieved via sealing (opaque
ascription).
A natural path to encode sealing is via subtyping between type contexts:
\[
\cdot, \hdef{\alpha}{\tint}, \alpha \rightarrow \alpha <: \cdot, \hdecl{\alpha}{\tstar}, \alpha \rightarrow \alpha
\]
where the manifest type information is hidden in the supertype. This
allows us to recover existential abstraction without introducing
explicit existential quantifiers. In other words, sealing becomes a subtyping
coercion, which would yield a more uniform account of
abstraction in \name. This change is not simple, as subtyping would replace type
equivalence. One immediate problem is that, unlike type equivalence,
subtyping is not symmetric. This loss of symmetry would require
substantial changes to the metatheory.

\section{Related Work}

\paragraph{First-class environments}
\citet{GelernterJL87} introduced first-class environments in the
research literature with Symmetric Lisp, a language that extends Lisp
by making environments first-class values that programmers can inspect
and manipulate at runtime. This pioneering work inspired a line of
research on first-class environments for dynamically typed
languages~\cite{Jag94,icclJagannathan94, MillerR91, Shutt2010Fexprs,
  QueinnecDeRoure1996}. There is also some work on first-class
environments for statically typed
calculi. \citet{SatoSakuraiBurstall2001} introduced a simply typed
calculus called \jlambdae with environments as first-class values,
under the setting of full reduction. \citet{Nishizaki1994} proposed
another typed calculus with first-class environments, called
\lambdaenv, which is based on the weak reduction of the
$\lambda\sigma$ calculus~\cite{DBLP:journals/jacm/CurienHL96}. The use
of explicit substitutions~\cite{abadiexplicit} allows the above
calculi to internalize substitution operations within the calculus
itself. However, with first-class environments, explicit substitutions
complicate both the dynamic semantics and the metatheory of these
calculi. More recently, \citet{tan2023dependent} proposed a simply typed calculus
\eical that supports dependent merges and first-class environments,
with a call-by-value semantics. In \eical, variables are encoded as
label names and thus alpha conversion is not allowed on variables
bound by a function. Following the \eical calculus, the \lambdae
calculus proposed by~\citet{DBLP:journals/pacmpl/TanO24} adopts de
Bruijn indices for variable representation for binders, simplifying
the representation of binding and the metatheory.
The design of \name follows the design of \lambdae,
but it tackles the addition of polymorphism and type components.

Among typed calculi with first-class environments there is limited work on polymorphism.
The most mature line of work is Nishizaki's extension of \lambdaenv,
$\mathit{ML}_\text{env}$ \citep{DBLP:journals/lisp/Nishizaki00}. His work is focused
on Hindley–Milner type schemes (rank-1 polymorphism) and type inference. The typing contexts
carry \emph{only} term-variable bindings.
Nishizaki's short paper on a second-order type theory~\cite{nishizaki2017second}
proposes a type system and reduction rules, but proves no metatheory.
Moreover, like $\mathit{ML}_\text{env}$, the calculus is based on explicit
substitutions. Its contexts and environments contain only term-variable entries. By contrast, \name brings
full System~F to a small-step, substitution-free environment semantics
in the \lambdae lineage \citep{DBLP:journals/pacmpl/TanO24}. Contexts
in \name mix term and type bindings, so terms may depend on types. At
the term level, \name supports \emph{manifest} type components, which $\mathit{ML}_\text{env}$
lacks.
\citet{amin17type} proposed a big-step,
call-by-value environment semantics for definitional interpreters with
polymorphism. However, their environments are not first-class values, and their
type system still relies on meta-level type substitution.

\paragraph{Modules with transparent type components}
Many mainstream languages provide transparent type components without ML-style
sealing. The distinction between transparent and opaque type components originates
with translucent sums~\cite{harper1994type} and manifest
types~\cite{LeroyPOPL94Manifest}. A transparent type component is exposed together
with its definition. A client may then use the type and its definition
interchangeably. Haskell's module system traditionally supports only transparent
type components through type synonyms~\cite{Marlow2010Haskell2L}. Its Backpack
extension is an exception, retrofitting the language with signatures and
sealing~\cite{DBLP:conf/popl/KilpatrickDJM14}. Rust and Go likewise
only provide
transparent type aliases in their module systems~\cite{RustLang,GoSpec}. Transparent type components are
therefore useful on their own. \name enables transparent components
via manifest types, and studies them in a substitution-free
setting. Thus, it could
serve as a foundation for module systems with transparent type components.

\paragraph{ML-style module systems}
ML-style module systems package terms and type members behind signatures and
functors~\cite{Milner1997Definition,MacQueen1984Modules}. They relate to \name
through two shared features: the transparent type components discussed above, and
first-class modules. \name obtains both directly from \emph{first-class
environments}, capturing transparency with manifest type bindings and type
equivalence.
First-class modules have likewise
been studied in ML, from early
proposals in the ML
literature~\cite{Russo2000FirstClassStructures,Russo1998TypesForModules}
to the unifying 1ML
perspective~\cite{Rossberg2018OneML}. In practice,
OCaml has provided first-class modules since
version~3.12~\cite{OCaml312RefMan2011,FrischGarrigue2010FirstClassModules,Yallop2010FirstClassModules}.

Beyond these shared features, \name rests on a different foundation. The ML-modules line explains modules by elaboration into a polymorphic core with a
substitution-based metatheory, from higher-order and applicative or generative
functors~\cite{MacQueenTofte1994HigherOrderFunctors,dreyer2005thesis} to unifying
accounts such as MixML~\cite{RossbergDreyerMixML}, the F-ing
method~\cite{ROSSBERG_2014}, and 1ML~\cite{Rossberg2018OneML}.
By contrast, following \lambdae~\cite{DBLP:journals/pacmpl/TanO24}, \name models modules as first-class environments, with no separate notion of signatures or functors required, which immediately yields first-class modules.
In \name, environments are \emph{values}, and module composition is the
\emph{dynamic}, run-time concatenation of dependency-carrying fragments. A record
calculus does not model this: an ordinary record, even a dependent one, fixes its
fields at definition time, whereas \name composes separately-built fragments at
run time while preserving dependency. A box additionally sandboxes a module from
ambient bindings. Thus, dynamic composition and sandboxing are the primary emphasis of \name, and
orthogonal to the ML-modules line.

The metatheory is moreover
\emph{substitution-free}: manifest type bindings together with a type-equivalence
judgment let type variables be replaced by their manifest definitions during type
checking.
This unifies two traditionally distinct mechanisms. ML module
calculi~\cite{LeroyPOPL94Manifest,harper1994type} typically require \emph{both}
type equivalence (or subtyping) and meta-level substitution, whereas \name relies solely on type
equivalence for both module components and polymorphic instantiation. This makes
\name especially apt when modules group related definitions behind
\emph{transparent} type components.

\name does not provide sealing~\cite{harper1994type,dreyer2005thesis,RossbergDreyerMixML},
subtyping, propagation of type equalities~\cite{harper1994type}, or recursive
modules~\cite{dreyer2005thesis,DBLP:conf/icfp/Dreyer07}. We treat these as optional
extensions toward ML-style abstraction, not requirements of our contribution, and
leave them to future work.

\paragraph{Explicit substitutions}
Explicit substitutions~\cite{abadiexplicit,DBLP:journals/jacm/CurienHL96} were
introduced to close the gap between the theory and implementation of
the lambda calculus. Unlike the standard lambda calculus, explicit
substitutions represent substitutions as syntactic entities, defined
mutually inductively with terms. Typically, these substitution
primitives include identity substitution ($\mathit{id}$), extension
($e \cdot s$), shifting ($\uparrow$), and composition ($s \circ t$).
\citet{abadiexplicit} showed that explicit substitutions calculus
$\lambda\sigma$ can be extended with a polymorphic type system. In
this setting, explicit substitutions also appear at the type level, and type
application instantiates a universal type by extending a substitution.
However, the mutual definition of types and substitutions
significantly complicates the formulation and metatheory. Type
equivalence must also be defined mutually with substitution
equivalence, and must account for the semantics of each substitution
primitive, resulting in a proliferation of equivalence rules. Our
calculus takes a more lightweight and straightforward approach. We
eliminate type-level substitution without introducing any substitution
primitives. Instead, we use manifest types as a single, uniform
mechanism for type instantiation. Our type equivalence judgment
handles manifest types simply by discharging them into the context as
manifest bindings, without needing to reason about various
substitution primitives. This design is both simple and easy
to mechanize.

\paragraph{Staged computation}
\citet{DBLP:journals/jacm/DaviesP01} proposed a type-theoretic
foundation for staged computation grounded in the intuitionistic modal
logic S4. The box modality $\square A$ denotes closed code to be
evaluated at a later stage. The construct $\mathsf{box}(e)$ encapsulates
$e$ for future execution, and requires the enclosed code to be closed.
Contextual modal type theory~\cite{nanevski2008contextual} generalizes
this to \emph{contextual} modalities. A contextual type $[\Psi]A$ types
code that may refer to the \emph{term} bindings listed in $\Psi$, through
the construct $\mathsf{box} \, \Psi. e$.
A parallel line of work uses \emph{environment
classifiers}~\citep{TahaNielsenPOPL03,TsukadaIgarashiLMCS10,CalcagnoMoggiTahaESOP04}
to regulate staging. Classifiers record the \emph{term} bindings that code may
rely on across stage boundaries. The construct $\mathsf{run}$ is permitted only
when those dependencies are satisfied.

Our boxes resemble both lines of work, since each carries an explicit environment
of bindings. The difference is what the environment holds. Contextual modalities
and classifiers track the \emph{term} bindings that code may use, and they
regulate when code runs. \name instead tracks \emph{type} information in box types. Our box
expression $\Delta \triangleright e$ packages a term-level environment, and box type
$\Gamma \triangleright A$ carries the \emph{type} context needed to type $A$. We use manifest
bindings and a two-sided type-equivalence judgment to internalize type
instantiation and thereby \emph{eliminate substitution}. Our boxes therefore do
not control when code runs. They make environments explicit at the value and type
level.

\section{Conclusion}
We have presented \name, showing that a polymorphic calculus with
first-class environments can be constructed without relying on the
meta-level substitution traditionally used in System F. A key
insight of this work is that \emph{manifest types} combined with
\emph{type equivalence} provide a sufficient mechanism to handle type
instantiation, effectively internalizing it within the
environment semantics. This shift simplifies mechanization of the
metatheory by replacing substitution and substitution lemmas with
instantiation lemmas, and is fully mechanized in Rocq.

On the practical side, \name demonstrates that the benefits of
first-class environments extend to languages with polymorphism. The
calculus naturally supports: polymorphic first-class modules with
type components that are transparently exposed to clients; robust
\emph{sandboxing}, where code execution is strictly isolated within
explicit local environments; and dynamic composition of environments and modules.
Future work incorporating subtyping
can extend this foundation to support features such as
opaque sealing, further bridging the gap between first-class
environments and full-scale module systems.

\section*{Data-Availability Statement}

The mechanization, implementation, and example code are available at
\url{https://github.com/tjhao/PolyFirstClassEnvs}.
Parts of the Rocq mechanization were carried out with the assistance of large
language models.

\bibliographystyle{ACM-Reference-Format}
\bibliography{reference}

\appendix
\newpage
\section{From Hint Notation to de Bruijn Notation}\label{sec:hints}
The examples in the paper use the hint notation. Deleting the gray binder annotations and replacing
each hinted variable by the corresponding index recovers the formal de~Bruijn term. The following table
lists the hinted examples used in the paper together with their de~Bruijn forms.

\begin{table}[h]
\centering
{%
\begin{tabular}{@{}c|c@{}}
\toprule
\textbf{hint notation} & \textbf{pure de~Bruijn notation} \\
\midrule
$\hLam{\alpha}{\mathsf{id}}$                          & $\Lambda \mathsf{id}$ \\
\midrule
$(\cdot, [\heq{\alpha}{\tint}]) \triangleright \mathsf{id}$                & $(\cdot, [\tint]) \triangleright \mathsf{id}$ \\
\midrule
$\eclos{\cdot, [\heq{\alpha}{\tint}]}{\mathsf{id}}$                                   & $\eclos{\cdot, [\tint]}{\mathsf{id}}$ \\
\midrule
$(\cdot, \hdef{\alpha}{\tint}) \triangleright (\alpha \rightarrow \alpha)$       & $(\cdot, [\tint]) \triangleright (0 \rightarrow 0)$ \\
\midrule
$(\hLam{{\alpha}_{2}}{(\hLam{{\alpha}_{1}}{\mathsf{id}})} \, {\alpha}_{2}) \, \tint$ & $(\Lambda (\Lambda \mathsf{id}) \, 0) \, \tint$ \\
\midrule
$\hmani{{\alpha}_{2}}{\tint}{\hmani{{\alpha}_{1}}{{\alpha}_{2}}{({\alpha}_{1} \rightarrow {\alpha}_{1})}}$ & $[\tint] [0] (0 \rightarrow 0)$ \\
\midrule
$\cdot, [\heq{{\alpha}_{2}}{\tint}], (\hLam{{\alpha}_{1}}{\hlam{a}{a}}) \, {\alpha}_{2}$ & $\cdot, [\tint], (\Lambda (\lambda 0)) \, 0$ \\
\midrule
$\hmani{{\alpha}_{1}}{{\alpha}_{2}}{({\alpha}_{1} \rightarrow {\alpha}_{1})}$ & $[0] (0 \rightarrow 0)$ \\
\midrule
$(\hLam{\alpha}{\hlam{a}{a}}) \, \tint$                                    & $(\Lambda (\lambda 0)) \, \tint$ \\
\midrule
$\hmani{\alpha}{\tint}{(\alpha \rightarrow \alpha)}$        & $[\tint] (0 \rightarrow 0)$ \\
\midrule
$\cdot, \hdef{\beta}{\tint}, \hdecl{\alpha}{\tstar} \vdash \beta \sim \alpha \dashv \cdot, \hdecl{\gamma}{\tstar}, \hdecl{\beta}{\tstar}, \hdef{\alpha}{\tint}$ & $\cdot, [\tint], \tstar \vdash 1 \sim 0 \dashv \cdot, \tstar, \tstar, [\tint]$ \\
\midrule
$\hmani{\alpha}{\tint}{\alpha}$           & $[\tint] 0$ \\
\midrule
$(\hLam{\alpha}{e} \, \alpha) \, \tint$           & $(\Lambda (e \, 0)) \, \tint$ \\
\midrule
$\cdot, [\heq{\beta}{\mathsf{Bool}}], [\heq{\alpha}{\tint}], \{ \mathit{l} = \mathsf{toBool} \}$         & $\cdot, [\mathsf{Bool}], [\tint], \{ \mathit{l} = \mathsf{toBool} \}$ \\
\midrule
$\cdot, \hdef{\beta}{\mathsf{Bool}}, \hdef{\alpha}{\tint}, \{ \mathit{l} : \alpha \rightarrow \beta \}$ & $\cdot, [\mathsf{Bool}], [\tint], \{ \mathit{l} : 0 \rightarrow 1 \}$ \\
\midrule
$\hmani{\beta}{\mathsf{Bool}}{\hmani{\alpha}{\tint}{(\alpha \rightarrow \beta)}}$ & $[\mathsf{Bool}] [\tint] (0 \rightarrow 1)$ \\
\midrule
$(\hlam{a}{a}) \, (\hlam{a}{a})$                      & $(\lambda 0) \, (\lambda 0)$ \\
\midrule
$\hforall{\alpha}{\alpha} \rightarrow \alpha$                                 & $\forall (0 \rightarrow 0)$ \\
\midrule
$\hLam{\alpha}{((\hlam{a}{a}) \, (\hlam{a}{a}))}$           & $\Lambda ((\lambda 0) \, (\lambda 0))$  \\
\midrule
$\cdot, \hdef{\alpha}{\tint}, \alpha \rightarrow \alpha$                              & $\cdot, [\tint], (0 \rightarrow 0)$ \\
\midrule
$\cdot, \hdecl{\alpha}{\tstar}, \alpha \rightarrow \alpha$           & $\cdot, \tstar, (0 \rightarrow 0)$ \\
\bottomrule
\end{tabular}}
\label{tab:hints}
\end{table}

\section{Auxiliary Definitions}\label{sec:aux}

This section gives the auxiliary operations omitted from
the main paper. For each, we point to the rules in which it is used.

\paragraph{Context membership} The membership condition $\mathit{N} \, \in \, \Gamma$ (used in \rref{Wft-var}) holds when the $N$-th type entry is an abstract type binding:
\begin{align*}
  &(1) \, \mathit{S} \, \mathit{N} \in \Gamma, \tstar \, \text{ if }  \, \mathit{N} \in \Gamma \quad
   (2) \, \mathit{S} \, \mathit{N} \in \Gamma, [A] \, \text{ if }  \, \mathit{N} \, \in \, \Gamma \quad
   (3) \, \alpha \in \Gamma, \tstar \quad
   (4) \, \mathit{N} \in \Gamma, A \, \text{ if }  \, \mathit{N} \, \in \, \Gamma
\end{align*}

\paragraph{Type-variable lookup} The lookup $\Gamma [\mathit{N}] = A$ (used in \rref{Wft-vareq}, \rref{Teq-eql}, and \rref{Teq-eqr}) holds when the $N$-th type entry is a concrete binding of type $A$. Note that $\tshift{A}$ denotes the shifting of type $A$, which is applied when a lookup descends past a
type binding.
\begin{align*}
  &(1) \, (\Gamma, [B])[ \mathit{S} \, \mathit{N} ] = \tshift{A} \, \text{ if }  \, \Gamma [\mathit{N}] = A  && (2) \, (\Gamma, \tstar)[ \mathit{S} \, \mathit{N} ] = \tshift{A} \, \text{ if }  \, \Gamma [\mathit{N}] = A \\
  &(3) \, (\Gamma, [B])[ \alpha ] = \tshift{B} &&(4) \, (\Gamma, B)[ \mathit{N} ] = A\, \text{ if }  \, \Gamma [\mathit{N}] = A
\end{align*}

\paragraph{Term-variable lookup} The lookup $\Gamma [\mathit{n}] = A$ (used in \rref{Typ-var}) retrieves the type $A$ of the $n$-th term entry. Similarly to type-variable lookup, the shifting operation is necessary when the context is extended with a type binding.
\begin{align*}
  &(1) \, (\Gamma, [B])[ \mathit{n} ] = \tshift{A} \, \text{ if }  \, \Gamma [\mathit{n}] = A  && (2) \, (\Gamma, \tstar)[ \mathit{n} ] = \tshift{A} \, \text{ if }  \, \Gamma [\mathit{n}] = A \\
  &(3) \, (\Gamma, B)[ a ] = B &&(4) \, (\Gamma, B)[ \mathit{S} \, \mathit{n} ] = A\, \text{ if }  \, \Gamma [\mathit{n}] = A
\end{align*}

\paragraph{Well-formed contexts} The judgment $\vdash \Gamma$ states that a context $\Gamma$ is well-formed. The empty context is well-formed; an abstract type binding $\tstar$ extends a well-formed context; and a term binding or a manifest type binding requires its type to be well-formed under the preceding context.
\begin{mathpar}
\drule{wfe-nil}{}{\vdash \cdot}\and
\drule{wfe-edef}{\Gamma \vdash A}{\vdash \Gamma, A}\and
\drule{wfe-star}{\vdash \Gamma}{\vdash \Gamma, \tstar}\and
\drule{wfe-tdef}{\Gamma \vdash A}{\vdash \Gamma, [A]}
\end{mathpar}

\paragraph{Context concatenation} The operation $\Gamma \envplus \Gamma_{\mathrm{1}}$ (used in well-formedness, typing, and type equivalence for environment types) appends the entries of $\Gamma_{\mathrm{1}}$ to $\Gamma$. It recurses on its right operand:
\begin{align*}
  &(1) \, \Gamma \envplus \cdot = \Gamma && (2) \, \Gamma \envplus (\Gamma_{\mathrm{1}}, A) = (\Gamma \envplus \Gamma_{\mathrm{1}}), A \\
  &(3) \, \Gamma \envplus (\Gamma_{\mathrm{1}}, \tstar) = (\Gamma \envplus \Gamma_{\mathrm{1}}), \tstar && (4) \, \Gamma \envplus (\Gamma_{\mathrm{1}}, [A]) = (\Gamma \envplus \Gamma_{\mathrm{1}}), [A]
\end{align*}

\paragraph{Rigidity} The rigidity judgment $\Gamma \vdash_{r} A$ is used for the well-formedness of box types (\rref{Wft-box}). A type is \emph{rigid} in a context $\Gamma$ at depth $d$ when each of its free type
variables is either bound by one of the innermost $d$ binders or pinned to a
concrete (manifest) definition that is itself rigid. Formally, the depth-indexed judgment $\Gamma \vdash^{d}_{r} A$ is defined as:
\begin{mathpar}
\inferrule[R-int]{ }{\Gamma \vdash^{d}_{r} \tint}
\and
\inferrule[R-avar]{ \mathit{N} \, \in \, \Gamma \\ \mathit{N} < d }{\Gamma \vdash^{d}_{r} \mathit{N}}
\and
\inferrule[R-cvar]{ \Gamma [\mathit{N}] = B \\ \Gamma \vdash^{d}_{r} B }{\Gamma \vdash^{d}_{r} \mathit{N}}
\and
\inferrule[R-arr]{ \Gamma \vdash^{d}_{r} A \\ \Gamma \vdash^{d}_{r} B }{\Gamma \vdash^{d}_{r} A \rightarrow B}
\and
\inferrule[R-rcd]{ \Gamma \vdash^{d}_{r} A }{\Gamma \vdash^{d}_{r} \{ \mathit{l} : A \}}
\and
\inferrule[R-mani]{ \Gamma, [A] \vdash^{d+1}_{r} B }{\Gamma \vdash^{d}_{r} [A] B}
\and
\inferrule[R-all]{ \Gamma, \tstar \vdash^{d+1}_{r} A }{\Gamma \vdash^{d}_{r} \forall A}
\and
\inferrule[R-edef]{ \Gamma \vdash^{d}_{r} \Gamma_{\mathrm{1}} \\ \Gamma \envplus \Gamma_{\mathrm{1}} \vdash^{d + \lvert \Gamma_{\mathrm{1}} \rvert}_{r} A }{\Gamma \vdash^{d}_{r} \Gamma_{\mathrm{1}}, A}
\and
\inferrule[R-tdef]{ \Gamma \vdash^{d}_{r} \Gamma_{\mathrm{1}} \\ \Gamma \envplus \Gamma_{\mathrm{1}} \vdash^{d + \lvert \Gamma_{\mathrm{1}} \rvert}_{r} A }{\Gamma \vdash^{d}_{r} \Gamma_{\mathrm{1}}, [A]}
\and
\inferrule[R-star]{ \Gamma \vdash^{d}_{r} \Gamma_{\mathrm{1}} }{\Gamma \vdash^{d}_{r} \Gamma_{\mathrm{1}}, \tstar}
\and
\inferrule[R-box]{ \Gamma_{\mathrm{1}} \vdash^{0}_{r} A }{\Gamma \vdash^{d}_{r} \Gamma_{\mathrm{1}} \triangleright A}
\end{mathpar}
Here $\lvert \Gamma \rvert$ is the number of type bindings in $\Gamma$.
We define $\Gamma \vdash_{r} A$ as $\Gamma \vdash^{0}_{r} A$, i.e.\ rigidity at
depth $0$.

\paragraph{Extraction} The operation $| \delta |$ (used in \rref{Step-tapp} and \rref{Step-tdef}) extracts the type information from a value environment:
\begin{center}
  $(1) \, | \cdot | = \cdot$ \quad \quad $(2) \, | \delta, v | = | \delta |$ \quad \quad $(3) \, | \delta, [A] | = | \delta |, [A]$
\end{center}

\paragraph{Environment concatenation} The operation $\delta \envplus \delta_{\mathrm{1}}$ (used in \rref{Step-edef} and \rref{Step-tdef}) appends the entries of environment $\delta_{\mathrm{1}}$ to $\delta$. Similar to $\Gamma \envplus \Gamma_{\mathrm{1}}$, it recurses on its right operand $\delta_{\mathrm{1}}$:
\begin{center}
  $(1) \, \delta \envplus \cdot = \delta$ \quad \quad $(2) \, \delta \envplus (\delta_{\mathrm{1}}, v) = (\delta \envplus \delta_{\mathrm{1}}), v$ \quad \quad $(3) \, \delta \envplus (\delta_{\mathrm{1}}, [A]) = (\delta \envplus \delta_{\mathrm{1}}), [A]$
\end{center}

\begin{figure}[t]
  \begin{align*}
  &\text{Expressions}
  &e &\Coloneqq \mathtt{i} ~|~ \mathit{n} ~|~ \lambda e ~|~ e_{\mathrm{1}} \, e_{\mathrm{2}} ~|~ \Lambda e
  ~|~e \, {\_} ~|~ \eclos{\delta}{\lambda e} ~|~ \eclos{\delta}{\Lambda e} ~|~ e_{\mathrm{1}} \triangleright e_{\mathrm{2}}
  ~|~ \cdot ~|~ e_{\mathrm{1}}, e_{\mathrm{2}} \\
  &\text{Environments} &\Delta
  &\Coloneqq \cdot ~|~ \Delta, e \\
  &\text{Values} &v
  &\Coloneqq \mathtt{i} ~|~ \eclos{\delta}{\lambda e} ~|~ \eclos{\delta}{\Lambda e} ~|~ \cdot ~|~ v_{\mathrm{1}}, v_{\mathrm{2}} \\
   &\text{Value environments} &\delta &\Coloneqq \cdot
  ~|~ \delta, v \\
  &\text{Frames} &F &\Coloneqq
  \eframe \, e ~|~ v \, \eframe ~|~ \eframe \triangleright e
  ~|~ \eframe , e
  \end{align*}
  \begin{drulepar}{\delta \vdash e \hookrightarrow_{|\mathsf{i}|} e'}{Type-erasing Semantics}
    \drule{Estep-eval}
      {\delta \vdash e \hookrightarrow_{|\mathsf{i}|} e'}
      {\delta \vdash F [e] \hookrightarrow_{|\mathsf{i}|} F [e']}

    \drule{Estep-var}
      {\delta [\mathit{n}] = v}
      {\delta \vdash \mathit{n} \hookrightarrow_{|\mathsf{i}|} v}

    \drule{Estep-box}
      {\delta_{\mathrm{1}} \vdash e_{\mathrm{2}} \hookrightarrow_{|\mathsf{i}|} e'_{\mathrm{2}}}
      {\delta \vdash \delta_{\mathrm{1}} \triangleright e_{\mathrm{2}} \hookrightarrow_{|\mathsf{i}|} \delta_{\mathrm{1}} \triangleright e'_{\mathrm{2}}}

    \drule{Estep-boxv}
      {}
      {\delta \vdash \delta_{\mathrm{1}} \triangleright v_{\mathrm{2}} \hookrightarrow_{|\mathsf{i}|} v_{\mathrm{2}}}

    \drule{Estep-dummy}
      {}
      {\delta \vdash \eclos{\delta_{\mathrm{1}}}{\Lambda e} \, {\_} \hookrightarrow_{|\mathsf{i}|} \delta_{\mathrm{1}} \triangleright e}

    \drule{Estep-beta}
      {}
      {\delta \vdash \eclos{\delta_{\mathrm{1}}}{\lambda e} \, v_{\mathrm{2}} \hookrightarrow_{|\mathsf{i}|} (\delta_{\mathrm{1}}, v_{\mathrm{2}}) \triangleright e}

    \drule{Estep-clos}
      {}
      {\delta \vdash \lambda e \hookrightarrow_{|\mathsf{i}|} \eclos{\delta}{\lambda e}}

    \drule{Estep-bclos}
      {}
      {\delta \vdash \Lambda e \hookrightarrow_{|\mathsf{i}|} \eclos{\delta}{\Lambda e}}

    \drule{Estep-edef}
      {\delta \envplus \delta_{\mathrm{1}} \vdash e \hookrightarrow_{|\mathsf{i}|} e'}
      {\delta \vdash \delta_{\mathrm{1}}, e \hookrightarrow_{|\mathsf{i}|} \delta_{\mathrm{1}}, e'}
  \end{drulepar}
  \caption{Syntax and reduction rules for the type-erasing semantics.}\label{fig:alt-sem}
\end{figure}

\section{Type-Erasing Semantics of \name}\label{sec:erase}
The syntax and reduction rules for the type-erasing semantics are given in Figure~\ref{fig:alt-sem}.
Type entries are removed from environments and value environments. Moreover, a dummy application $e \, {\_}$ replaces the original type application $e \, A$. Hence, there is no type information in the erased syntax.
To be distinguishable, we use
the subscript $|\mathsf{i}|$ to represent the type-erasing semantics.
The reduction rules for the type-erasing semantics are mostly the same as in the original semantics.
Nevertheless, \rref{Step-tapp} is replaced with~\rref{Estep-dummy}.
In this rule, the dummy application consumes the type abstraction closure, and simply returns a boxed expression $e$ along with the environment $\delta_{\mathrm{1}}$ captured by the closure. Furthermore, \rref{Step-tdef} that performs runtime type computation for environments is removed, since in the erased syntax there is no type entries in environments.

Importantly, we show that the type-erasing semantics reduces to
the same (erased) values as the type-passing semantics.
The erasure function $\bra{e}$ is defined as:

\begin{minipage}{.3\linewidth}
  \centering
  \begin{align*}
    \bra{\mathtt{i}} &= \bra{\mathtt{i}} \\
  \bra{\mathit{n}} &= \bra{\mathit{n}} \\
  \bra{\cdot} &= \bra{\cdot} \\
  \bra{\lambda e} &= \lambda \bra{e}
\end{align*}
\end{minipage}
\begin{minipage}{.3\linewidth}
  \centering
  \begin{align*}
  \bra{e_{\mathrm{1}} \, e_{\mathrm{2}}} &= \bra{e_{\mathrm{1}}} \bra{e_{\mathrm{2}}} \\
  \bra{\Lambda e} &= \Lambda \bra{e} \\
    \bra{\eclos{v}{\lambda e}} &= \eclos{\bra{v}}{\lambda \bra{e}} \\
  \bra{\eclos{v}{\Lambda e}} &= \eclos{\bra{v}}{\Lambda \bra{e}}
\end{align*}
\end{minipage}
\begin{minipage}{.3\linewidth}
  \centering
  \begin{align*}
  \bra{e \, A} &= \bra{e} \, {\_} \\
  \bra{e_{\mathrm{1}}, e_{\mathrm{2}}} &= \bra{e_{\mathrm{1}}}, \bra{e_{\mathrm{2}}} \\
  \bra{e, [A]} &= \bra{e} \\
  \bra{e_{\mathrm{1}} \triangleright e_{\mathrm{2}}} &= \bra{e_{\mathrm{1}}} \triangleright \bra{e_{\mathrm{2}}}
\end{align*}
\end{minipage}

This erasure function is similar to the one used in TAPL~\citep{DBLP:books/daglib/0005958} to show the erasure of System F terms.
In particular, the erasure of type application $\bra{e \, A}$ is defined as $\bra{e} \, {\_}$, which is a dummy application that does not carry any type information.
Moreover, in our setting, we also erase types from environments ($\bra{e, [A]} = \bra{e}$), since environments in \name are first-class and they can contain type entries.

\section{Measures for Deciding Type Equivalence}\label{sup:decide}
To establish the decidability of our type equivalence relation, we define
measures for types and typing contexts, both parameterized by a
measure context defined as $\Psi \Coloneqq \cdot ~|~ \Psi, k ~|~
\Psi, [k]$.
Intuitively, the measure context $\Psi$ keeps track of
the sizes of bindings bound in the typing context, where $k$ is for
term variable bindings and $[k]$ is for type variable
bindings. The definitions of measures are given next.

\begin{definition}[Measures for Decidability] The measure $\size{A}{\Psi}$ for types and $\mkctx{\Gamma}{\Psi}$ for contexts are:
\[
\begin{minipage}{.45\linewidth}
  \centering
  \begin{align*}
  \size{\tint}{\Psi} &= 1 \\
  \size{\mathit{N}}{\Psi} &= \Psi(\mathit{N}) + 1 \\
  \size{A \rightarrow B}{\Psi} &= \size{A}{\Psi} + \size{B}{\Psi} + 1 \\
  \size{\forall A}{\Psi} &= \size{A}{\Psi, [\mathit{1}]} + 1 \\
  \size{\Gamma \triangleright A}{\Psi} &= \size{A}{\mkctx{\Gamma}{\cdot}} + 1
\end{align*}
\end{minipage}%
\begin{minipage}{.45\linewidth}
  \centering
  \begin{align*}
  \size{[A] B}{\Psi} &= \texttt{let } k \Coloneqq \size{A}{\Psi} \texttt{ in } k + \size{B}{\Psi, [k]} + 1 \\
  \size{\cdot}{\Psi} &= 1 \\
  \size{\Gamma, A}{\Psi} &= \size{\Gamma}{\Psi} + \size{A}{\Psi \texttt{+} \mkctx{\Gamma}{\Psi}} + 1 \\
  \size{\Gamma, \tstar}{\Psi} &= \size{\Gamma}{\Psi} + 1 \\
  \size{\Gamma, [A]}{\Psi} &= \size{\Gamma}{\Psi} + \size{A}{\Psi \texttt{+} \mkctx{\Gamma}{\Psi}} + 1
\end{align*}
\end{minipage}
\]
\begin{alignat*}{2}
  \mkctx{(\cdot)}{\Psi} &= \cdot &\qquad
  \mkctx{(\Gamma, A)}{\Psi} &= \texttt{let } k \Coloneqq \size{A}{\Psi \texttt{+} \mkctx{\Gamma}{\Psi}} \texttt{ in } \mkctx{\Gamma}{\Psi}, k \\
  \mkctx{(\Gamma, \tstar)}{\Psi} &= \mkctx{\Gamma}{\Psi}, [\mathit{1}] &\qquad
  \mkctx{(\Gamma, [A])}{\Psi} &= \texttt{let } k \Coloneqq \size{A}{\Psi \texttt{+} \mkctx{\Gamma}{\Psi}} \texttt{ in } \mkctx{\Gamma}{\Psi}, [k]
\end{alignat*}
\end{definition}

The measure for types $\size{A}{\Psi}$ is a natural
number that is the size of type $A$ with respect to the measure context $\Psi$. Moreover, measure $\mkctx{\Gamma}{\Psi}$ is producing a list of natural numbers, which is essentially a measure context. These two measures are defined mutually.

Unlike System F, a context in \name can be used as a first-class environment type, and thus it can depend on the global context. Hence, $\mkctx{\Gamma}{\Psi}$ is equipped with a measure context $\Psi$, which acts as the global measure context. To compute the measure of context $\Gamma, A$ with $\Psi$, the measure of $\Gamma$ is first computed by $\mkctx{\Gamma}{\Psi}$. Moreover, since $A$ can depend on $\Gamma$, the size of $A$ should be computed with $\Psi \envplus \mkctx{\Gamma}{\Psi}$, the concatenation of global measure context and the measure for $\Gamma$. Moreover, for an abstract type variable binding $\tstar$, the measure context is extended with $[\mathit{1}]$. For a concrete type variable binding $[A]$, the measure is computed similarly to term variable bindings, except that the measure context is extended with $[k]$, signaling that the size is for a type variable binding.

The size measure for types $\size{A}{\Psi}$ is defined according to the structure of types. The size of a type variable $\mathit{N}$ is determined by looking up the measure context $\Psi$ at index $\mathit{N}$, which gives the size of the type bound to that type variable. The size of a manifest type $[A] B$ is computed by first computing the size of $A$, which is a natural number $k$, and then computing the size of $B$ with the measure context extended with $[k]$.
Moreover, for a box type $\Gamma \triangleright A$, the size is computed by computing the size of $A$, with respect to the measure context $\mkctx{\Gamma}{\cdot}$ derived from $\Gamma$ with an empty measure context, since the local context $\Gamma$ in a box is closed. Finally, when computing the size of an environment type, the size of a term binding ($\Gamma, A$) or a concrete type binding ($\Gamma, [A]$) is computed with respect to the measure context extended with $\mkctx{\Gamma}{\Psi}$.
\newcommand{\klen}[1]{\lvert #1 \rvert}
\newcommand{\scons}[2]{#1 \mathbin{::} #2}
\newcommand{\stl}{\mathsf{tl}}
\newcommand{\Vint}[1]{\mathcal{V}\bra{#1}}
\newcommand{\Greal}[1]{\mathcal{G}\bra{#1}}
\newcommand{\Ppin}[1]{\mathcal{F}\bra{#1}}

\section{Semantic Typing and Normalization for \name}\label{sec:sem_sound}
We prove normalization of \name through a unary logical
relation that interprets each type as a set of values (a reducibility
candidate). The relation yields a notion of semantic typing. By the
fundamental theorem, every well-typed term is semantically well-typed, and
normalization follows as a corollary. The development is mechanized in Rocq.

A candidate $R$ is a set of values, and serves as the semantic interpretation of types.
A semantic environment $\rho$ (defined as $\mathbb{N}\to R$) assigns a
candidate to each de~Bruijn type variable.
The logical relation relies on two auxiliary operations on candidates and environments: $\scons{R}{\rho} \triangleq \lambda k.\ \text{if } k = 0 \text{ then } R \text{ else } \rho(k{-}1)$, which conses a candidate $R$ at slot $0$ of $\rho$, and $\stl(\rho) \triangleq \lambda k.\ \rho(k{+}1)$, which drops slot $0$ of $\rho$.

\begin{figure}[t]
\small
\[
\begin{array}{l@{\;\;}c@{\;\;}l}
\multicolumn{3}{l}{\textbf{Type interpretation}\quad v \in \Vint{A}_\rho:}\\
\Vint{\tint}_\rho & = & \{\, \mathtt{i} \,\} \quad \text{i.e. set of all integer literals}\\
\Vint{\mathit{N}}_\rho & = & \rho(\mathit{N})\\
\Vint{A \rightarrow B}_\rho & = & \{\, \eclos{\delta}{\lambda e} : \forall v\in\Vint{A}_\rho .\, \exists v' .\,
   (\delta, v) \vdash e \hookrightarrow^* v',\ v'\in\Vint{B}_\rho \,\}\\
\Vint{[A] B}_\rho & = & \{\, v : \exists R.\ (\forall v'.\ v' \in R \Leftrightarrow v'\in\Vint{A}_\rho) \wedge v\in\Vint{B}_{\scons{R}{\rho}} \,\}\\
\Vint{\forall B}_\rho & = & \{\, \eclos{\delta}{\Lambda e} : \forall R,\Gamma, A.\,  \exists v' .\,
   (\delta, [\Gamma \triangleright A]) \vdash e \hookrightarrow^* v' \wedge v'\in\Vint{B}_{\scons{R}{\rho}} \,\}\\
\Vint{\Gamma \triangleright A}_\rho & = & \{\, v : \exists \rho'.\ \Ppin{\Gamma}_{\rho'}\wedge v\in\Vint{A}_{\rho'} \,\} \\
\Vint{\{ \mathit{l} : A \}}_\rho & = & \{\, \{ \mathit{l} = v \} : v\in\Vint{A}_\rho \,\}\\
\Vint{\cdot}_\rho & = & \{\, \cdot \,\}\\
\Vint{\Gamma, A}_\rho & = & \{\, (\delta, v) : \exists \rho'.\  \delta \in \Greal{\Gamma}_{\rho'}
   \wedge (\forall k.\ \rho'(k{+}\klen{\Gamma})=\rho(k)) \wedge v\in\Vint{A}_{\rho'} \,\}\\
\Vint{\Gamma, [A]}_\rho & = & \{\, (\delta, [\Gamma_{\mathrm{1}} \triangleright B]) : \exists \rho'.\
   \delta \in \Greal{\Gamma}_{\rho'} \wedge (\forall k.\ \rho'(k{+}\klen{\Gamma})=\rho(k)) \,\}\\
\Vint{\Gamma, \tstar}_\rho & = & \{\, (\delta, [\Gamma_{\mathrm{1}} \triangleright B]) : \exists \rho'.\
    \delta \in \Greal{\Gamma}_{\rho'} \wedge (\forall k.\ \rho'(k{+}\klen{\Gamma})=\rho(k)) \,\}\\[8pt]

\multicolumn{3}{l}{\textbf{Context interpretation}\quad \delta \in \Greal{\Gamma}_\rho :}\\
\Greal{\cdot}_\rho & = & \{\, \cdot \,\}\\
\Greal{\Gamma, A}_\rho & = & \{\, (\delta, v) : \delta \in \Greal{\Gamma}_\rho \wedge v\in\Vint{A}_\rho \,\}\\
\Greal{\Gamma, [A]}_\rho & = & \{\, (\delta, [\Gamma_{\mathrm{1}} \triangleright B]) : \delta \in \Greal{\Gamma}_{\stl(\rho)}
   \wedge (\forall v.\ v \in \rho(0) \Leftrightarrow v\in\Vint{A}_{\stl(\rho)}) \,\}\\
\Greal{\Gamma, \tstar}_\rho & = & \{\, (\delta, [\Gamma_{\mathrm{1}} \triangleright B]) : \delta \in \Greal{\Gamma}_{\stl(\rho)} \,\}\\[8pt]

\multicolumn{3}{l}{\textbf{Realizer-free interpretation}\quad \Ppin{\Gamma}_\rho :}\\
\Ppin{\cdot}_\rho & = & \mathsf{True} \\
\Ppin{\Gamma, A}_\rho & = & \Ppin{\Gamma}_\rho\\
\Ppin{\Gamma, [A]}_\rho & = & \Ppin{\Gamma}_{\stl(\rho)} \wedge (\forall v.\ v \in \rho(0)\Leftrightarrow v\in\Vint{A}_{\stl(\rho)})\\
\Ppin{\Gamma, \tstar}_\rho & = & \Ppin{\Gamma}_{\stl(\rho)}
\end{array}
\]
\caption{The logical relation, comprising the type interpretation
$v \in \Vint{A}_\rho$, context interpretation $\delta \in \Greal{\Gamma}_\rho$, and realizer-free interpretation $\Ppin{\Gamma}_\rho$.}
\label{fig:lr}
\end{figure}

Figure~\ref{fig:lr} shows the definition of the logical relation.
Judgment $v \in \Vint{A}_\rho$ is the type interpretation: $A$ is a type formed
in a context interpreted by the semantic environment $\rho$.
The interpretation for integers, type variables, and function types is standard. In the clause of manifest types $[A] B$, the type variable $\alpha \, (\triangleq 0)$ is assigned the interpretation $\Vint{A}_\rho$, and the body $B$ is read with $\Vint{A}_\rho$ at slot $0$. In the $\forall B$
clause, a polymorphic value $\eclos{\delta}{\Lambda e}$ is reducible iff, for \emph{every}
candidate $R$, the body produces a value in $\Vint{B}_{\scons{R}{\rho}}$.
Hence, the
abstract variable is interpreted by the arbitrary $R$, and the type argument
actually stored in the runtime environment is irrelevant. In particular, $R$
ranges over the \emph{empty} candidate, so $\hforall{\alpha}{\alpha}$ is uninhabited.

The uninhabitedness of $\hforall{\alpha}{\alpha}$ is why the box clause
$\Vint{\Gamma \triangleright A}_\rho$ is \emph{realizer-free}: rather than demanding a value
environment that realizes the local context $\Gamma$, it reads the body $A$
under some environment $\rho'$ in realizer-free agreement with
$\Gamma$, written $\Ppin{\Gamma}_{\rho'}$. Realizer-free interpretation constrains only
the manifest type entries of $\Gamma$: a manifest binding fixes slot $0$ to the
interpretation of its definition.
For abstract type bindings, the interpretation is unconstrained because of the rigidity requirement: a rigid type $A$ formed in context $\Gamma$ cannot refer to the abstract bindings.
For term bindings, realization demands an inhabitant
and so can fail: type equivalence identifies box types whose local context has no
realizer, for instance $\cdot \vdash (\cdot, \hforall{\alpha}{\alpha}) \triangleright \tint \sim \tint \dashv \cdot$, where
realizing $\cdot, \hforall{\alpha}{\alpha}$ would require a value of the uninhabited
$\hforall{\alpha}{\alpha}$, yet the box type must stay inhabited like $\tint$.
With realizer-free interpretation, such box types whose local context contains uninhabited types remain inhabited.

For an environment type, its own internal type bindings are not entries of the ambient environment $\rho$, so the clauses for environment types existentially provide a combined environment $\rho'$ whose low slots $[0,\klen{\Gamma})$ carry $\Gamma$'s own type bindings (ensured by the context interpretation $\delta \in \Greal{\Gamma}_{\rho'}$) and whose high part is $\rho$ (encoded as $\forall k.\ \rho'(k{+}\klen{\Gamma})=\rho(k)$).
The manifest ($\Gamma, [A]$) and abstract ($\Gamma, \tstar$) clauses are identical: in both, the newly stored entry is a type definition $[\Gamma_{\mathrm{1}} \triangleright B]$, on which the type interpretation imposes no condition. The definition $A$ only fixes how the type variable it introduces is interpreted, and that variable is read only when the environment is used as a context, by the later types formed under it.

Judgment $\delta \in \Greal{\Gamma}_\rho$ is the interpretation for contexts: it holds when the environment $\delta$ realizes $\Gamma$ entry by entry under $\rho$.
The low slots $[0,\klen{\Gamma})$ of $\rho$ hold $\Gamma$'s type bindings, one per slot, and the recursive clauses reach them by recursing under $\stl(\rho)$. Because these slots already carry $\Gamma$'s bindings, a type well-formed in $\Gamma$ is interpreted at this same $\rho$: its de~Bruijn indices into $\Gamma$ point to exactly these slots.

\paragraph{Semantic typing and normalization} With the interpretation of types and contexts, we define semantic typing $\Gamma\models e : A$ as follows: for every value environment $\delta$ that interprets $\Gamma$, the evaluation of $e$ under $\delta$ produces a value in the interpretation of $A$.
\[
\Gamma\models e : A \;\triangleq\; \forall \rho, \delta.\ \delta \in \Greal{\Gamma}_\rho
  \;\Rightarrow\; \exists v'.\ \delta \vdash e \hookrightarrow^* v' \wedge v'\in\Vint{A}_\rho.
\]

We prove the following fundamental theorem, which states that every syntactically well-typed term is also semantically well-typed.

\begin{theorem}[Fundamental Theorem]
If $\Gamma \vdash e : A$ then $\Gamma\models e : A$.
\end{theorem}
The proof is by induction on the typing derivation, with one
compatibility lemma per typing rule.
In particular, because the typing judgment contains a type-equivalence rule, we need to prove the following compatibility lemma, which shows that equivalent types have the same interpretation in the semantic typing:

\begin{lemma}[Soundness of Type Equivalence]
If $\Gamma\models e : A$ and $\Gamma \vdash A \sim B \dashv \Gamma$ then $\Gamma\models e : B$.
\end{lemma}

With the fundamental theorem, we can prove the normalization theorem as a corollary by instantiating the typing judgment with the empty context.

\begin{corollary}[Normalization]
If $\cdot \vdash e : A$, then $\cdot \vdash e \hookrightarrow^* v$ for some value $v$.
\end{corollary}

\section{A Source Language and its Elaboration}\label{sec:source}

This appendix gives the full source language and elaboration outlined in
Section~\ref{subsec:module-elaboration}, formalizing the module constructs
illustrated in Section~\ref{subsec:modules}.
Figure~\ref{fig:source} gives its syntax, the elaboration rules for module expressions
and declarations, and the expansion of merge operands; Theorem~\ref{thm:elab} states
that elaboration preserves typing.

\begin{figure}[t]
\setlength{\jot}{0pt}
\begin{align*}
  &\text{Module expressions} & M &\Coloneqq \mathit{n}.\mathit{l} ~|~ \lambda A.\, M
    ~|~ \Lambda M ~|~ M_1 \, M_2 ~|~ M \, A
    ~|~ \sstruct{D} ~|~ \ssbox{M} ~|~ M_1 + M_2 \\
  &\text{Declarations} & D &\Coloneqq \cdot ~|~ D, \mathit{l} = e
    ~|~ D, [A] ~|~ D, \mathit{l} = M
\end{align*}
\begin{mathpar}
\inferrule[Elab-var]{ \vdash \Gamma \\ \Gamma [\mathit{n}] = A \\ \mathit{l} : B \in A }
  {\selab{\Gamma}{\mathit{n}.\mathit{l}}{\mathit{n}.\mathit{l}}{B}}
\and
\inferrule[Elab-fun]{ \selab{\Gamma, A}{M}{e}{B} }
  {\selab{\Gamma}{\lambda A.\, M}{\lambda e}{A \rightarrow B}}
\and
\inferrule[Elab-tfun]{ \selab{\Gamma, \tstar}{M}{e}{B} }
  {\selab{\Gamma}{\Lambda M}{\Lambda e}{\forall B}}
\and
\inferrule[Elab-app]{ \selab{\Gamma}{M_1}{e_{\mathrm{1}}}{A \rightarrow B} \\
                      \selab{\Gamma}{M_2}{e_{\mathrm{2}}}{A} }
  {\selab{\Gamma}{M_1 \, M_2}{e_{\mathrm{1}} \, e_{\mathrm{2}}}{B}}
\and
\inferrule[Elab-tapp]{ \selab{\Gamma}{M}{e}{\forall B} \\ \Gamma \vdash A }
  {\selab{\Gamma}{M \, A}{e \, A}{[A] B}}
\and
\inferrule[Elab-eq]{ \selab{\Gamma}{M}{e}{A} \\ \Gamma \vdash A \sim B \dashv \Gamma }
  {\selab{\Gamma}{M}{e}{B}}
\and
\inferrule[Elab-box]{ \selab{\cdot}{M}{e}{A} \\ \vdash \Gamma }
  {\selab{\Gamma}{\ssbox{M}}{\cdot \triangleright e}{\cdot \triangleright A}}
\and
\inferrule[Elab-struct]{ \selab{\Gamma}{D}{\Delta}{\Gamma_{\mathrm{1}}} }
  {\selab{\Gamma}{\sstruct{D}}{\Delta}{\Gamma_{\mathrm{1}}}}
\and
\inferrule[Elab-merge]{
  \selab{\Gamma}{M_1}{e_{\mathrm{1}}}{\Gamma_{\mathrm{1}}} \\ \selab{\Gamma}{M_2}{e_{\mathrm{2}}}{\Gamma_{\mathrm{2}}} \\
  \cdot \vdash \Gamma_{\mathrm{1}} \\ \cdot \vdash \Gamma_{\mathrm{2}} \\\\
  \sig{\Gamma_{\mathrm{1}}} \\ \sig{\Gamma_{\mathrm{2}}} \\ \uniq{\Gamma_{\mathrm{1}}} \\ \uniq{\Gamma_{\mathrm{2}}} \\\\
  \Gamma_{\mathrm{3}} = \Gamma, \Gamma_{\mathrm{1}}, \Gamma_{\mathrm{2}} \\
  \sexp{\Gamma_{\mathrm{3}}}{1}{\Gamma_{\mathrm{1}}}{\cdot}{\cdot}{\Gamma_{\mathrm{1}}}{\Delta_{\mathrm{1}}}{\Gamma_{\mathrm{4}}} \\
  \sexp{\Gamma_{\mathrm{3}}}{0}{\Gamma_{\mathrm{2}}}{\Delta_{\mathrm{1}}}{\Gamma_{\mathrm{4}}}{\Gamma_{\mathrm{2}}}{\Delta}{\Gamma_{\mathrm{5}}} }
  {\selab{\Gamma}{M_1 + M_2}{(\lambda\, \lambda\, \Delta) \; e_{\mathrm{1}} \; e_{\mathrm{2}}}{\Gamma_{\mathrm{5}}}}
\end{mathpar}
\begin{mathpar}
\inferrule[Elabd-nil]{ \vdash \Gamma }
  {\selab{\Gamma}{\cdot}{\cdot}{\cdot}}
\and
\inferrule[Elabd-val]{ \selab{\Gamma}{D}{\Delta}{\Gamma_{\mathrm{1}}} \\ \Gamma \envplus \Gamma_{\mathrm{1}} \vdash e : A }
  {\selab{\Gamma}{D, \mathit{l} = e}{\Delta, \{ \mathit{l} = e \}}{\Gamma_{\mathrm{1}}, \{ \mathit{l} : A \}}}
\and
\inferrule[Elabd-typ]{ \selab{\Gamma}{D}{\Delta}{\Gamma_{\mathrm{1}}} \\ \Gamma \envplus \Gamma_{\mathrm{1}} \vdash A }
  {\selab{\Gamma}{D, [A]}{\Delta, [A]}{\Gamma_{\mathrm{1}}, [A]}}
\and
\inferrule[Elabd-mod]{ \selab{\Gamma}{D}{\Delta}{\Gamma_{\mathrm{1}}} \\
                       \selab{\Gamma \envplus \Gamma_{\mathrm{1}}}{M}{e}{A} }
  {\selab{\Gamma}{D, \mathit{l} = M}{\Delta, \{ \mathit{l} = e \}}{\Gamma_{\mathrm{1}}, \{ \mathit{l} : A \}}}
\end{mathpar}
\begin{mathpar}
\inferrule[Exp-nil]{ }
  {\sexp{\Gamma}{k}{\Gamma_{\mathrm{1}}}{\Delta}{\Gamma_{\mathrm{2}}}{\cdot}{\Delta}{\Gamma_{\mathrm{2}}}}
\and
\inferrule[Exp-rcd]{
  \sexp{\Gamma}{k}{\Gamma_{\mathrm{1}}}{\Delta}{\Gamma_{\mathrm{2}}}{\Gamma_{\mathrm{3}}}{\Delta_{\mathrm{1}}}{\Gamma_{\mathrm{4}}} \\
  \mathit{l} : B \in \Gamma_{\mathrm{1}} }
  {\sexp{\Gamma}{k}{\Gamma_{\mathrm{1}}}{\Delta}{\Gamma_{\mathrm{2}}}{\Gamma_{\mathrm{3}}, \{ \mathit{l} : A \}}
        {\Delta_{\mathrm{1}}, \{\, \mathit{l} = (\numv{\Gamma_{\mathrm{4}}} + k).\mathit{l} \,\}}{\Gamma_{\mathrm{4}}, \{ \mathit{l} : B \}}}
\and
\inferrule[Exp-tdef]{
  \sexp{\Gamma}{k}{\Gamma_{\mathrm{1}}}{\Delta}{\Gamma_{\mathrm{2}}}{\Gamma_{\mathrm{3}}}{\Delta_{\mathrm{1}}}{\Gamma_{\mathrm{4}}} \\
  \Gamma \envplus \Gamma_{\mathrm{4}} \vdash A }
  {\sexp{\Gamma}{k}{\Gamma_{\mathrm{1}}}{\Delta}{\Gamma_{\mathrm{2}}}{\Gamma_{\mathrm{3}}, [A]}
        {\Delta_{\mathrm{1}}, [A]}{\Gamma_{\mathrm{4}}, [A]}}
\end{mathpar}
\caption{The source language: its syntax, the elaboration of module expressions
$\selab{\Gamma}{M}{e}{A}$ and of declarations $\selab{\Gamma}{D}{\Delta}{\Gamma_{\mathrm{1}}}$, and the
expansion of a merge operand (from top to bottom).}\label{fig:source}
\end{figure}

\begin{lemma}[Closed Types are Shift-Invariant]
If $\cdot \vdash A$, then ${\uparrow}^{\mathit{N}}\!\!A = A$.
\end{lemma}

\begin{lemma}[Environment Growth]
If $\Gamma [\mathit{n}] = A$ and $\cdot \vdash A$, then $(\Gamma \envplus \Gamma_{\mathrm{1}})[\,\numv{\Gamma_{\mathrm{1}}} + \mathit{n}\,] = A$.
\end{lemma}

\begin{lemma}[Existence of Label Lookup]
If $\decl{\Gamma}{\mathit{l}}{A}$, $\sig{\Gamma}$ and $\uniq{\Gamma}$,
then $\mathit{l} : B \in \Gamma$ for some $B$.
\end{lemma}

\begin{lemma}[Soundness of Expansion]
If $\sexp{\Gamma}{k}{\Gamma_{\mathrm{1}}}{\Delta}{\Gamma_{\mathrm{2}}}{\Gamma_{\mathrm{3}}}{\Delta_{\mathrm{1}}}{\Gamma_{\mathrm{4}}}$,
$\cdot \vdash \Gamma_{\mathrm{1}}$, $\Gamma[\,k\,] = \Gamma_{\mathrm{1}}$ and $\Gamma \vdash \Delta : \Gamma_{\mathrm{2}}$,
then $\Gamma \vdash \Delta_{\mathrm{1}} : \Gamma_{\mathrm{4}}$.
\end{lemma}

\begin{theorem}[Elaboration Preserves Typing]\label{thm:elab}
If $\selab{\Gamma}{M}{e}{A}$, then $\Gamma \vdash e : A$;
and if $\selab{\Gamma}{D}{\Delta}{\Gamma_{\mathrm{1}}}$, then $\Gamma \vdash \Delta : \Gamma_{\mathrm{1}}$.
\end{theorem}

\end{document}